\documentclass[12pt,a4paper]{article}
\usepackage{amssymb}
\usepackage[english]{babel}
\usepackage{amsmath}
\usepackage[fbox]{mathtools} 
\usepackage{empheq}
\usepackage{amsfonts}
\usepackage{amssymb}
\usepackage{graphicx}
\usepackage{framed}
\usepackage{threeparttable}
\usepackage{braket}
\usepackage{framed}
\usepackage{graphics}
\usepackage{esint}
\usepackage[left=2cm,right=2cm,top=2cm,bottom=2cm]{geometry}
\usepackage[numbers,sort&compress]{natbib}
\usepackage{amssymb}
\usepackage{graphicx}
\usepackage{threeparttable}
\usepackage{xcolor}
\usepackage[countmax]{subfloat}
\usepackage{xcolor}
\usepackage{relsize}
\usepackage{mathrsfs}
\usepackage[export]{adjustbox}
\usepackage{float}
\usepackage{imakeidx}
\makeindex[columns=3, title=Alphabetical Index, intoc]
\usepackage{setspace}
\usepackage{empheq}
\usepackage{hyperref}
\usepackage{amsmath}

\title{Comprehensive Analysis of Geometric Phase for SU(3) Representations }
\author{Abhirup Chatterjee\thanks{\texttt{ph24resch11005@iith.ac.in}},\thanks{\texttt{chattopadhyayabhirup65@gmail.com}}\\
 \\ Department of Physics \\ {\it Indian Institute of Technology Hyderabad}\\Dr. Sobhan Kumar Sounda\thanks{\texttt{sobhan.physics@presiuniv.ac.in}},\thanks{\texttt{shounda6@gmail.com}}\\ Department of Physics \\ Presidency University, Kolkata}
\date{November 2025}
\begin{document}
\maketitle
\setcounter{equation}{0}
\renewcommand{\theequation}{2.\arabic{equation}}
\begin{abstract}
Geometric Phase in Quantum Mechanics is generally formulated entirely in terms of geometric structure of the Complex Hilbert Space. We will exploit this fact in case of mixed states for three level open systems undergoing depolarization using the eight dimensional Poincare sphere in the SU(2) Polarisation picture and non unit vector rays in $H\textsuperscript{3}$ within the limit of pure state approach may be found to be in agreement with the Pancharatnam Phase, Berry Phase and Aharonov-Anandan Phase.
\end{abstract}\vspace{0.5cm}\newpage 

\tableofcontents
\newpage
\section{\textbf{Introduction}}
A Pure Quantum State retains a memory of its evolution in terms of Geometric Phase when it undergoes an evolution in the Parameter Space.The phase factor has its origin which is purely geometric in nature and it can arise even under the most general conditions where the system is undergoing an non-unitary evolution corresponding to the Hamiltonian which does not satisfy the criterion of adiabeticity and cyclicity. Although when geometric phase comes into the picture we talk about Berry's framework \cite{berry1984quantal} where he assumed the evolution to be cyclic in some parameter space and the Hamiltonian obeys the adiabeticity and cyclicity condition but Pancharatnam's experimental work \cite{fowles1989introduction} on interference gives us the idea about the existence of such phase factor which is known as pancharatnam connection which states that if we consider any three mutually nonorthogonal vectors in a certain hilbert space say $\ket{\psi_1},\ket{\psi_2} and \ket{\psi_3}$ so
that,$\braket{\psi_m|\psi_n}\neq 0 \forall m,n=1,2,3$ then if $\ket{\psi_1}$ is in phase with $\ket{\psi_2}$ and $\ket{\psi_2}$ is in phase with $\ket{\psi_3}$ then $\ket{\psi_1}$ may not be in phase with $\ket{\psi_3}$.The relative phase difference between any two non-orthogonal vectors is defined as follows.
\\ \indent Let us consider two vectors $\ket{\Psi}$ and $\ket{\Phi}$ such that they have a non zero value of inner product which is in general complex i.e.$\braket{\Psi|\Phi}\neq0$ and any complex number can be represented in the polar coordinate so with $\braket{\Psi|\Phi}=re^{i\theta}$ and here $\theta$ denotes the relative phase difference between the vectors. Now in Pancharatnams framework the three vectors were the three different electric fields $\Vec{E_1},\Vec{E_2},\Vec{E_3}$ such that $\Vec{E_i}\cdot\Vec{E_j}\neq0,\text{ where }i\neq j$ say, chosen for the interferrometry experiment and experimentally the Pancharatnams connection was found with the fact that when the two electric fields are in phase with each other it will correspond to the interference maximum and when the relative phase difference between them be $\pi$ it will correspond to the minimum in the interference pattern i.e. the intensities will be maximum and minimum when they are in phase or out of phase.Apparently there is no quantum mechanics or Schrodinger equation is involved in the Pancharatnams framework as it comes form the result of experimental classical optics but it has a connection in the Three dimensional Poincare sphere representation of the manifold. But after Berry's discovery of the geometric phase lots of attempts has been done to generalize the idea of the geometric phase and Aharonov and Anandan \cite{aharonov1987phase} showed that if we lift up the condition of adiabeticity still we can define the geometric phase and the total phase can be written as a sum of the geometric phase and dynamical phase keeping the condition of cyclical evolution of the quantum system and the geometric phase in their framework is called Aharonov-Anandan Phase.
\\ \indent After Aharonov Anandan's discovery Samuel and Bhandari \cite{samuel1988general} further relaxed the condition of the cyclicity as the condition of adiabaticity has already been lifted in Aharonov-Anandans framework and gives the definition of the geometric phase for the non cyclic and non adiabatic evolution of the quantum system.All the approaches of the generalization of the Geometric phase requires the time dependent Schrodinger equation so these are known as the differential approach but Mukunda and simon \cite{mukunda1993quantum} defined the geometric phase from the kinematic point of view which does not require the schrodinger equation hence the evolution of the quantum system can be non-unitary, thereby proving that the geometric phase is a Ray space quantity  and showed that for three arbitrary non-orthogonal  state vectors $\ket{\phi_1},\ket{\phi_2},\ket{\phi_3}$ the non-vanishing geometric phase is given by the Bargmann Invariant of order 3 abbreviated by BI(3).
Hence, $\Phi_{geo}=-arg(\Delta_3)=-arg[Tr(\rho_1\rho_2\rho_3)]=-arg[(\phi_1,\phi_2)(\phi_2,\phi_3)(\phi_3,\phi_1)]\neq0$ with $\rho_1,\rho_2,\rho_3$ are the pure state density matrices corresponding to the states  $\ket{\phi_1},\ket{\phi_2},\ket{\phi_3}$
and by definition of the density or projection operators we have $\rho_m=\ket{\phi_m}\bra{\phi_m}\forall{m=1,2,3}$ and $Tr(\rho_m)=1 \forall m=1,2,3$.So far all the calculations on the geometric phase has been done for the pure states now it can be generalized for the mixed states too.In the present context of the project work we try to find out an expression of the Geometric phase for a three level open quantum system by the SU(3) representations using the fact that the mixed states lies on the interior of the eight dimensional Bloch sphere.Previously progress has been made on the calculation of the geometric phases for two level open quantum systems and there are works on geometric phases for mixed states using the Schmidt's purification method which lifts the mixed states to pure states.
\\ \indent Geometric phase for the three level quantum system of the mixed states can be defined by establishing in connecting the density matrix with the non-unit vector ray in a three dimensional complex Hilbert space. Because the Geometric Phase depends only on the smooth curve on this space,it is formulated entirely in terms of geometric structures.Under the limiting pure state, our approach is in agreement with the Berry's phase, Pancharatnam phase and Aharonov and Anandan Phase.We find that Berry phase of mixed state correlated to population inversions of three level open quantum system \cite{jiang2010geometric}. More precisely,the amplitudes of wave functions are mapped onto given points on the Poincare sphere for a pure state \cite{khanna1997geometric}.

  Application of geometric phases in quantum computation \cite{wang2009geometric} has motivated their studies under more realistic situations \cite{uhlmann1986parallel} \cite{tong2004kinematic}. In a real system, it is unavoidable interaction of a quantum system with its surrounding environment. The interaction may lead to an irreversible loss of information on the system so the process limits the ability to maintain pure quantum states in quantum information. Therefore, it is necessary to include the effect of decoherence. Up to now,
however, the definition of the geometric phase for the open
system is still a controversial issue. It is therefore extremely
important to understand all aspects of the geometric phase in
open system \cite{filipp2009experimental}.
\par\indent It is known that a d-dimensional qudit possesses a much more complex but richer structure than the ordinary qubit. In quantum information science, thus, information processing tasks can not only be implemented using two-dimensional qubits but also are sometimes more efficiently performed using the qudits as carriers of information \cite{howell2002experimental}. It has been found that the qudits are better adapted for certain purposes, such as quantum cryptography.
\\ \indent It is known that there have been many proposals tackling
the geometric phase of two-level open systems from different
generalizations of the parallel transport condition \cite{dattoli1990geometrical}. However, it may be difficult to expand it to the three-level open system. The generalizations especially are not unique so as to give out different results. In addition, a general belief is that the Berry phases are geometric in their nature, i.e., proportional to the solid angle enclosed by the closed curve in parameter space. Therefore, we will express the geometric phase \cite{wang2007effects} for the three-level open system in terms of geometric structures on a three-dimensional complex Hilbert space.
\\ \indent We have calculated the geometric phase for a Three level quantum system incorporating the effect of \textbf{Depolarization} in the SU(2) Polarisation Picture. Polarization of light is a key concept that has deserved a lot of attention over the years. Apart from its fundamental significance, it is also of interest in several active technological fields.One of the interesting phenomena is the Depolarization. When a polarised electromagnetic beam propagates through a certain medium due to the interaction of the beam with its surrounding's the degree of polarization of the polarised electromagnetic beam decreases and this is known as Depolarization \cite{brosseau1998fundamentals}.
In classical optics, this depolarization is ascribed, broadly
speaking, either to birefringence (as it usually happens, e.g., in
optical fibers \cite{karlsson1998polarization}) or to scattering by randomly distributed particles \cite{hulst1981light}. In both cases the net result is an effective anisotropy that leads to a decorrelation of the phases of the electric field vector. 

\indent In quantum optics, a sensible approach to deal with this
decorrelation is through the notion of decoherence, by which
we loosely understand the appearance of irreversible and uncontrollable quantum correlations when a system interacts
with its environment \cite{zurek2003decoherence}. Usually, decoherence is accompanied by dissipation, i.e., a net exchange of energy with the
environment. However, we are interested in the case of pure
decoherence (also known as dephasing), for which the process of energy dissipation is negligible. 

\indent The purpose of this project work is to address the SU(2) Polarization picture along with the  quantum theory of SU(2) Depolarization and to model the depolarization dynamics by Lindblad type master equation, to solve the master equation for few photonic systems and the calculation of the geometric phase.
First we will discuss very fundamental model for the description of the Depolarization along with its drawbacks and the alternative approaches to describe the depolarization.

\section{\textbf{Generalized Bloch Sphere representation of three level systems}}
Because a higher-dimensional Hilbert space may associate with quantum cryptography and entanglement, an increasing interest is to study the geometric phase of the three-level open system. Now,in order to deal with the quantum states (both, the pure and the mixed states) for a three level quantum system we need to represent them in a suitable manner i.e. some sort of mapping is required just like in classical polarisation optics the completely polarised states are mapped to the surface of the Poincare sphere and the partially polarised states are being mapped in the interior of the Poincare sphere.
\\ \indent So, in order to understand the representation of the states for the three level Quantum system we first introduce the idea  for the representation of the pure and mixed states of a two level quantum system using a three dimensional Bloch Sphere and then extend this idea for the three level system, roughly speaking, the Bloch sphere with radius unity is sometimes called the Poincare Sphere; the terminology is being used in the context of polarisation optics. In the present context we will call it a Bloch sphere. In, the Classical Polarisation optics the  idea behind that is to map the polarisation states of light on the surface of the sphere, the polarisation states with degree of polarisation being unity are mapped on the surface of the sphere and the partially polarised states being mapped inside the sphere so for a sphere of radius unity all the normalised completely polarised states can be represented by using the polar angle $\theta\in[0,\pi]$ and azimuthal angle $\phi\in[0,2\pi]$ only in a three dimensional Spherical polar Coordinate System. For the description of the partially polarised light we require three variables $r,\theta,\phi$ respectively.

\subsection{Bloch Sphere Representation of the Two-Level Quantum System}
\setcounter{equation}{0}
\renewcommand{\theequation}{2.1.\arabic{equation}}
Now, let us do the similar thing as discussed above in the context of Classical Polarisation Optics for the description of the polarised states using Quantum mechanics  for a two level quantum system and the most simple system will be a Qubit.A qubit is defined as a two level quantum system having two states represented as,$\ket{0}$ and $\ket{1}$ which forms the orthonormal basis of the underlying two dimensional Hilbert Space $\mathcal{H}^2$ and can be represented by $2\times1$ column matrices so that the general state of the system $\ket{\psi}$ can be written as a linear combination of $\ket{0}=\begin{pmatrix} 1 \\ 0\end{pmatrix}$ and $\ket{1}=\begin{pmatrix} 0 \\ 1\end{pmatrix}$ with $\braket{m|n}=\delta_{mn}\forall m,n=1,2$.So we have,
\begin{equation}
\ket{\psi}=\alpha\ket{0}+\beta\ket{1};\alpha,\beta\in\mathbb{C} 
\end{equation}
Let us write, $\alpha=a_1+ib_1$ and $\beta=a_2+ib_2$,where $a_1,a_2,b_1,b_2\in\Re$ the normalaisation of the state vector $\ket{\psi}$ i.e. $\braket{\psi|\psi}=1$ requires $|\alpha|^2+|\beta|^2=a_1^2+a_2^2+a_3^2+a_4^2=1$.It will be better to use the polar coordinate representation of the complex numbers $\alpha,\beta\in\mathbb{C}$ so that, $\alpha=r_0e^{i\theta_0}$ and $\beta=r_1e^{i\theta_1}$ we get,
\begin{equation}
\begin{split}
\ket{\psi}&=r_0e^{i\theta_0}\ket{0}+r_1e^{i\theta_1}\ket{1} \\
&= e^{i\theta_0}\{r_0\ket{0}+r_1e^{i(\theta_1-\theta_0)}\ket{1}\} \\
&= e^{i\theta_0}\{r_0\ket{0}+r_1e^{i\phi}\ket{1}\}, [\because(\theta_1-\theta_0)=\phi]
\end{split}
\end{equation}
As, the two states differing by a phase factor does not represent two different physical states we can write,
\begin{equation}
\ket{\psi}=r_0\ket{0}+r_1e^{i\phi}\ket{1}
\end{equation}
the normalaisation of the state vector requires $|\alpha|^2+|\beta|^2=r_0^2+r_1^2=1$. One of the possible choice of $r_0$ and $r_1$ that will satisfy the condition is $r_0=\sin{\frac{\theta}{2}}$ and $r_1=\cos{\frac{\theta}{2}}$ such that $r_0^2+r_1^2=1$.With this substitutions we get,
\begin{equation}
\ket{\psi}=\cos{\frac{\theta}{2}}\ket{0}+e^{i\phi}\sin{\frac{\theta}{2}},[0\leq\theta\leq\pi,0\leq\phi\leq2\pi]
\end{equation}
in the above equation $\theta$ and $\phi$ are the usual polar and azimuthal angle for the spherical polar coordinate system.So, we can represent the general state of the qubit by using a sphere of radius unity where all the pure states i.e. $\ket{\psi}$ can be mapped to the surface of the 3 dimensional sphere of unit radius.
\\ \indent We know that the basis states of the qubit are the pure states and any state which can be written as a linear combination of the basis states will also be a pure state and they will lie on the surface of the 3 dimensional sphere also, known as \textbf{Bloch Sphere} which can be represented by the pair of angular coordinates $\theta,\phi$. Let, us consider some simple situations.
For, $\theta=0$ and $\phi=0$ we get, $\ket{\psi}=\ket{0}$ ; corresponds to the state $\ket{0}$ being mapped to the north pole. Similarly,for $\theta=\pi$ and $\phi=0$ we get, $\ket{\psi}=\ket{1}$ ; corresponds to the state $\ket{1}$ being mapped to the diametrically opposite position i.e. south pole and let us also check that for $\theta=\frac{\pi}{2}$ and $\phi=0$ we get, $\ket{\psi}=\frac{1}{\sqrt{2}}(\ket{0}+\ket{1})=\ket{+}$ which corresponds to the homogeneous linear superposition of the basis states being mapped at the equator.
But, here we are dealing so far with the pure states which lies on the surface of the Bloch Sphere but it is known that the mixed states for the two level quantum system can be mapped to the interior region of the Bloch sphere. But there is a problem that the mixed states can not be represented as a linear combination of the pure states so the alternative approach is to use the density operator to represent the mixed states.First we introduce the Density Operator and Density Matrix Formalism required to represent the Mixed states of a Quantum system but it can be equally applicable for the pure states too. 
\pagebreak
\subsubsection{\textbf{Density Matrix and the Density Operator Formalism}}\label{sec:2.1.1}
Let us consider a system and large number of virtual copies of the  system which constitutes an ensemble.The time dependent schrodinger equation for the $k$th member of the ensemble is given by,
\begin{eqnarray}
\hat{H}\ket{\psi^{(k)}(t)}=i\hbar\frac{\partial}{\partial{t}}\ket{\psi^{(k)}(t)}\nonumber \\
\ket{\psi^k(t)}=\sum_{n}a^k_n(t)\ket{\phi_n}
\end{eqnarray}
here, $\ket{\phi_n}$ be the orthonormal basis and can be choosen as the eigenbasis of the hamiltonian satisfying the condition $\braket{\phi_m|\phi_n}=\delta_{mn}\forall{m,n}$ and $\sum_n\ket{\phi_n}\bra{\phi_n}=I$. Then the matrix elements of the density operator and the operator itself is defined as follows,
\begin{eqnarray}
\rho_{mn}(t)=\frac{1}{N}\sum_{k=1}^{N}a^{k}_{m}(t)a^{k*}_{n}(t)\nonumber \\
\hat{\rho}=\frac{1}{N}\sum_k\ket{\psi^{(k)}(t)}\bra{\psi^{(k)}(t)}
\end{eqnarray}
so, we can define the matrix element of the density operator $\hat{\rho}$ in the eigenbasis of the Hamiltonian  as, $\rho_{mn}=\braket{\phi_m|\hat{\rho}|\phi_n}$ and exploiting the completeness relation of the normalised eigenbasis of the hamiltonian we van write,
\begin{equation}
\begin{split}
\hat{\rho}&=\hat{I}\hat{\rho}\hat{I} \\
&= \sum_m\ket{\phi_m}\bra{\phi_m}\hat{\rho}\sum_n\ket{\phi_n}\bra{\phi_n} \\
&= \sum_m\sum_n\braket{\phi_m|\hat{\rho}|\phi_n}\ket{\phi_m}\bra{\phi_n} \\
&=\sum_m\sum_n\rho_{mn}\ket{\phi_m}\bra{\phi_n}
\end{split}
\end{equation}
where,$\rho_{mn}=\braket{\phi_m|\hat{\rho}|\phi_n}$ defines  the matrix elements of the density operator.
The time evolution of the density operator is governed by the Liouville equation of Quantum Statistical Mechanics given by,
\begin{equation}
i\hbar\frac{\partial}{\partial{t}}\hat{\rho}(t)=[\hat{\rho},\hat{H}]
\end{equation}
in statistical equilibrium $\frac{\partial{\rho}}{\partial{t}}=\Dot{\rho}=0\Rightarrow{[\hat{\rho},\hat{H}]=0}$ so,the density operator and the Hamiltonian has the simultaneous eigenstates i.e.$\{\ket{\phi_n}\}$ which allows us to write $\rho_{mn}=\braket{\phi_m|\hat{\rho}|\phi_n}=\rho_n\braket{\phi_m|\phi_n}=\rho_n\delta_{mn}$ where, $\rho_n$ be the eigenvalues of the density operator and the density matrix will be diagonal in its own eigenbasis. It is easy to check that the density operator is hermitian i.e.$\hat{\rho}^\dag=\hat{\rho}$ So, finally we have,
\begin{equation}
\hat{\rho}=\sum_n\rho_n\ket{\phi_n}\bra{\phi_n}
\end{equation}
The expectation value of any observable $\hat{A}$ represented by a hermitian operator can be expressed as,
\begin{equation}
\begin{split}
\braket{\hat{A}}&=\frac{1}{N}\sum_{k=1}^N\braket{\psi
^{(k)}(t)|\hat{A}|\psi^{(k)}(t)} \\ 
&= \frac{Tr[\hat{\rho}\hat{A}]}{Tr[\hat{\rho}]}
\end{split}
\end{equation}
\\ \indent The above process is called the double averaging process in quantum statistical mechanics i.e. the quantum mechanical expectation value followed by the classical ensemble average.For simplicity if we assume the state vector to be normalized i.e. $\braket{\psi^{(k)}(t)|\psi^{(k)}(t)}=1$ then $Tr[\hat{\rho}]=1$ which leads to the form,
\begin{equation}
\braket{\hat{A}}=Tr[\hat{\rho}\hat{A}]
\end{equation}
The above equation gives the closed mathematical form of the Density operator corresponding to the mixed state and as a special case of this we can obtain the density of state corresponding to a pure state $\ket{\phi}$ say, $\hat{\rho}(\phi)=\ket{\phi}\bra{\phi}=\phi\phi^\dag$ and if $\ket{\phi}$ be normalized then the pure state density operator will satisfy $\hat{\rho}^2=\hat{\rho}$, $\hat{\rho}^\dag=\hat{\rho}$ and $Tr[\hat{\rho}]=1$.
Now, for a two level quantum system the density matrix will be of dimension $2\times2$ hermitian matrix with $Tr[\hat{\rho}]=1$ and if it corresponds to a pure state then $\hat{\rho}^2=\hat{\rho}$.
\\ \indent As we know that the Generators of the $SU(2)$ are the Pauli spin matrices which are hermitian and traceless i.e. $\sigma_i^\dag=\sigma_i,i=x,y,z$ and $Tr[\sigma_i]=0$ denoted by $\hat{\Vec{\sigma}}$ having three components $\sigma_i,i=x,y,z$ spans over the two dimensional matrix spaces then any $2\times2$ traceless hermitian matrix can be written as a linear combination of the three pauli matrices. Any $2\times2$ hermitian matrix can be written as a linear combination of the three pauli matrices and one identity matrix $I_{2\times2}$. The density matrix corresponding to the two level quantum system being hermitian  can be written as a linear combination of one $2\times2$ Identity matrix i.e. $I_{2\times2}$ and the three Pauli Matrices $\sigma_i$.
The pauli Matrices are given by,
\begin{equation}
\sigma_x=\begin{pmatrix} 0 & 1\\ 1 & 0\end{pmatrix},\sigma_y=\begin{pmatrix} 0 & -i\\ i & 0\end{pmatrix},\sigma_z=\begin{pmatrix} 1 & 0\\ 0 & -1\end{pmatrix}
\end{equation}
So, we can write that,
\begin{equation}
\label{eq:2.1.13}
\begin{split}
\hat{\rho}&=A\hat{I}+\Vec{B}\cdot\hat{\Vec{\sigma}} \\
&= A\hat{I}+B_i\hat{\sigma}_i
\end{split}
\end{equation}
Now, recalling the Lie algebra of the $SU(2)$ we can write the following properties of the pauli matrices $\sigma_x,\sigma_y,\sigma_z$ as follows,
\begin{eqnarray}
[\hat{\sigma}_i,\hat{\sigma}_j]=\hat{\sigma}_i\hat{\sigma}_j-\hat{\sigma}_j\hat{\sigma}_i=2i\epsilon_{ijk}\hat{\sigma}_k\nonumber \\
\{\hat{\sigma}_i,\hat{\sigma}_j\}=\hat{\sigma}_i\hat{\sigma}_j+\hat{\sigma}_j\hat{\sigma}_i=2\delta_{ij}\nonumber \\
\hat{\sigma}_i\hat{\sigma}_j=i\epsilon_{ijk}\hat{\sigma}_k+\delta_{ij}\forall i,j=1,2,3\nonumber \\ 
Tr[\hat{\sigma}_i\hat{\sigma}_j]=2\delta_{ij}\forall i,j=1,2,3
\end{eqnarray}
Using the above mentioned properties of the pauli matrices we get,
\begin{eqnarray}
\hat{\rho}=A\hat{I}+B_i\hat{\sigma}_i\nonumber \\
Tr[\hat{\rho}]=ATr[\hat{I}]+B_iTr[\hat{\sigma}_i]\Rightarrow{A=\frac{1}{2}}
\end{eqnarray}
and,
\begin{eqnarray}
\hat{\rho}\hat{\sigma}_j=A\hat{\sigma}_j+B_i\hat{\sigma}_i\hat{\sigma}_j\nonumber \\
Tr[\hat{\rho}\hat{\sigma}_j]=A Tr[\hat{\sigma}_j]+B_iTr[\hat{\sigma}_i\hat{\sigma}_j]\nonumber \\
Tr[\hat{\rho}\hat{\sigma}_j]= 2B_i\delta_{ij}\nonumber \\
Tr[\hat{\rho}\hat{\sigma}_j]= 2B_j\Rightarrow{B_i=\frac{1}{2}Tr[\hat{\rho}\hat{\sigma}_i]}
\end{eqnarray}
Putting the values of $A$ and $B_i$ in the equation~[\ref{eq:2.1.13}] we get,
\begin{equation}
\label{eq:2.1.17}
\begin{split}
\rho&=A\hat{I}+B_i\hat{\sigma}_i \\
&= \frac{1}{2}\hat{I}+\frac{1}{2}Tr(\hat{\rho}\hat{\sigma}_i)\hat{\sigma}_i \\
&= \frac{1}{2}\Big(\hat{I}+Tr(\hat{\rho}\hat{\sigma}_i)\hat{\sigma}_i\Big) \\
&= \frac{1}{2}\Big(\hat{I}+\Vec{n}\cdot\hat{\Vec{\sigma}}\Big)
\end{split}
\end{equation}
where, we have defined the three dimensional  Bloch Vector $\Vec{n}\in\mathcal{R}^3$ having three components enlisted as, $n_i=Tr[\hat{\rho}\hat{\sigma}_i],\text{ where },{i=1,2,3}$.If we represent a pure state by the density operator $\rho$ then it can be easily shown that $n_1^2+n_2^2+n_3^2=1$. Let us check using the properties of the pauli matrices as follows;
\begin{equation}
\begin{split}
\hat{\rho}^2&=\hat{\rho}\hat{\rho} \\
&= \frac{1}{2}\Big(\hat{I}+\Vec{n}\cdot\hat{\Vec{\sigma}}\Big)\frac{1}{2}\Big(\hat{I}+\Vec{n}\cdot\hat{\Vec{\sigma}}\Big) \\ 
&= \frac{1}{4}\Big(\hat{I}+\Vec{n}\cdot\Vec{\sigma}+\Vec{n}\cdot\Vec{\sigma}+(\Vec{n}\cdot\Vec{\sigma})^2\Big) \\
&= \frac{1}{4}\Big(\hat{I}+n_i\hat{\sigma}_i+n_i\hat{\sigma}_i+n_in_j\hat{\sigma}_i\hat{\sigma}_j\Big) \\
&= \frac{1}{4}\Big(\hat{I}+2n_i\hat{\sigma}_i+n_in_j\hat{\sigma}_i\hat{\sigma}_j\Big) \\
&= \frac{1}{4}\Big(\hat{I}+2n_i\hat{\sigma}_i+\Big[n_in_j\delta_{ij}+i\epsilon_{ijk}n_in_j\hat{\sigma_k}\Big]\Big) \\
&= \frac{1}{4}\Big(\hat{I}+2n_i\hat{\sigma}_i+\Big[\sum_in_in_i+i(n\times n)_k\hat{\sigma}_k\Big]\Big) \\
&= \frac{1}{4}\Big(\hat{I}+2n_i\hat{\sigma}_i+\sum_in_i^2\Big) 
\end{split}
\end{equation}
Now, in order to have $\hat{\rho}^2=\hat{\rho}$ we must have $\sum _in_i^2=1$, the maximum number of independent Bloch Vector components are 2 but the equation~[\ref{eq:2.1.17}] holds for both the pure and Mixed states.
using the fact that $=\braket{\hat{A}}=Tr[\hat{\rho}\hat{A}]$; for a pure state $\hat{\rho}=\ket{\psi}\bra{\psi}$ the components of the bloch vector can be written as,
\begin{equation}
n_i=Tr[\rho\sigma_i]=\braket{\sigma_i}=\braket{\psi|\sigma_i|\psi};\hspace{0.1cm}\forall{i=1,2,3}
\end{equation}
\subsubsection{\textbf{Extension for the Three Level Quantum System}}
The next task is to generalize the previous idea for a three level Quantum system. The similar idea of the density operator and density matrix formalism works well in this context and the density matrix $\hat{\rho}$ for the three level system will be a $3\times3$ hermitian matrix and just like the case where any $2\times2$ hermitian matrix can be written as a linear combination of an $2\times2$ unit matrix i.e. $I_{2\times2}$ and three pauli matrices which are the generators of the $SU(2)$ group, similarly any $3\times3$ hermitian matrix can be written as linear combination of an $3\times3$ unit matrix i.e. $I_{3\times3}$ and the eight traceless hermitian  Gell-Mann Matrices $\lambda_{i}$ i=1,2,..,8 which are the generators of the $SU(3)$ group.\vspace{0.2cm}
The Gell-Mann Matrices $(\lambda_i)$ are given by,
\[\lambda_1=\begin{pmatrix}0 & 1 & 0\\1 & 0 & 0\\0 & 0 & 0\end{pmatrix} ; \lambda_2=\begin{pmatrix}0 & -i & 0\\i & 0 & 0\\0 & 0 & 0\end{pmatrix} ; \lambda_3=\begin{pmatrix}1 & 0 & 0\\0 & -1 & 0\\0 & 0 & 0\end{pmatrix} ;\lambda_4=\begin{pmatrix}0 & 0 & 1\\0 & 0 & 0\\1 & 0 & 0\end{pmatrix}\]
\[\lambda_5=\begin{pmatrix}0 & 0 & -i\\0 & 0 & 0\\i & 0 & 0\end{pmatrix} ; \lambda_6=\begin{pmatrix}0 & 0 & 0\\0 & 0 & 1\\0 & 1 & 0\end{pmatrix} ; \lambda_7=\begin{pmatrix}0 & 0 & 0\\0 & 0 & -i\\0 & i & 0\end{pmatrix} ; \lambda_8=\frac{1}{\sqrt{3}}\begin{pmatrix}1 & 0 & 0\\0 & 1 & 0\\0 & 0 & -2\end{pmatrix}\]
Some important properties of $\lambda_i$'s ; the generators of  $SU(3)$ are listed below,
\begin{gather}
Tr[\lambda_i]=0\nonumber \\
[\lambda_r,\lambda_s]=2if_{rst}\lambda_t\nonumber \\
f_{123}=1,f_{458}=f_{678}=\frac{\sqrt{3}}{2},f_{147}=f_{246}=f_{257}=f_{345}=f_{516}=f_{637}=\frac{1}{2};\nonumber \\
\{\lambda_r,\lambda_s\}=\frac{4}{3}\delta_{rs}+2d_{rst}\lambda_{t}\nonumber \\
d_{118}=d_{228}=d_{338}=-d_{888}=\frac{1}{\sqrt{3}},d_{448}=d_{558} =d_{668}=d_{778}=-\frac{1}{2\sqrt{3}},\nonumber \\
d_{146}=d_{157}=-d_{247}=d_{256}=d_{344}=d_{355}=-d_{366}=-d_{377}=\frac{1}{2};\nonumber \\
\lambda_r\lambda_s=\frac{2}{3}\delta_{rs}+if_{rst}\lambda_{t}+d_{rst}\lambda_{t},Tr[\lambda_r\lambda_s]=2\delta_{rs}
\end{gather}
Their commutators, anti-commutators and products involve two different three-index symbols\footnote{The commutation relation of the Pauli matrices involves only a single type of antisymmetric tensor i.e.levi-civita but in case of SU(3) it involves one completely symmetric and one completely antisymmetric tensor.} or invariant tensors respectively $f_{ijk},d_{rst},\delta_{rs}$.Only the independent components of the completely anti-symmetric f's and the completely symmetric d's have been listed above so that, $d_{ijk}=d_{jik}$ which is completely symmetric and $f_{ijk}=-f_{jik}$ which is completely antisymmetric and the kronecker delta $\delta_{rs}$ is completely symmetric unlike the case of the pauli matrices whose commutation and anti-commutation relations only involve a completely antisymmetric levicivita symbol i.e. $\epsilon_{ijk}$ or, an invariant tensor of rank 3.
So, we can write the density matrix as follows,
\begin{equation}
\hat{\rho}=A\hat{I}+\Vec{B}\cdot\hat{\Vec{\lambda}}
\end{equation}
Here, the unknown quantities i.e. $A$ is a constant and $\Vec{B}$ is an octet vector having eight components are yet to be determined and this can be done by using the above trace relations as follows,
\begin{gather}
\hat{\rho}=A\hat{I}+B_{i}\hat{\lambda}_{i}\nonumber \\
Tr[\hat{\rho}]=ATr[\hat{I}]+B_iTr[\hat{\lambda}_{i}]\nonumber \\
Tr[\hat{\rho}]=1=3A\Rightarrow{A=\frac{1}{3}}\nonumber \\
Tr[\hat{\rho}\hat{\lambda}_j]=ATr[\hat{\lambda}_j]+B_iTr[\hat{\lambda}_i\hat{\lambda}_j]=2B_iTr[\delta_{ij}]=2B_i\delta_{ij}=2B_j\Rightarrow{B_i=\frac{1}{2}Tr[\hat{\rho}\hat{\lambda}_i]}\nonumber \\
\hat{\rho}=\frac{1}{3}\hat{I}+\frac{1}{2}Tr[\hat{\rho}\hat{\lambda}_i]\hat{\lambda}_i\nonumber \\
\hat{\rho}=\frac{1}{\sqrt{3}}\Big(I+\frac{\sqrt{3}}{2}\Vec{n}\cdot\hat{\Vec{\lambda}}\Big)
\end{gather}
where, we have defined the eight dimensional Bloch Vector $\Vec{n}\in\mathbb{R}^8$ having eight components $n_i$,i=1,2,..,8 defined as, $n_i=\frac{\sqrt{3}}{2}Tr[\hat{\rho}\hat{\lambda}_i]$ along with $n_i^*=n_i,n_i\in\mathbb{R}$.The equation of the eight dimensional Bloch sphere is defined as $\Vec{n}\cdot\Vec{n}=\sum_{i}n_i^2=r^2\leq1$ where, $r$ being the radius of the Bloch sphere.
In short we can define a finite set of cardinality 8 such that,
\begin{equation}
\mathcal{S}=\{\textbf{n}\in\mathbb{R}^8|\Vec{\textbf{n}}\cdot\Vec{\textbf{n}}=\textbf{r}^2,\textbf{n}^*=\textbf{n}\},
\end{equation}
is an analog of the generalized Bloch Sphere with eight dimensions for the three-level system.
All the pure states being mapped to the surface of the Bloch sphere and we know that the mixed states will lie in the interior of the eight dimensional Bloch sphere\footnote{In the limit of pure states the points lying inside the sphere will shift towards the surface but its still a problem that where the quantum states will mapp which lie in between the completely pure and completely mixed states.}
Now, we can easily enumerate the components of the Bloch Vectors using equation(2.1.22) and can find the equation of the Bloch Sphere.The components are as follows,
Let us assume the $3\times3$ density matrix having elements $\rho_{ij};\hspace{0.1cm}\forall{(i,j)=1(1)3}$ so that,
\[\hat{\rho}=\begin{pmatrix} \rho_{11} & \rho_{12} & \rho_{13} \\ \rho_{21} & \rho_{22} & \rho_{23} \\ \rho_{31} & \rho_{32} & \rho_{33} \end{pmatrix}\]
\begin{gather}
\label{eq:2.1.24}
n_1=\frac{\sqrt{3}}{2}Tr[\hat{\rho}\hat{\lambda}_1]=\frac{\sqrt{3}}{2}Tr\Bigg[\begin{pmatrix} \rho_{11} & \rho_{12} & \rho_{13} \\ \rho_{21} & \rho_{22} & \rho_{23} \\ \rho_{31} & \rho_{32} & \rho_{33} \end{pmatrix}\begin{pmatrix}0 & 1 & 0\\1 & 0 & 0\\0 & 0 & 0\end{pmatrix}\Bigg]=\frac{\sqrt{3}}{2}(\rho_{12}+\rho_{21})\nonumber \\
n_2=\frac{\sqrt{3}}{2}Tr[\hat{\rho}\hat{\lambda}_2]=\frac{\sqrt{3}}{2}Tr\Bigg[\begin{pmatrix} \rho_{11} & \rho_{12} & \rho_{13} \\ \rho_{21} & \rho_{22} & \rho_{23} \\ \rho_{31} & \rho_{32} & \rho_{33} \end{pmatrix}\begin{pmatrix}0 & -i & 0\\i & 0 & 0\\0 & 0 & 0\end{pmatrix}\Bigg]=i\frac{\sqrt{3}}{2}(\rho_{12}-\rho_{21})\nonumber \\
n_3=\frac{\sqrt{3}}{2}Tr[\hat{\rho}\hat{\lambda}_3]=\frac{\sqrt{3}}{2}Tr\Bigg[\begin{pmatrix} \rho_{11} & \rho_{12} & \rho_{13} \\ \rho_{21} & \rho_{22} & \rho_{23} \\ \rho_{31} & \rho_{32} & \rho_{33} \end{pmatrix}\begin{pmatrix}1 & 0 & 0\\0 & -1 & 0\\0 & 0 & 0\end{pmatrix}\Bigg]=\frac{\sqrt{3}}{2}(\rho_{11}-\rho_{22})\nonumber \\
n_4=\frac{\sqrt{3}}{2}Tr[\hat{\rho}\hat{\lambda}_4]=\frac{\sqrt{3}}{2}Tr\Bigg[\begin{pmatrix} \rho_{11} & \rho_{12} & \rho_{13} \\ \rho_{21} & \rho_{22} & \rho_{23} \\ \rho_{31} & \rho_{32} & \rho_{33} \end{pmatrix}\begin{pmatrix}0 & 0 & 1\\0 & 0 & 0\\1 & 0 & 0\end{pmatrix}\Bigg]=\frac{\sqrt{3}}{2}(\rho_{13}+\rho_{31})\nonumber
\end{gather}
similarly, we can find the remaining components of the Bloch vector as follows,
\begin{gather}
n_5=\frac{\sqrt{3}}{2}Tr[\hat{\rho}\hat{\lambda}_5]=\frac{\sqrt{3}}{2}Tr\Bigg[\begin{pmatrix} \rho_{11} & \rho_{12} & \rho_{13} \\ \rho_{21} & \rho_{22} & \rho_{23} \\ \rho_{31} & \rho_{32} & \rho_{33} \end{pmatrix}\begin{pmatrix}0 & 0 & -i\\0 & 0 & 0\\i & 0 & 0\end{pmatrix}\Bigg]=i\frac{\sqrt{3}}{2}(\rho_{13}-\rho_{31})\nonumber \\
n_6=\frac{\sqrt{3}}{2}Tr[\hat{\rho}\hat{\lambda}_6]=\frac{\sqrt{3}}{2}Tr\Bigg[\begin{pmatrix} \rho_{11} & \rho_{12} & \rho_{13} \\ \rho_{21} & \rho_{22} & \rho_{23} \\ \rho_{31} & \rho_{32} & \rho_{33} \end{pmatrix}\begin{pmatrix}0 & 0 & 0\\0 & 0 & 1\\0 & 1 & 0\end{pmatrix}\Bigg]=\frac{\sqrt{3}}{2}(\rho_{23}+\rho_{32})\nonumber \\
n_7=\frac{\sqrt{3}}{2}Tr[\hat{\rho}\hat{\lambda}_7]=\frac{\sqrt{3}}{2}Tr\Bigg[\begin{pmatrix} \rho_{11} & \rho_{12} & \rho_{13} \\ \rho_{21} & \rho_{22} & \rho_{23} \\ \rho_{31} & \rho_{32} & \rho_{33} \end{pmatrix}\begin{pmatrix}0 & 0 & 0\\0 & 0 & -i\\0 & i & 0\end{pmatrix}\Bigg]=i\frac{\sqrt{3}}{2}(\rho_{23}-\rho_{32})\nonumber \\
n_8=\frac{\sqrt{3}}{2}Tr[\hat{\rho}\hat{\lambda}_8]=\frac{\sqrt{3}}{2}Tr\Bigg[\begin{pmatrix} \rho_{11} & \rho_{12} & \rho_{13} \\ \rho_{21} & \rho_{22} & \rho_{23} \\ \rho_{31} & \rho_{32} & \rho_{33} \end{pmatrix}\frac{1}{\sqrt{3}}\begin{pmatrix}1 & 0 & 0\\0 & 1 & 0\\0 & 0 & -2\end{pmatrix}\Bigg]=\frac{1}{2}(\rho_{11}+\rho_{22}-2\rho_{33})
\end{gather}
Then, the equation of the Bloch Sphere can be written as,
\begin{equation}
\label{eq:2.1.25}
\begin{split}
r^2&=\Vec{n}\cdot\Vec{n} \\
&= \sum_{i=1}^{8}n_{i}^2 \\
&= \frac{3}{4}\Big[(\rho_{12}+\rho_{21})^2-(\rho_{12}-\rho_{21})^2+(\rho_{11}-\rho_{22})^2+ {}\\
& (\rho_{13}+\rho_{31})^2-(\rho_{13}-\rho_{31})^2+(\rho_{23}+\rho_{32})^2- {}\\
& (\rho_{23}-\rho_{32})^2+\frac{1}{3}(\rho_{11}+\rho_{22}-2\rho_{33})^2\Big] \\
&= \frac{3}{4}\Big[(\rho_{12}+\rho_{21})^2-(\rho_{12}-\rho_{21})^2+(\rho_{11}-\rho_{22})^2+ {}\\
& (\rho_{13}+\rho_{31})^2-(\rho_{13}-\rho_{31})^2+(\rho_{23}+\rho_{32})^2- {}\\
& (\rho_{23}-\rho_{32})^2\Big]+\frac{1}{4}(\rho_{11}+\rho_{22}-2\rho_{33})^2
\end{split}
\end{equation}
This describes the mixed state of a three level open system and the equation of the eight dimensional Bloch sphere written in terms of the components of the density matrix.Now, the components of the Bloch vector may be redefined by introducing,
\begin{gather}
u_{mn}=(\rho_{mn}+\rho_{nm}), (m,n)=1,2,3;m\neq n\nonumber \\
v_{mn}=i(\rho_{mn}-\rho_{nm}), m<n=1,2,3\nonumber \\
w_2=(\rho_{11}-\rho_{22}),
w_3=\frac{1}{2}(\rho_{11}+\rho_{22}-2\rho_{33})
\end{gather}
so that the bloch vector components can also be written as,
\begin{subequations}
\begin{align}
n_1=\frac{\sqrt{3}}{2}u_{12},n_2=\frac{\sqrt{3}}{2}v_{12},n_3=\frac{\sqrt{3}}{2}w_2,n_4=\frac{\sqrt{3}}{2}u_{13}\\
n_5=\frac{\sqrt{3}}{2}v_{13},n_6=\frac{\sqrt{3}}{2}u_{23},n_7=\frac{\sqrt{3}}{2}v_{23},n_8=w_3
\end{align}
\end{subequations}
clearly the components of the bloch vector represented by the symmetric object $u_{mn}$ and the antisymmetric objects $v_{mn}$ measures the amount of overlap between the mth and the nth energy levels and the components $w_2$,$w_3$ corresponds to the population inversion. In general the off diagonal matrix elements of the density operator gives us the nonzero overlap between the mth and nth energy level.
\subsubsection{\textbf{Parametraisation of the Bloch Sphere for the Three Level system}}
The components of the Bloch vectors can be written as,
\begin{equation}
n_i=\frac{\sqrt{3}}{2}Tr[\hat{\rho}\hat{\lambda}_i]=\frac{\sqrt{3}}{2}\braket{\psi|\hat{\lambda}_i|\psi}
\end{equation}
, where $\rho=\ket{\psi}\bra{\psi}$ be the pure state density operator corresponding to some pure state $\ket{\psi}$. For a two level quantum system the parametraisation is bit simpler because we have a three dimensional sphere which can be parametrised by one radial coordinate $r$ and two angular coordinates i.e. polar angle $\theta$ and azimuthal angle $\phi$ respectively, very similar to a three dimensional spherical polar coordinate system. In case of the two level system as mentioned before the pure states are mapped to the surface of the sphere of radius unity leads to the following condition that the components of the bloch vector will satisfy the condition,$n_1^2+n_2^2+n_3^2=1$ so the parametraisation is as follows with any point on the surface of the sphere is designated by the pair of variables $(\theta,\phi)$ and the mixed states being mapped inside the sphere can be designated by the three variables $(r,\theta,\phi)$ so that we have,
\begin{equation}
n_1=\sin{\theta}\cos{\phi},n_2=\sin{\theta}\sin{\phi},n_3=\cos{\theta} \text{ for } 0\leq\theta\leq\pi \text{ and }0\leq\phi\leq 2\pi.
\end{equation}
the above choice of $n_1,n_2,n_3$ satisfies the condition $n_1^2+n_2^2+n_3^2=1$ but this is not the only possible choice for the bloch vector components so, the parametraisation is not \textbf{Unique}.Any possible set of choices of the bloch vector components satisfying the equation of the bloch sphere can be a possible parametraisation.Now, let us generalize it for the Bloch sphere in eight dimension. In case of the generalization from the $3$d bloch sphere to a $8$d bloch sphere first of all we require a set of eight parameters more specifically one radial parameter and seven angular parameters or, coordinates. Let, the set of parameters be $(r,\theta,\phi,\alpha,\beta,\gamma,\chi,\xi)$ with $r$ being the radial coordinate and the others being the angular coordinates. It can be also viewed as the case that in order to designate the point on the surface of the sphere we require $2$ angular coordinates respectively $(\theta,\phi)$ in case of a $3$D Bloch sphere here it requires seven angular coordintes to denote the position of a point on the surface of the eight dimensional Bloch Sphere i.e.$(\theta,\phi,\alpha,\beta,\gamma,\chi,\xi)$ respectively.
The Bloch Sphere Parametraisation equations are defined as follows,
\begin{subequations}
\label{eq:2.1.30a-2.1.30e}
\begin{align}
\cos{\theta}=\frac{1}{\sqrt{3}}\Big[1-\frac{\sqrt{3}}{2r}(\rho_{11}+\rho_{22}-2\rho_{33})\Big]^{\frac{1}{2}},\\
\cos{\phi}=\frac{1}{\sqrt{2}}\Big[1+\frac{\rho_{11}-\rho_{22}}{r\sin^2{\theta}}\Big]^\frac{1}{2},\\
\tan(\beta-\chi-\alpha+\gamma)=\frac{n_2}{n_1}=i\frac{\rho_{12}-\rho_{21}}{\rho_{12}+\rho_{21}},\\
\tan(\alpha-\gamma-\xi)=\frac{n_5}{n_4}=i\frac{\rho_{13}-\rho_{31}}{\rho_{13}+\rho_{31}},\\
\tan(\beta-\chi+\xi)=\frac{n_7}{n_6}=i\frac{\rho_{23}-\rho_{32}}{\rho_{23}+\rho_{32}}.
\end{align}
\end{subequations}
The parametraisation relations are chosen in such a way that they satisfy the equation of the Bloch sphere.The next task is to express the Bloch vector components in terms of the Bloch Parameters which will satisfy the equation of the eight dimensional Bloch Sphere.
Now, using the set of equations~[\ref{eq:2.1.24}] along with the Bloch Sphere Parametraisation equations  we can express the bloch vector components as follows,
\begin{subequations}
\label{eq:2.1.31}
\begin{align}
n_1=\sqrt{3}r\sin^2{\theta}\sin{\phi}\cos{\phi}\cos{(\beta-\chi-\alpha+\gamma)} \\
n_2=\sqrt{3}r\sin^2{\theta}\sin{\phi}\cos{\phi}\sin{(\beta-\chi-\alpha+\gamma)} \\
n_3=\sqrt{3}r\Big\{\frac{1}{2}(\cos^2{\phi}-\sin^2{\phi})\Big\}\sin^2{\theta} \\
n_4=\sqrt{3}r\sin{\theta}\cos{\theta}\cos{\phi}\cos{(\alpha-\gamma-\xi)} \\
n_5=-\sqrt{3}r\sin{\theta}\cos{\theta}\cos{\phi}\sin{(\alpha-\gamma-\xi)} \\
n_6=\sqrt{3}r\sin{\theta}\cos{\theta}\sin{\phi}\cos{(\beta-\chi+\xi)} \\
n_7=-\sqrt{3}r\sin{\theta}\cos{\theta}\sin{\phi}\sin{(\beta-\chi+\xi)} \\
n_8=\sqrt{3}r\frac{(1-3\cos^2{\theta})}{2\sqrt{3}}
\end{align}
\end{subequations}
the above set of equations given by~[\ref{eq:2.1.31}] satisfies the equation of the bloch sphere $\sum_in_i^2=r^2\leq1$ and the components of the bloch vector being real.Keeping an analogy with the result we have obtained in the case of two level quantum system that if the radius of the Bloch sphere is unity i.e. $r=1$ we will call it a poincare sphere and the three level system will be in the pure state.Thus the bloch vector is pointed on the surface of the bloch sphere. But if $r\leq0$ then the bloch vector $\textbf{n}$ is in interior of this unit poincare sphere.Therefore, the mixed states in the three level system may be identified with the interior points of this generalized unit poincare sphere.
\section{\textbf{Non Unit State Vector Mapping For the Mixed State of Three Level System}}
\setcounter{equation}{0}
\renewcommand{\theequation}{3.\arabic{equation}}
It is known that the pure state is a special case of the mixed state. It is very easy to unify them by using the representation of the poincare sphere, where the pure states are being mapped on the surface of the sphere of radius unity and the mixed states lying inside the sphere.
\\\indent In the case of two level open system, a mapping of the nonunit vector states in the two dimensional complex hilbert space $\mathcal{H}^2$ onto the interior points in the three dimensional poincare sphere.As we know that for the two level quantum system the bloch vector components are given by, $n_i=Tr[\hat{\rho}\hat{\sigma}_i]=\braket{\hat{\sigma}_i}=\braket{\psi|\hat{\sigma}_i|\psi}\text{ i=1,2,3 }$ where, $\psi$ be the vector which can be choosen as,$\ket{\psi}=\sqrt{r}\begin{pmatrix}\sin{\theta} \\ \cos{\theta}e^{i\phi}\end{pmatrix}$ or even we can in general associate an overall phase factor with $\ket{\psi}$ because both will not differ physically. This ansatz will satisfy the equation of the Bloch sphere in the three dimension in case of the two level quantum system i.e. $n_1^2+n_2^2+n_3^2=r^2$.
\\ \indent The next task will be to generalize the above mentioned idea for the three level quantum system where the nonunit vector ray i.e. $\ket{\psi}$ will be a $3\times1$ column matrix which will obey the equation $n_i=\frac{\sqrt{3}}{2}Tr[\rho\lambda_i]=\frac{\sqrt{3}}{2}\braket{\lambda_i}=\frac{\sqrt{3}}{2}\braket{\psi|\lambda_i|\psi}$, giving back the bloch vector components enlisted by the set of equations (2.1.31).
We begin with an ansatz that,
\begin{equation}
\ket{\psi}=\sqrt{r}\begin{pmatrix} e^
{i(\alpha-\gamma)}\sin{\theta}\cos{\phi} \\  e^
{i(\beta-\chi)}\sin{\theta}\sin{\phi} \\  e^{i\xi}\cos{\theta}\end{pmatrix}
\end{equation}
It leads to the following equations after some steps of mathematical calculations.
\begin{gather*}
 n_1=r\frac{\sqrt{3}}{2}\begin{pmatrix} e^{-i(\alpha-\gamma)}\sin{\theta}\cos{\phi} & e^{-i(\beta-\chi)}\sin{\theta}\sin{\phi} & e^{-i\xi}\cos{\theta}\end{pmatrix}\begin{pmatrix}0 & 1 & 0\\1 & 0 & 0\\0 & 0 & 0\end{pmatrix}\begin{pmatrix} e^
{i(\alpha-\gamma)}\sin{\theta}\cos{\phi} \\  e^
{i(\beta-\chi)}\sin{\theta}\sin{\phi} \\  e^{i\xi}\cos{\theta}\end{pmatrix} {}\\
=\sqrt{3}r\sin^2{\theta}\sin{\phi}\cos{\phi}\cos{(\beta-\chi-\alpha+\gamma)} \\
n_2=r\frac{\sqrt{3}}{2}\begin{pmatrix} e^{-i(\alpha-\gamma)}\sin{\theta}\cos{\phi} & e^{-i(\beta-\chi)}\sin{\theta}\sin{\phi} & e^{-i\xi}\cos{\theta}\end{pmatrix}\begin{pmatrix}0 & -i & 0\\i & 0 & 0\\0 & 0 & 0\end{pmatrix}\begin{pmatrix} e^
{i(\alpha-\gamma)}\sin{\theta}\cos{\phi} \\  e^
{i(\beta-\chi)}\sin{\theta}\sin{\phi} \\  e^{i\xi}\cos{\theta}\end{pmatrix} {}\\
=\sqrt{3}r\sin^2{\theta}\sin{\phi}\cos{\phi}\sin{(\beta-\chi-\alpha+\gamma)} 
\end{gather*}\begin{gather*}
n_3=r\frac{\sqrt{3}}{2}\begin{pmatrix} e^{-i(\alpha-\gamma)}\sin{\theta}\cos{\phi} & e^{-i(\beta-\chi)}\sin{\theta}\sin{\phi} & e^{-i\xi}\cos{\theta}\end{pmatrix}\begin{pmatrix}1 & 0 & 0\\0 & -1 & 0\\0 & 0 & 0\end{pmatrix}\begin{pmatrix} e^
{i(\alpha-\gamma)}\sin{\theta}\cos{\phi} \\  e^
{i(\beta-\chi)}\sin{\theta}\sin{\phi} \\  e^{i\xi}\cos{\theta}\end{pmatrix} {}\\
  =\sqrt{3}r\Big\{\frac{1}{2}(\cos^2{\phi}-\sin^2{\phi})\Big\}\sin^2{\theta} \\
n_4=r\frac{\sqrt{3}}{2}\begin{pmatrix} e^{-i(\alpha-\gamma)}\sin{\theta}\cos{\phi} & e^{-i(\beta-\chi)}\sin{\theta}\sin{\phi} & e^{-i\xi}\cos{\theta}\end{pmatrix}\begin{pmatrix}0 & 0 & 1\\0 & 0 & 0\\1 & 0 & 0\end{pmatrix}\begin{pmatrix} e^
{i(\alpha-\gamma)}\sin{\theta}\cos{\phi} \\  e^
{i(\beta-\chi)}\sin{\theta}\sin{\phi} \\  e^{i\xi}\cos{\theta}\end{pmatrix} {}\\
=\sqrt{3}r\sin{\theta}\cos{\theta}\cos{\phi}\cos{(\alpha-\gamma-\xi)} \\
n_5=r\frac{\sqrt{3}}{2}\begin{pmatrix} e^{-i(\alpha-\gamma)}\sin{\theta}\cos{\phi} & e^{-i(\beta-\chi)}\sin{\theta}\sin{\phi} & e^{-i\xi}\cos{\theta}\end{pmatrix}\begin{pmatrix}0 & 0 & -i\\0 & 0 & 0\\i & 0 & 0\end{pmatrix}\begin{pmatrix} e^
{i(\alpha-\gamma)}\sin{\theta}\cos{\phi} \\  e^
{i(\beta-\chi)}\sin{\theta}\sin{\phi} \\  e^{i\xi}\cos{\theta}\end{pmatrix} {}\\
=-\sqrt{3}r\sin{\theta}\cos{\theta}\cos{\phi}\sin{(\alpha-\gamma-\xi)} \\
n_6=r\frac{\sqrt{3}}{2}\begin{pmatrix} e^{-i(\alpha-\gamma)}\sin{\theta}\cos{\phi} & e^{-i(\beta-\chi)}\sin{\theta}\sin{\phi} & e^{-i\xi}\cos{\theta}\end{pmatrix}\begin{pmatrix}0 & 0 & 0\\0 & 0 & 1\\0 & 1 & 0\end{pmatrix}\begin{pmatrix} e^
{i(\alpha-\gamma)}\sin{\theta}\cos{\phi} \\  e^
{i(\beta-\chi)}\sin{\theta}\sin{\phi} \\  e^{i\xi}\cos{\theta}\end{pmatrix} {}\\
=\sqrt{3}r\sin{\theta}\cos{\theta}\sin{\phi}\cos{(\beta-\chi+\xi)} \\
n_7=r\frac{\sqrt{3}}{2}\begin{pmatrix} e^{-i(\alpha-\gamma)}\sin{\theta}\cos{\phi} & e^{-i(\beta-\chi)}\sin{\theta}\sin{\phi} & e^{-i\xi}\cos{\theta}\end{pmatrix}\begin{pmatrix}0 & 0 & 0\\0 & 0 & -i\\0 & i & 0\end{pmatrix}\begin{pmatrix} e^
{i(\alpha-\gamma)}\sin{\theta}\cos{\phi} \\  e^
{i(\beta-\chi)}\sin{\theta}\sin{\phi} \\  e^{i\xi}\cos{\theta}\end{pmatrix} {}\\
= -\sqrt{3}r\sin{\theta}\cos{\theta}\sin{\phi}\sin{(\beta-\chi+\xi)} \\ 
n_8=r\frac{\sqrt{3}}{2}\begin{pmatrix} e^{-i(\alpha-\gamma)}\sin{\theta}\cos{\phi} & e^{-i(\beta-\chi)}\sin{\theta}\sin{\phi} & e^{-i\xi}\cos{\theta}\end{pmatrix}\begin{pmatrix}1 & 0 & 0\\0 & 1 & 0\\0 & 0 & -2\end{pmatrix}\begin{pmatrix} e^
{i(\alpha-\gamma)}\sin{\theta}\cos{\phi} \\  e^
{i(\beta-\chi)}\sin{\theta}\sin{\phi} \\  e^{i\xi}\cos{\theta}\end{pmatrix} {}\\
=\sqrt{3}r\frac{\left(1-3\cos^2{\theta}\right)}{2\sqrt{3}}.
\end{gather*}

\vspace{0.5cm}
We retrieve the results for the components of the bloch vectors and this confirms that our ansartz is correct.Now in order to define the geometric phase, therefore, a reasonable approach is to seek for the mapping between the eight-dimensional poincare sphere and the three dimensional Hilbert space.Now we are going to implement the $\mathcal{H}-\mathcal{B}-\mathcal{R}$ framework used in the fibre bundle approach towards the generalaization of the approach given by Aharonov Anandan and Samuel Bhandari.Where, the state space is defined in such a way that it consists of the subset of Nonunit vectors lying in the hilbert space $\mathcal{H}^3$ and at each point of the state space there is a vector $\ket{\psi}$. Now the mapping between the state space and the Ray space i.e. $\mathcal{B}\rightarrow\mathcal{R}$ allows us to map the points associated to some vectors $\ket{\psi}$ in the state space $\mathcal{B}$ to the corresponding density operators in the Ray space $\mathcal{R}$, so at each point in the Ray space we can associate a density operator $\rho$ and the inverse mapping  between the Ray space and State Space.Let, us define it mathematically,
\begin{equation}
\Pi(\mathcal{B}\rightarrow\mathcal{R}):\Big\{\rho=\Pi\ket{\psi}\text{ and the inverse mapping }\ket{\psi}=\Pi^{-1}\rho\Big\}
\end{equation}
where, $\hat{\rho}=\ket{\psi}\bra{\psi}=\psi\psi^{\dag}$.So, for a given nonunit vector ray $\ket{\psi}\in\mathcal{H}^3$ the components of the bloch vector are given by, $n_i=Tr[\hat{\rho}\hat{\lambda}_i]=\braket{\hat{\lambda}_i}=\braket{\psi|\hat{\lambda}_i|\psi}$, \text{      i=1,2,..,8       }
Now, we wnat to find how the components of the bloch vector changes under the unitary transformation of the nonunit vector ray in the hilbert space.Again we bring back the idea for the two level quantum system and try to extend it for the three level.
\\ \indent In case of two level quantum system the density matrix being $2\times2$ the bloch vector has the three components which can be written as the expectation value of the pauli spin matrices with respect to the nonunit vector ray in the two dimensional hilbert space. If we consider a $SU(2)$ transformation of the nonunit vector ray $\ket{\psi}\in\mathcal{H}^2$ such that, $\ket{\psi}^\prime=u\ket{\psi}$ where, $u\in SU(2)$ then the components of the bloch vector changes to $n_i^\prime$ such that $n_i^\prime=R_{ik}(u)n_k$ where,$R_{ik}$ being the matrix elements of a $3\times3$ matrix $R(u)\in S0(3)$ and this becomes possible because of the homomorphism between the groups $SU(2)$ and $SO(3)$ i.e. an unitary transformation corresponds to the orthogonal transformation. We have,
\begin{equation}
n_i^\prime=R_{ik}n_k\Rightarrow{\Vec{\textbf{n}}^\prime=\textbf{R}\Vec{\textbf{n}}}.
\end{equation}
We know that the components of a vector transforms in the similar way as the coordinates transforms under the passive transformation and as a result the components of the bloch vector will obey the same coordinate transformation rule under the $SO(3)$ group as we know in case of the Euclidean coordinate system.
where, the matrix elements $R_{ik}$ are given by,
\begin{gather}
n_i^\prime=\braket{\psi^\prime|\hat{\sigma}_i|\psi^\prime}=\braket{\psi|u^\dag\hat{\sigma}_iu|\psi}\nonumber \\
n_i^\prime=R_{ik}n_k=R_{ik}\braket{\psi|\hat{\sigma}_k|\psi}=\braket{\psi|R_{ik}\hat{\sigma}_k|\psi}\nonumber \\
R_{ik}\hat{\sigma}_{k}=u^\dag\hat{\sigma}_iu\nonumber \\
R_{im}\hat{\sigma}_m\hat{\sigma}_k=u^\dag\hat{\sigma}_iu\hat{\sigma}_k\nonumber \\
\Rightarrow{R_{im}Tr[\hat{\sigma}_m\hat{\sigma}_k]=Tr[u^\dag\hat{\sigma}_iu\hat{\sigma}_k]=Tr[\hat{\sigma}_iu\hat{\sigma}_ku^\dag]}\nonumber \\
\Rightarrow{2R_{im}\delta_{mk}=Tr[\hat{\sigma}_iu\hat{\sigma}_ku^\dag]}\nonumber \\
\Rightarrow{R_{ik}=\frac{1}{2}Tr[\hat{\sigma}_iu\hat{\sigma}_ku^\dag]}
\end{gather}
We will just extend the idea of the homomorphism of the groups $SU(2)$ and $SO(3)$ in case of $SU(3)$ and $SO(8)$ for the three level quantum systems. In the present situation the bloch vector being an eight dimensional vector with eight components if we condider an unitary transformation of the nonunit vector ray $\ket{\psi}\in\mathcal{H}^3$ the components of the bloch vector will, transform under the $SO(8)$ transformation.So, let us define an $SU(3)$ transformation of the nonunit vector ray $\ket{\psi}^\prime=u\ket{\psi}$ where, $u\in SU(3)$.Then, the componenets of the bloch vector will transform to $n_i^\prime$ such that,
\begin{equation}
n_i=R_{ik}(u)n_k\Rightarrow{\Vec{\textbf{n}}^\prime=\textbf{R}\Vec{\textbf{n}}}.
\end{equation}
where,$R_{ik}$ being the matrix elements of a $8\times8$ matrix $R(u)\in SO(8)$.With SU(3) the situation is more intricate. We have a particular real eight-dimensional adjoint or octet representation of SU(3) given by certain $8\times8$ real orthogonal unimodular matrices, which is irreducible. All real orthogonal unimodular $8\times8$ matrices taken together form the 28 dimensional group SO(8); the matrices of the octet representation of SU(3) are a `very small' eight-dimensional subset of SO(8), in fact a subgroup.
Just like we have calculated the matrix elements of the unimodular orthogonal matrix $R\in SO(3)$ in case of the two level quantum system here we will proceed in the similar way to calculate the transformation matrix $R\in SO(8)$ for the components of the bloch vector.
\begin{gather}
n_i^\prime=\braket{\psi^\prime|\hat{\lambda}_i|\psi^\prime}=\braket{\psi|u^\dag\hat{\lambda}_iu|\psi}\nonumber \\
n_i^\prime=R_{ik}n_k=R_{ik}\braket{\psi|\hat{\lambda}_k|\psi}=\braket{\psi|R_{ik}\hat{\lambda}_k|\psi}\nonumber \\
R_{ik}\hat{\lambda}_{k}=u^\dag\hat{\lambda}_iu\nonumber \\
R_{im}\hat{\lambda}_m\hat{\lambda}_k=u^\dag\hat{\lambda}_iu\hat{\lambda}_k\nonumber \\
\Rightarrow{R_{im}Tr[\hat{\lambda}_m\hat{\lambda}_k]=Tr[u^\dag\hat{\lambda}_iu\lambda_k]=Tr[\hat{\lambda}_iu\hat{\lambda}_ku^\dag]}\nonumber \\
\Rightarrow{2R_{im}\delta_{mk}=Tr[\hat{\lambda}_iu\hat{\lambda}_ku^\dag]}\nonumber \\
\Rightarrow{R_{ik}(u)=\frac{1}{2}Tr[\hat{\lambda}_iu\hat{\lambda}_ku^\dag]}
\end{gather}
Similarly one can verify that for two different transformation matrices $u,u^\prime\in SU(3)$ we have  $R(u)R(u^\prime)=R(uu^\prime)$.In Short we can summarise the results as, 
\begin{gather}
u\in\text{SU(3)}\rightarrow R_{ik}(u)=
\frac{1}{2}Tr[\hat{\lambda}_iu\hat{\lambda}_ku^\dag],R(u)\in SO(8);\nonumber  \\
R(u)R(u^\prime)=R(uu^\prime)
\end{gather}
Now, given any two 'octet vectors' $\textbf{a},\textbf{b}\in\mathbb{R}^8$, we can form one scalar product and two diffrent octet vectors from them defined as follows:
\begin{subequations}
\begin{align}
\textbf{a}\cdot\textbf{b}=a_rb_r;\\
\big(\textbf{a}\wedge\textbf{b}\big)=-\big(\textbf{b}\wedge\textbf{a}\big)=f_{rst}a_sb_t;\\
\big(\textbf{a}*\textbf{b}\big)=\big(\textbf{b}*\textbf{a}\big)=\sqrt{3}d_{rst}a_sb_t; \\
R(u)\textbf{a}\wedge R(u)\textbf{b}=R(u)\big(\textbf{a}\wedge\textbf{b}\big); \\
R(u)\textbf{a}* R(u)\textbf{b}=R(u)\big(\textbf{a}*\textbf{b}\big)
\end{align}
\end{subequations}
\section{\textbf{Geometric Phase For Three Level Open Quantum System}}
\setcounter{equation}{0}
\renewcommand{\theequation}{4.\arabic{equation}}
Let us consider an open curve $\mathcal{C}=\ket{\psi(t)}$ defined in the manifold $\mathcal{B}$ which is also known as the state space and the curve is smoothly parametrised by some parameter $t$ lies entirely inside the State space so that at each points on the curve corresponds to some $\ket{\psi(t)}$. Because of the mapping exist between State space and the Ray Space i.e. $\mathcal{B}\rightarrow\mathcal{R}$ the point on the curve $\mathcal{C}$ will be mapped to their corresponding density operators $\hat{\rho}(t)=\ket{\psi(t)}\bra{\psi(t)}=\psi(t)\psi(t)^\dag$  so that there will be an image curve in the Ray Space $\mathcal{R}$ defined by, $C=\Big\{\hat{\rho}(t)=\ket{\psi(t)}\bra{\psi(t)}\Big\}$ on each point of the curve there is a density operator or, projection operator $\hat{\rho}(t)=\Pi\ket{\psi(t)}$ satisfying the conditions $Tr[\hat{\rho}(t)]=1,\hat{\rho}^\dag(t)=\hat{\rho}(t)$ and the inverse mapping leads to a lift of the curve $C$ in the state space $\mathcal{B}$ which can be defined as,
\begin{eqnarray}
\mathcal{C}\rightarrow C=\ket{\psi(t)}\rightarrow\hat{\rho}(t)=\Pi(\ket{\psi(t)})=\ket{\psi(t)}\bra{\psi(t)} \\
C\rightarrow\mathcal{C} =\Big\{\ket{\psi(t)}=\Pi^{-1}\hat{\rho}(t)\Big\}
\end{eqnarray}
Let the curve $\mathcal{C}$ describes the evolution of the quantum state $\ket{\psi(t)}$ parametrized by a smoothly varying parameter $t$ although we can consider either an open curve or a closed curve.Let us divide the curve into small segments and the points of subdivisions are $t_0,t_1,t_2,...,t_N$. The state vectors at that points of subdivision are respectively $\ket{\psi(t_0)},\ket{\psi(t_1)},\ket{\psi(t_2)},...,\ket{\psi(t_N)}$ or, in general $\ket{\psi_i}=\ket{\psi(t=t_i)}=\ket{\psi(t_i)},i=1,2,3,...,N$ and the corresponding density operators in the image curve be $\rho(t_i)=\rho(t=t_i)=\ket{\psi(t_i)}\bra{\psi(t_i)}$ and $\ket{\psi_i}=\ket{\psi(t=t_i)}=\Pi^{-1}\rho(t_i)\in\mathcal{B}$.Then, each trajectory can be described by a discrete sequence of quantum states $\Big\{\ket{\psi_0},\ket{\psi_1},\ket{\psi_2},..,\ket{\psi_N}\Big\}$.
\\ \indent Then the equation of the geometric phase for the three level open quantum mixed state is given by the Pancharatnam Formulae.In order to understand the idea let us work out some simple mathematical steps. We have subdivided the entire curve $\mathcal{C}$ into some small segments and let us call those segments as $\mathcal{C}_0,\mathcal{C}_1,\mathcal{C}_2,...,\mathcal{C}_N$ and the corresponding image segments of the curve in the Ray space are $C_0,C_1,C_2,...,C_N$ respectively where,$\mathcal{C}_0$ joins $\ket{\psi_0}$ and $\ket{\psi_1}$,$\mathcal{C}_1$ joins $\ket{\psi_1}$ and $\ket{\psi_2}$ and so on. The same goes for the curve segments $C_0,C_1,C_2,...,C_N$ which joins the density operators corresponding to the points of subdivions associated with the vectors $\ket{\psi_0},\ket{\psi_1},\ket{\psi_2},..,\ket{\psi_N}$. Now, the geometric phase along the union of those curve segments in the Ray space will be as follows,
\begin{gather}
\gamma_{g}\Big[C_0\cup C_1\cup C_2\cup ...\cup C_N\Big]=\gamma_g[C_0]+\gamma_g[C_1]+\gamma_g[C_2]+...+\gamma_g[C_N]-B_{N}[\psi_0,\psi_1,\psi_2,...,\psi_N]
\end{gather}
where,
\begin{equation}
\begin{split}
B_N&=\arg\text{Tr}[\rho_0\rho_1\rho_2...\rho_N] \\
&= \arg\Big\{\braket{\psi_0|\psi_1}\braket{\psi_1|\psi_2}\braket{\psi_2|\psi_3}...\braket{\psi_N|\psi_0}\Big\} \\
&= \arg\Big\{\Delta_N(\psi_0,\psi_1,\psi_2,...\psi_N)\Big\}
\end{split}
\end{equation}
with, $\Delta_N(\psi_0,\psi_1,\psi_2,...\psi_N)$ be the Bargmann Invariant of order N i.e. BI(N).
In the continuum limit i.e. $N\rightarrow\infty$ if the geometric phases vanishes on individual segment curves i.e. $\gamma_g[C_i]=0\forall{i=1(1)N}$  the formulae for the Geometric Phase $\gamma_{g}$ becomes,
\begin{equation}
\gamma_{g}=-\lim_{N\rightarrow\infty}\arg\text{Tr}[\rho_0\rho_1\rho_2...\rho_N]=-\lim_{N\rightarrow\infty}\arg\Big\{\braket{\psi_0|\psi_1}\braket{\psi_1|\psi_2}\braket{\psi_2|\psi_3}...\braket{\psi_N|\psi_0}\Big\}
\end{equation}
Now, as we know that the Geometric Phase can be written as, $\gamma_{g}(C)=\varphi_{tot}(\mathcal{C})-\varphi_{dyn}(\mathcal{C})$ the former being the Total phase that depends on the initial and final points of the curve $\mathcal{C}\subset\mathcal{B}$ and the later being the Dynamical Phase and the striking feature of the geometric phase is that it does not depend on the System Hamiltonian rather it only depends on the curve joining the end points lying in the Ray Space $\mathcal{R}$.
\\ \indent Using the fact that the total phase can be written as the sum of the Dynamical phase and the Geometric phase the Pancharatnam's formulae in the continuum limit leads to,
\begin{equation}
\label{eq:4.6}
\begin{split}
\gamma_{g}&=-\mathcal{L}t_{N\rightarrow{\infty}} arg\left(\braket{\psi_0|\psi_1}\braket{\psi_1|\psi_2}....\braket{\psi_{N-1}|\psi_N}\braket{\psi_N|\psi_0}\right) \\
&= \arg\braket{\psi(t_0)|\psi(t)}-Im\left(\int_{t_{0}}^{t} d\tau\frac{\braket{\psi(\tau)|\frac{d}{d\tau}|\psi(\tau)}}{\braket{\psi(\tau)|\psi(\tau)}}\right) 
\end{split}
\end{equation}
where, the total phase $\varphi_{tot}$ and the dynamic phase $\varphi_{dyn}$ is given by,
\begin{subequations}
\begin{align}
\varphi_{tot}[\mathcal{C}]=\arg\braket{\psi(t_0)|\psi(t)} \\
\varphi_{dyn}[\mathcal{C}]=-Im\left(\int_{t_{0}}^{t} d\tau\frac{\braket{\psi(\tau)|\frac{d}{d\tau}|\psi(\tau)}}{\braket{\psi(\tau)|\psi(\tau)}}\right)
\end{align}
\end{subequations}
From ,the previous ansartz we have $\ket{\psi}=\sqrt{r}\begin{pmatrix} e^
{i(\alpha-\gamma)}\sin{\theta}\cos{\phi} \\  e^
{i(\beta-\chi)}\sin{\theta}\sin{\phi} \\  e^{i\xi}\cos{\theta}\end{pmatrix}$ which satisfies the equation $n_i=\frac{\sqrt{3}}{2}\braket{\psi|\lambda_i|\psi}$ and now the time dependence of $\ket{\psi}$ will be translated through the angular bloch parameters i.e.$(\theta,\phi,\alpha,\beta,\gamma,\chi,\xi)$. For convenience we define $\theta(t)=\theta_t,\alpha(t)=\alpha_t,\beta(t)=\beta_t,\gamma(t)=\gamma_t,\phi(t)=\phi_t,\chi(t)=\chi_t,\xi(t)=\xi_t$ and $\theta(t_0)=\theta_0,\alpha(t_0)=\alpha_0,\beta(t_0)=\beta_0,\gamma(t_0)=\gamma_0,\phi(t_0)=\phi_0,\chi(t_0)=\chi_0,\xi(t_0)=\xi_0$ so that we can write,
\begin{equation}
\ket{\psi(t)}=\sqrt{r}\begin{pmatrix} e^
{i(\alpha_t-\gamma_t)}\sin{\theta_t}\cos{\phi_t} \\  e^
{i(\beta_t-\chi_t)}\sin{\theta_t}\sin{\phi_t} \\  e^{i\xi_t}\cos{\theta_t}\end{pmatrix}
\end{equation}
Then, the Total phase $\varphi_{tot}[\mathcal{C}]$ is calculated as follows,
\begin{equation}
\begin{split}
\braket{\psi(t_0)|\psi(t)}& =r\begin{pmatrix} e^{-i(\alpha_0-\gamma_0)}\sin{\theta_0}\cos{\phi_0} & e^{-i(\beta_0-\chi_0)}\sin{\theta_0}\sin{\phi_0} & e^{-i\eta_0}\cos{\theta_0}\end{pmatrix}\begin{pmatrix} e^
{i(\alpha_t-\gamma_t)}\sin{\theta_t}\cos{\phi_t} \\  e^
{i(\beta_t-\chi_t)}\sin{\theta_t}\sin{\phi_t} \\  e^{i\xi_t}\cos{\theta_t}\end{pmatrix} \\
&= X+iY
\end{split}
\end{equation}
where, $X$ and $Y$ being the real and imaginary part of $\braket{\psi(t_0)|\psi(t)}$ which are given by,
\begin{gather}
X=r\Big\{\cos{(\alpha_t-\gamma_t-\alpha_0+\gamma_0)}\cos{\theta_t}\cos{\phi_t}\cos{\theta_0}\cos{\phi_0}+ {}\\
\sin{(\beta_t-\chi_t-\beta_0+\chi_0)}\sin{\theta_t}\sin{\phi_t}\sin{\theta_0}\sin{\phi_0}+\sin{(\eta_t-\eta_0)}\cos{\theta_0}\cos{\theta_t}\Big\} \\
Y=r\Big\{\sin{(\alpha_t-\gamma_t-\alpha_0+\gamma_0)}\sin{\theta_t}\cos{\phi_t}\sin{\theta_0}\cos{\phi_0}+ {}\\
\sin{(\beta_t-\chi_t-\beta_0+\chi_0)}\sin{\theta_t}\sin{\phi_t}\sin{\theta_0}\sin{\phi_0}+\sin{(\eta_t-\eta_0)}\cos{\theta_0}\cos{\theta_t}\Big\}
\end{gather}
Then, the Total phase $\varphi_{tot}=\arg\braket{\psi(t_0)|\psi(t)}=\tan^{-1}(\frac{Y}{X})$
where, X and Y are the real and imaginary part of $\braket{\psi(t_0)|\psi(t)}$ and it leads to 
\begin{equation}
\boxed{\varphi_{tot}=\tan^{-1}\frac{\Big\{\sin{(\alpha_t-\gamma_t-\alpha_0+\gamma_0)}A+\sin{(\beta_t-\chi_t-\beta_0+\chi_0)}B+\sin{(\eta_t-\eta_0)}C\Big\}}{\Big\{\cos{(\alpha_t-\gamma_t-\alpha_0+\gamma_0)}A+\cos{(\beta_t-\chi_t-\beta_0+\chi_0)}B+\cos{(\eta_t-\eta_0)}C\Big\}}}
\end{equation}
where, we have introduced three factors $A,B,C$ given by, $A=\sin{\theta(t)}\cos{\phi(t)}\sin{\theta(t_0)}\cos{\phi(t_0)}=\sin{\theta_t}\cos{\phi_t}\sin{\theta_0}\cos{\phi_0}$,$B=\sin{\theta_t}\sin{\phi_t}\sin{\theta_0}\sin{\phi_0}=\sin{\theta(t)}\sin{\phi(t)}\sin{\theta(t_0)}\sin{\phi(t_0)}$
,$C=\cos{\theta(t)}\cos{\theta(t_0)}=\cos{\theta_t}\cos{\theta_0}$, these terms appear in both the real and imaginary part of the inner product $\braket{\psi(t_0)|\psi(t)}$.
The expression of the geometric phase given by the equation~[\ref{eq:4.6}] is gauge and reparametrisation invariant, therefore $\gamma_g$ is a geometric phase associated with a three level open quantum system. 
Now, it is interesting to note that under a specific  local $U(1)$ Gauge transformation defined as, $\ket{\psi^\prime(t)}=\exp{(-i\arg\braket{\psi(t_0)|\psi(t)})}\ket{\psi(t)}$, the geometric phase takes the exact same form we obtained in the context of Aharonov-Anandan's Generalization.For simplicity let us consider the phase involves in the gauge transformation be $\theta(t)=i\arg\braket{\psi(t_0)|\psi(t)})$, so that we can write, $\ket{\psi^\prime(t)}=\exp{(-i\theta(t))}\ket{\psi(t)}$.
This can be easily achieved by considering the following mathematical steps,
\begin{equation}
\begin{split}
\frac{\braket{{\psi^\prime(t)}|d|{\psi^\prime(t)}}}{\braket{{\psi^\prime(t)}|{\psi^\prime(t)}}}&=\frac{\braket{{\psi^\prime(t)}|\frac{d}{dt}|{\psi^\prime(t)}}}{\braket{{\psi^\prime(t)}|{\psi^\prime(t)}}}dt \\
&= \frac{\braket{\psi(t)|d|\psi(t)}-id\theta(t)\braket{\psi(t)|\psi(t)}}{\braket{\psi(t)|\psi(t)}} \\
&= \frac{\braket{\psi(t)|d|\psi(t)}}{\braket{\psi(t)|\psi(t)}}-id\theta(t)
\end{split}
\end{equation}
Now, let us consider the following integral,
\begin{equation}
\begin{split}
\int_{t_0}^{t}\frac{\braket{\psi^\prime(t)|d|\psi^\prime(t)}}{\braket{\psi^\prime(t)|\psi^\prime(t)}}&=\int_{t_0}^{t}\Bigg[\frac{\braket{\psi(t)|d|\psi(t)}}{\braket{\psi(t)|\psi(t)}}-id\theta(t)\Bigg] \\
&= \int_{t_0}^{t}\frac{\braket{\psi(t)|d|\psi(t)}}{\braket{\psi(t)|\psi(t)}}-i\int_{t_0}^{t}d\theta(t) \\
&=  \int_{t_0}^{t}\frac{\braket{\psi(t)|d|\psi(t)}}{\braket{\psi(t)|\psi(t)}}-i[\theta(t)-\theta(t_0)] \\
& \Rightarrow -Im\left(\int_{t_0}^{t}\frac{\braket{\psi^\prime(t)|d|\psi^\prime(t)}}{\braket{\psi^\prime(t)|\psi^\prime(t)}}\right)=-Im\left(\int_{t_0}^{t}\frac{\braket{\psi(t)|d|\psi(t)}}{\braket{\psi(t)|\psi(t)}}\right)+[\theta(t)-\theta(t_0)] \\
&= -Im\left(\int_{t_0}^{t}\frac{\braket{\psi(t)|d|\psi(t)}}{\braket{\psi(t)|\psi(t)}}\right)+\arg\braket{\psi(t_0)|\psi(t)}
\end{split}
\end{equation}
here, we have used the fact that $\theta(t_0)=\arg\braket{\psi(t_0)|\psi(t
_0)}=0$ as, $\braket{\psi(t_0)|\psi(t_0)}$ is purely real.
Then the Geometric phase can be represented by the formulae,
\begin{equation}
\boxed{\gamma_{g}^{A}=-Im\left(\int_{t_0}^{t}\frac{\braket{\psi^\prime(t)|d|\psi^\prime(t)}}{\braket{\psi^\prime(t)|\psi^\prime(t)}}\right)}
\end{equation}
The above formulae is the generalization of the Aharonov-Anandan's phase for the pure state with the condition that the vector ray is normalized i.e. $\braket{\psi^\prime(t)|\psi^\prime(t)}=\braket{\psi(t)|\psi(t)}=1$ so that,
\begin{equation}
\gamma_g^{A}=-Im\left(\int_{t_0}^{t}\braket{\psi^\prime(t)|\frac{d}{dt}|\psi^\prime(t)}dt\right)
\end{equation}
In the terminology of the fibre bundle approach $\ket{\psi^\prime(t)}$ can be viewed as the lift of the closed curve in the Ray space for a cyclic evolution of the quantum system in the Aharonov-Anandan Framework.Here $\gamma_g^A$ can be viewed as the generalized Aharonov Anandan Phase for the mixed state.In other other words the geometric phase can be expressed as the integral of a gauge invariant one-form defined on $\mathcal{B}$ defined by,
\begin{equation}
\boxed{K=\Im\frac{\braket{\overline{\psi}(t)|d|\overline{\psi}(t)}}{\braket{\overline{\psi}(t)|\overline{\psi}(t)}}}
\end{equation}
Now, let is consider an interaction between the physical system and the surrounding, in this case we can't expect a cyclic evolution of the quantum system and as a result of the system is not undergoing a cyclic evolution the Bloch parameters, the components of the bloch vector and the matrix elemets of the density operator will be decaying function of time with the time evolution of the density operator of the system will be described by the Lindblad type of Master equation.The nonunit state vectors will also involve deacying exponent.When the system is isolated from the environment, it can be considered that the system is undergoing a quasicyclic evolution so that the total phase $\varphi_{tot}=\arg\braket{\psi(t_0)|\psi(t)}$ becomes $2\pi$ which can be dropped out in quantum computation as it's a constant term and in general we identify the geometric phase term by dropping the overall phase or, total phase and the geometric phase is found as the negative of the dynamic phase.
Thus we obtained the geometric phase under the quasicyclic evolution of the quantum system with $\ket{\psi}=\sqrt{r}\begin{pmatrix} e^
{i(\alpha-\gamma)}\sin{\theta}\cos{\phi} \\  e^
{i(\beta-\chi)}\sin{\theta}\sin{\phi} \\  e^{i\xi}\cos{\theta}\end{pmatrix}$ is given by,
\begin{equation}
\label{eq:4.20}
\begin{split}
\gamma_g(\mathcal{C})&=-\Im\oint_{\mathcal{C}}\frac{\braket{\psi|d|\psi}}{\braket{\psi|\psi}} \\
&= -\Im\oint_{\mathcal{C}}\Bigg[i\frac{\begin{pmatrix} e^{-i(\alpha-\gamma)}\sin{\theta}\cos{\phi} & e^{-i(\beta-\chi)}\sin{\theta}\sin{\phi} & e^{-i\xi}\cos{\theta}\end{pmatrix}\begin{pmatrix} e^
{i(\alpha-\gamma)}d(\alpha-\gamma)\sin{\theta}\cos{\phi} \\  e^
{i(\beta-\chi)}d(\beta-\xi)\sin{\theta}\sin{\phi} \\  e^{i\xi}d\xi\cos{\theta}\end{pmatrix}}{\begin{pmatrix} e^{-i(\alpha-\gamma)}\sin{\theta}\cos{\phi} & e^{-i(\beta-\chi)}\sin{\theta}\sin{\phi} & e^{-i\xi}\cos{\theta}\end{pmatrix}\begin{pmatrix} e^
{i(\alpha-\gamma)}\sin{\theta}\cos{\phi} \\  e^
{i(\beta-\chi)}\sin{\theta}\sin{\phi} \\  e^{i\xi}\cos{\theta}\end{pmatrix}}\Bigg] \\
&= -\Im\oint_\mathcal{C}i\Big\{d(\alpha-\gamma)\sin^2{\theta}\cos^2{\phi}+d(\beta-\chi)\sin^2{\theta}\sin^2{\phi}+d\xi\cos^2{\theta}\Big\} \\
&=  -\Im\oint_\mathcal{C}i\Big\{\sin^2{\theta}[d(\alpha-\gamma)\cos^2{\phi}+d(\beta-\chi)\sin^2{\theta}]+\cos^2{\theta}d\xi\Big\} \\
&= -\oint_{\mathcal{C}}\Big\{\sin^2{\theta}[d(\alpha-\gamma)\cos^2{\phi}+d(\beta-\chi)\sin^2{\theta}]+\cos^2{\theta}d\xi\Big\}
\end{split}
\end{equation}
Here, due to the spherical symmetry we have considered the variations of the bloch parameters $\alpha,\gamma,\beta,\chi,\xi$ only with the other parameters fixed.

\indent The equation is known as the Berry phase of mixed state, while $\mathcal{C}$ is a closed circular arc of the eight dimensional bloch sphere for the three level system.It is mentioned before that the geometric phase $\gamma_g$ can be expressed by an integral around the closed curve $\mathcal{C}$ of the one form defined in the manifold $\mathcal{B}$ given by $K=\Im\frac{\braket{\overline{\psi}(t)|d|\overline{\psi}(t)}}{\braket{\overline{\psi}(t)|\overline{\psi}(t)}}$ which is also known as the Mead-Berry connection one form. We want to show that the expression of the geometric phase is invariant under the gauge transformation.
\\ \indent Let us consider a gauge transform defined as, $\ket{\overline{\psi}^*(t)}=e^{i\beta(t)}\ket{\overline{\psi}(t)}$.
So. under the transformation if the one form $K$ changes to $K^\prime$ i.e. $K\rightarrow k^\prime$ and along with $\braket{\overline{\psi}(t)|\overline{\psi}(t)}=\braket{\overline{\psi}^*(t)|\overline{\psi}^*(t)}$  then we have,
\begin{equation}
\begin{split}
K^*&=\Im\frac{\braket{\overline{\psi}^*(t)|d|\overline{\psi}^*(t)}}{\braket{\overline{\psi}^*(t)|\overline{\psi}^*(t)}} \\
&= \Im\Big[\frac{\braket{\overline{\psi}(t)|d|\overline{\psi}(t)}}{\braket{\overline{\psi}(t)|\overline{\psi}(t)}}+\frac{id\beta(t)\braket{\overline{\psi}(t)|\overline{\psi}(t)}}{\braket{\overline{\psi}(t)|\overline{\psi}(t)}}\Big] \\
&=  \Im\Big[\frac{\braket{\overline{\psi}(t)|d|\overline{\psi}(t)}}{\braket{\overline{\psi}(t)|\overline{\psi}(t)}}\Big]+d\beta(t) \\
&\Rightarrow{ K^*=\Im\frac{\braket{\overline{\psi}^*(t)|d|\overline{\psi}^*(t)}}{\braket{\overline{\psi}^*(t)|\overline{\psi}^*(t)}}=K+d\beta(t)}
\end{split}
\end{equation}
We can even consider the effect of the reparametraisation along with the Gauge transformation by replacing $t$ by $t^\prime$ such that,$\overline{\psi}(t)=\overline{\psi^\prime}(t^\prime)$, a smooth monotonic parametraisation is defined as,$\overline{\psi}(t)=\overline{\psi^\prime}(t^\prime),t^\prime=f(t)|\frac{df(t)}{dt}\geq0$.The parametraisation leads to the transformation $\mathcal{C}\rightarrow\mathcal{C}^\prime$ and also the similar kind of transformation in terms of the image of the curve in the Ray space. Now,it is evident under that transformation the geometric phase $\gamma_{g}$ and the Mead-Berry connection one form remains invariant.
So, under the gauge transformation we have $K\rightarrow K^*$ and assume that the geometric phase $\gamma_g$ changes to $\gamma_{g^*}$ which implies that,
\begin{equation}
\begin{split}
\gamma_{g^*}&=-\Im\oint_{\mathcal{C}}\frac{\braket{\overline{\psi}^*(t)|d|\overline{\psi}^*(t)}}{\braket{\overline{\psi}^*(t)|\overline{\psi}^*(t)}} \\
&= -\oint_{\mathcal{C}}K^* \\
&= -\oint_{\mathcal{C}}\left(K+d\beta(t)\right) \\
&= -\Big[\oint_{\mathcal{C}}K+\oint_{\mathcal{C}}d\beta(t)\Big] \\
&= -\oint_{\mathcal{C}}K=-\Im\oint_{\mathcal{C}}\frac{\braket{\overline{\psi}(t)|d|\overline{\psi}(t)}}{\braket{\overline{\psi}(t)|\overline{\psi}(t)}}=\gamma_{g}
\end{split}
\end{equation}
So, we have $K^*=K+d\beta(t)$ and as a consequence $dK^*=dK$ The closed integral of the total differential $\beta(t)$ over the path $\mathcal{C}$ vanishes.This establishes the gauge invariance of the Berry phase for a three level open quantum system of the mixed states i.e. $\gamma_{g^*}=\gamma_{g}$. Thus the Berry phase in the equation (4.20) is gauge invariant and the invariance condition holds for any arbitrary choice of $\beta(t)$ for a given closed curve $\mathcal{C}\subset\mathcal{B}$ under the quasicyclic evolution.
\section{\textbf{Lindblad Master Equation For Dissipative System}}
\setcounter{equation}{0}
\renewcommand{\theequation}{5.\arabic{equation}}
As mentioned earlier in order to find out the geometric phase given by the equation (4.20) we require the time dependent bloch sphere parameters which depends on the matrix elements of the density operator which are time dependent, when interaction comes into the picture in case of the open system between the system and the surroundings the time evolution of the density operator is governed by the Lindblad type of Mater Equation. As an example, let us consider a three-level system interacting with environment.When a relevant dynamical time scale of the open quantum system is long compared to the time for the environment to the forgetting quantum information, the evolution of system is effectively local in time (the Markovian approximation) and may be described by the Lindblad’s master equation \cite{lindblad1976generators} given by,
\begin{equation}
\label{eq:5.1}
\boxed{\frac{\partial{\rho}}{\partial t}=-\frac{i}{\hbar}[\hat{\mathcal{H}},\rho]+\sum_{i=1}^8\left(\Gamma_i^\dag\rho\Gamma_i-\frac{1}{2}\{\Gamma_i^\dag\Gamma_i,\rho\}\right)}
\end{equation}
The first term of the master equation is a usual schr$\ddot{o}$dinger term which generates an unitary evolution which appear in the well known Von-Neumann Liouville equation of quantum statistical mechanics and the remaining part of the equation 
describe all possible transitions that the open system may undergo due to the interaction with the reservior.The operators $\Gamma_{i}$(i=1,2,3,...,8) are called the Lindblad Operators or the quantum jump operators. It can be readily checked that $\Dot{\rho}$ is hermitian and $Tr(\Dot{\rho})=0$, which implies that that he Lindblad Master equation preserves the positivity of the density operator $\rho(t)$ for the open system.
In the present work the Lindblad operators are choosen as $\Gamma_{i}=\sqrt{\eta_i(t)}\lambda_{i}$ which represents the coupling to the environement where,$\lambda_i,i=1,2,...,8$ are the standard Gell-Mann matrices which are Hermitian and traceless.The Decoherence time is approximately given by $\frac{1}{\Gamma_i(t)}$.
\\ \indent Now, for simplicity we will consider a simple situation where the dephasing noise term or, the system and environment coupling term is represented by only a single type of Lindblad Operator $\Gamma=\sqrt{\eta}\lambda_{3}$ is applied to the three level quantum system in the present context with the Hamiltonian $\mathcal{H}=\frac{1}{2}\hbar\Omega\lambda_3$.The next task is to find the solution to the Master Equation, the matrix element of the Density operator which has been represented by a $3\times3$ matrix with, $\rho=\begin{pmatrix} \rho_{11} & \rho_{12} & \rho_{13} \\ \rho_{21} & \rho_{22} & \rho_{23} \\ \rho_{31} & \rho_{32} & \rho_{33} \end{pmatrix}$ and all the matrix elements are time dependent i.e. $\rho_{ij}(t)$.
\\ \indent Now, taking the trace of both sides of equation~[\ref{eq:5.1}] we get,
\begin{equation}
\begin{split}
Tr[\Dot{\rho}(t)]&=Tr[\mathcal{H}\rho-\rho\mathcal{H}]+\sum_{i=1}^8\Big\{Tr[\Gamma_i^\dag\rho\Gamma_i-\frac{1}{2}\Gamma_i^\dag\rho\Gamma_i-\frac{1}{2}\rho\Gamma_i^\dag\Gamma_i]\Big\} \\
&= \sum_{i=1}^8\Big\{Tr[\Gamma_i^\dag\rho\Gamma_i]-\frac{1}{2}Tr[\Gamma_i^\dag\rho\Gamma_i]-\frac{1}{2}Tr[\rho\Gamma_i^\dag\Gamma_i]\Big\} \\
&= \sum_{i=1}^8\Big\{Tr[\Gamma_i^\dag\rho\Gamma_i]-\frac{1}{2}Tr[\rho\Gamma_i^\dag\Gamma_i]-\frac{1}{2}Tr[\rho\Gamma_i^\dag\Gamma_i]\Big\} \\
&= \sum_{i=1}^8\Big\{Tr[\rho\Gamma_i\Gamma_i^\dag]-Tr[\rho\Gamma_i^\dag\Gamma_i]\Big\}=0
\end{split}
\end{equation}
Where we  have used the Cyclical Invariance properties of Trace operation $Tr[AB]=Tr[BA]$, $Tr[ABC]=Tr[BCA]=Tr[CAB]$ and in the last step we have used the Hermiticity condition of the Lindblad Operators i.e. $\Gamma_i^\dag=\Gamma_i,i=1,2,...,8$. As a consequence of the following result $Tr[\Dot{\rho}(t)=0$ we can say that the Lindblad Master Equation preserves the positivity of the Density Operator with ellapsation of time.
\subsection{\textbf{The Solution Of The Master Equation For The Three Level Quantum System}}
\setcounter{equation}{0}
\renewcommand{\theequation}{5.1.\arabic{equation}}
We will solve the Lindblad equation with the special case mentioned before where the dephasing noise term or, the system and environment coupling term is represented by only a single type of Lindblad Operator $\Gamma=\sqrt{\eta}\lambda_{3}$ is applied to the three level quantum system in the present context with the Hamiltonian $\mathcal{H}=\frac{1}{2}\hbar\Omega\lambda_3$.
So, in our case the Lindblad Master equation will be,
\begin{equation}
\frac{\partial{\rho}}{\partial t}=-\frac{i}{\hbar}[\hat{\mathcal{H}},\rho]+\left(\Gamma^\dag\rho\Gamma-\frac{1}{2}\{\Gamma^\dag\Gamma,\rho\}\right)
\end{equation}
And we have the following results, 
\begin{gather}
\frac{1}{i\hbar}[\mathcal{H},\rho]=\frac{-i\Omega}{2}\begin{pmatrix}0& 2\rho_{12} & \rho_{13} \\ -2\rho_{21} & 0 & -\rho_{23} \\ -\rho_{31} & \rho_{32} & 0\end{pmatrix}\nonumber \\
\Gamma^\dag\rho\Gamma-\frac{1}{2}\{\Gamma^\dag\Gamma,\rho\}=\eta\begin{pmatrix} 0 & -2\rho_{12} & -\frac{\rho_{13}}{2} \\ -2\rho_{21} & 0 & -\frac{\rho_{23}}{2} \\ -\frac{\rho_{31}}{2} & -\frac{\rho_{32}}{2} & 0\end{pmatrix}
\end{gather}
Subsituting them back in equation (5.3) we get,
\begin{equation}
\label{eq:5.1.3}
\begin{split}
\frac{d}{dt}\begin{pmatrix} \rho_{11}(t) & \rho_{12}(t) & \rho_{13}(t) \\ \rho_{21}(t) & \rho_{22}(t) & \rho_{23}(t) \\ \rho_{31}(t) & \rho_{32}(t) & \rho_{33}(t) \end{pmatrix}&=\begin{pmatrix} \Dot{\rho_{11}}(t) & \Dot{\rho_{12}}(t) & \Dot{\rho_{13}}(t) \\ \Dot{\rho_{21}}(t) & \Dot{\rho_{22}}(t) & \Dot{\rho_{23}}(t) \\ \Dot{\rho_{31}}(t) & \Dot{\rho_{32}}(t) & \Dot{\rho_{33}}(t) \end{pmatrix}\nonumber
\end{split}
\end{equation}
\[=\Bigg[\frac{-i\Omega}{2}\begin{pmatrix}0 & 2\rho_{12}(t) & \rho_{13}(t) \\ -2\rho_{21}(t) & 0 & -\rho_{23}(t) \\ -\rho_{31}(t) & \rho_{32}(t) & 0\end{pmatrix}+\eta\begin{pmatrix} 0 & -2\rho_{12}(t) & -\frac{\rho_{13}(t)}{2} \\ -2\rho_{21}(t) & 0 & -\frac{\rho_{23}(t)}{2} \\ -\frac{\rho_{31}(t)}{2} & -\frac{\rho_{32}(t)}{2} & 0\end{pmatrix}\Bigg]\tag{5.1.3}\]
Now, on compairing the both side of the matrix equation~[\ref{eq:5.1.3}] elementwise we obtain the following set of ordinary coupled first order diffetential equations involving the rate of change of the matrix elements $\rho_{ij}(t)$ as follows,
\begin{subequations}
\begin{align}
\Dot{\rho_{11}}(t)=\Dot{\rho_{22}}(t)=\Dot{\rho_{33}}(t)=0 \\
\Dot{\rho_{12}}(t)=-\Omega\rho_{12}(t)-2\eta\rho_{12}(t)=-\rho_{12}(t)(i\Omega+2\eta) \\
\Dot{\rho_{21}}(t)=\rho_{21}(t)(i\Omega-2\eta) \\
\Dot{\rho_{31}}(t)=\rho_{31}(t)(\frac{i\Omega-\eta}{2}) \\
\Dot{\rho_{23}}(t)=\rho_{23}(t)(\frac{i\Omega-\eta}{2}) \\
\Dot{\rho_{32}}(t)=\rho_{32}(t)(\frac{-i\Omega-\eta}{2}) \\
\Dot{\rho_{13}}(t)=\rho_{13}(t)(\frac{-i\Omega-\eta}{2})
\end{align}
\end{subequations}
Solving the above equations we get the solution of the Lindblad Master Equation as follows,
\begin{subequations}
\label{eq:5.1.4a-5.1.4i}
\begin{align}
\rho_{11}(t)=\rho_{11}(0), \\
\rho_{22}(t)=\rho_{22}(0), \\
\rho_{33}(t)=\rho_{33}(0), \\
\rho_{12}(t)=\rho_{12}(0)\exp{(-i\Omega-2\eta)t}, \\
\rho_{21}(t)=\rho_{21}(0)\exp{(i\Omega-2\eta)t},\\
\rho_{13}(t)=\rho_{13}(0)\exp{(\frac{-i\Omega}{2}-\frac{\eta}{2})t}, \\
\rho_{31}(t)=\rho_{31}(0)\exp{(\frac{i\Omega}{2}-\frac{\eta}{2})t}, \\
\rho_{23}(t)=\rho_{23}(0)\exp{(\frac{i\Omega}{2}+\frac{\eta}{2})t}, \\
\rho_{32}(t)=\rho_{32}(0)\exp{(\frac{-i\Omega}{2}-\frac{\eta}{2})t}
\end{align}
\end{subequations}
For the unique solution of the Lindblad Master Equation we require the initial condition i.e. $\rho_{ij}(0),\hspace{0.1cm} (i,j)=1,2,3.$.Let, us consider that the initial pure state $\ket{\psi(t=0)}=\ket{\psi(0)}$ given by $\ket{\psi(0)}=\delta_1\ket{1}+\delta_2\ket{2}+\delta_3\ket{3}$ with $\delta_1,\delta_2,\delta_3\in\mathbb{C}$ where,$\delta_i\hspace{0.1cm}(i=1,2,3)$ are independent of the evolving time and may be tuned by the external conditions and the canonical Basis vectors $\ket{1},\ket{2},\ket{3}$ are given by,$\ket{1}=\begin{pmatrix} 1\\ 0\\0\end{pmatrix},\ket{2}=\begin{pmatrix} 0\\ 1\\0\end{pmatrix},\ket{3}=\begin{pmatrix} 0\\ 0\\1\end{pmatrix}$ respectively and then the initial state vector will be $\ket{\psi(0)}=\begin{pmatrix}\delta_1\\ \delta_2\\ \delta_3\end{pmatrix}$.Then the density matrix at $t=0$ will be,$\hat{\rho}(0)=\ket{\psi(0)}\bra{\psi(0)}$. Let us denote the matrix elements of the density matrix at initial time be $\rho_{ij}(0)$ and it will be a $3\times3$ matrix.
So, the density matrix at the initial time i.e. $t=0$ is given by,
\begin{equation}
\begin{split}
\hat{\rho}(0)&=\ket{\psi(0)}\bra{\psi(0)} \\
&= \begin{pmatrix} \rho_{11}(0) & \rho_{12}(0) & \rho_{13}(0) \\ \rho_{21}(0) & \rho_{22}(0) & \rho_{23}(0) \\ \rho_{31}(0) & \rho_{32}(0) & \rho_{33}(0) \end{pmatrix} \\
&= \begin{pmatrix}\delta_1\\ \delta_2\\ \delta_3\end{pmatrix}\begin{pmatrix} \delta_1^* & \delta_2^* & \delta_3^*\end{pmatrix} \\
&= \begin{pmatrix} |\delta_1|^2 & \delta_1\delta_2^* & \delta_1\delta_3^*\\ \delta_2\delta_1^* & |\delta_2|^2 & \delta_2\delta_3^*\\ \delta_3\delta_1^* & \delta_3^*\delta_2 & |\delta_3|^2\end{pmatrix}
\end{split}
\end{equation}
So, from the initial condition we get,
\begin{subequations}
\begin{align}
\rho_{11}(0)=|\delta_1|^2,\rho_{22}(0)=|\delta_2|^2 \\
\rho_{33}(0)=|\delta_3|^2,\rho_{12}(0)=\delta_1\delta_2^* \\
\rho_{21}(0)=\delta_1^*\delta_2,\rho_{13}(0)=\delta_1\delta_3^* \\
\rho_{31}(0)=\delta_1^*\delta_3,\rho_{23}(0)=\delta_2\delta_3^*,\rho_{32}(0)=\delta_3\delta_2^*
\end{align}
\end{subequations}
So, the solution of the lindblad's master equation can be obtained by putting $\rho_{ij}(0);\hspace{0.1cm}\forall{(i,j=1,2,3}$ in equations~[\ref{eq:5.1.4a-5.1.4i}] and it leads to,
\begin{subequations}
\label{eq:5.1.7a-5.1.7i}
\begin{align}
\rho_{11}(t)=|\delta_1|^2, \\
\rho_{22}(t)=|\delta_2|^2, \\
\rho_{33}(t)=|\delta_3|^2, \\
\rho_{12}(t)=\delta_1\delta_2^*\exp{(-i\Omega-2\eta)t}, \\
\rho_{21}(t)=\delta_1^*\delta_2\exp{(i\Omega-2\eta)t},\\
\rho_{13}(t)=\delta_1\delta_3^*\exp{(\frac{-i\Omega}{2}-\frac{\eta}{2})t}, \\
\rho_{31}(t)=\delta_1^*\delta_3\exp{(\frac{i\Omega}{2}-\frac{\eta}{2})t}, \\
\rho_{23}(t)=\delta_2\delta_3^*\exp{(\frac{i\Omega}{2}+\frac{\eta}{2})t}, \\
\rho_{32}(t)=\delta_3\delta_2^*\exp{(\frac{-i\Omega}{2}-\frac{\eta}{2})t}
\end{align}
\end{subequations}
As, we know that the Bloch parameters depend on the matrix elements of the density operator then the time dependent bloch parameters will be obtained by substituting the matrix elements $\rho_{ij}(t);\hspace{0.1cm}\forall{(i,j)=1,2,3}$ in the set of equations~[\ref{eq:2.1.30a-2.1.30e}] we get the time dependent angular bloch parameters $(\theta,\phi,\xi,\chi,\alpha,\beta,\gamma)$ and then using the formulae for the Geometric phase $\gamma_g$ given by equation~[\ref{eq:4.20}] we will get the expression of the geometric phase.
\\ \indent In the equations~[\ref{eq:5.1.7a-5.1.7i}] the dephasing factors $\exp{(-2\eta t)}$ and $\exp{(\frac{-\eta t}{2})}$parametrize the amount of decoherence.The effect of dephasing are to decrease the size of the nondiagonal elements of the density matrix in a basis determined by the dephasing interaction with the environment.

\section{\textbf{SU(2) Polarization Picture}}
\setcounter{equation}{0}
\renewcommand{\theequation}{6.\arabic{equation}}
Polarization is a property applying to transverse waves that specifies the geometrical orientation of the oscillations. In a transverse wave, the direction of the oscillation is perpendicular to the direction of motion of the wave and for light which is an electromagnetic wave with a certain direction of propagation the vibration of the electric field is confined to a plane which is perpendicular to the direction of propagation. The plane at which the electric filed oscillates is known as the plane of polarization.
Now,the amount of polarization of the electromagnetic beam is quantified by Degree of polarization.The set of parameters required to describe the polarization state of electromagnetic radiation. They were defined by George Gabriel Stokes in 1852 \cite{stokes1851composition}.
\\ \indent We can define four stokes parameter classically .Let us consider two plane waves orthogonal to each other at a point in the space (x,y,z) with $E_{0x}(t)$,$E_{0y}(t)$ being their instantaneous amplitudes and $\delta_x(t)$,$\delta_y(t)$ being their instantaneous phase factors with intensities $I_1=|E_{0x}|^2$ and $I_2=|E_{0y}|^2$. Then the stokes parameters are defined as follows,

\begin{gather}
S_0=|E_{0x}|^2+|E_{0y}|^2=I_1+I_2 \\
S_1=|E_{0x}|^2-|E_{0y}|^2 =I_1-I_2\\
S_2=2|E_{0x}||E_{0y}|\cos{\delta}=2\sqrt{I_1I_2}\cos{\delta} \\
S_3=2|E_{0x}||E_{0y}|\sin{\delta}=2\sqrt{I_1I_2}\sin{\delta}
\end{gather}
where, $\delta=\delta_y-\delta_x$We can identify the first stokes parameter i.e. $S_0$ as the sum of intensities of the two electromagnetic waves or,the total intensity and parameter $S_1$ as the difference of intensity. Now, the degree of polarization of the electromagnetic beam or the polarized light can be defined as,
\begin{equation}
    \mathbb{P}=\frac{\sqrt{\sum_{i=1}^3S_i^2}}{S_0}
\end{equation}
and, we have $\mathbb{P}=1$ for Completely polarized light,the degree of polarization vanishes for unpolarized light and for partially polarized light $0<\mathbb{P}<1$.Alternatively one can define the degree of polarization of a polarized beam from the polarization matrix which will be introduced later.Now, we will introduce the notion of the quantum theory of polarization and the concept of the stokes operators which can be seen as an extension of the Classical theory of polarization and stokes parameters.
\\ \indent Let, us consider a plane polarized monochromatic Electromagnetic wave is propagating along the $z$ direction.Then the oscillation of the electric field must be confined to the plane perpendicular to the propagation direction i.e. $xy$ plane is the plane of polarization. The electric field is a two dimensional vector lying in the $xy$ plane.As a result we can have two different polarization states say $\{+,-\}$ and the general state of polarization is a linear combination of the two mutually orthogonal polarization states.The two mode field can be described by two complex amplitude operators.
Associating two sets of creation and annihilation operators corresponding to two different polarization states i.e.$\{\hat{a_+},\hat{a_{+}^\dag}\}$ and,$\{\hat{a_-},\hat{a_{-}^\dag}\}$ the quantum stokes operators can be defined by the standard Jordan-Schwinger mapping \cite{mota2004jordan}.
\\ \indent Before we proceed further let us introduce the Jordan-Schwinger mapping briefly.Jordan Schwinger mapping is the consequence of the Oscillator model proposed by Erwin Schwinger which enables us to establish the one to one correspondence between the States of a two dimensional isotropic harmonic oscillator with the angular momentum states of a quantum particle.
Let us consider the Hamiltonian of a two dimensional isotropic harmonic oscillator as follows,
\begin{equation}
\label{eq:6.6}
    \hat{H}=\frac{\hat{p_1}^2}{2m}+\frac{\hat{p_2}^2}{2m}+\frac{1}{2}m\omega^2\hat{x_1}^2+\frac{1}{2}m\omega^2\hat{x_2}^2
\end{equation}
The Hamiltonian is the sum of two uncoupled harmonic oscillators and it can be rewritten in terms of two different sets of bosonic creation and annihilation operators say,$\{\hat{a_1},\hat{a_1}^\dag\}$ and $\{\hat{a_2},\hat{a_2}^\dag\}$  for individual Oscillators.Let us define,
\[\hat{a_1}=\frac{1}{\sqrt{2m\hbar\omega}}\big(i\hat{p_1}+m\omega\hat{x_1}\big),\hat{a_2}=\frac{1}{\sqrt{2m\hbar\omega}}\big(i\hat{p_2}+m\omega\hat{x_2}\big)\]
\[\hat{a_1}^\dag=\frac{1}{\sqrt{2m\hbar\omega}}\big(-i\hat{p_1}+m\omega\hat{x_1}\big),\hat{a_2}^\dag=\frac{1}{\sqrt{2m\hbar\omega}}\big(-i\hat{p_2}+m\omega\hat{x_2}\big)\]
Along with this the Hamiltonian [\ref{eq:6.6}] becomes,
\begin{equation}
    \hat{H}=\big(\hat{a_1}^\dag\hat{a_1}+\hat{a_2}^\dag\hat{a_2}+\hat{I}\big)\hbar\omega
\end{equation}
Here the standard bosonic commutation relation leads to,$\big[\hat{a_i},\hat{a_j}^\dag\big]=\delta_{ij}$, (i,j=1,2) and $\big[\hat{a_i},\hat{a_j}\big]=0$,$\big[\hat{a_i}^\dag,\hat{a_j}^\dag\big]=0$.
\\ \indent Here we have two types of number operator respectively $\hat{N}_1=\hat{a_1}^\dag\hat{a_1}$ and $\hat{N}_{2}=\hat{a_2}^\dag\hat{a_2}$, so we can define the total number operator $\hat{N}=\big(\hat{a_1}^\dag\hat{a_1}+\hat{a_2}^\dag\hat{a_2}\big)$ and the total Hamiltonian commutes with the total number operator i.e. $\big[\hat{H},\hat{N}\big]=0$. Recalling the algebra of the bosonic creation and annihilation operators $\{\hat{a}_{i},\hat{a}_{i}^\dag\}$ for i=1,2 we can write, \[\hat{a}_{i}\ket{n_{i}}=\sqrt{n_{i}}\ket{n_i-1}\] \[\hat{a}_{i}^\dag\ket{n_{i}}=\sqrt{n_{i}+1}\ket{n_{i}+1}\]
\[\hat{a}_{i}^\dag\hat{a}_{i}\ket{n_{i}}=n_i\ket{n_{i}}\]
Along with, $\big[\hat{N}_{i},\hat{a}_{i}\big]=-\hat{a}_{i}$ and $\big[\hat{N}_{i},\hat{a}_{i}^\dag\big]=\hat{a}_{i}^\dag$ for i=1,2.
\\ \indent Now the eigenstates of the harmonic oscillator can be represented by the two mode fock state i.e. $\ket{n_1,n_2}=\ket{n_1}\otimes\ket{n_2}$ and we have, $\hat{H}\ket{n_1,n_2}=(n_1+n_2+1)\hbar\omega$ and $\hat{N}\ket{n_1,n_2}=(n_1+n_2)\ket{n_1,n_2}$.
Now, the angular momentum operators are defined by the Schwinger map as follows,
\begin{equation}
    \hat{{J}_i}=\frac{1}{2}\big[\hat{\textbf{a}}^\dag\sigma_{i}\hat{\textbf{a}}\big], i=1,2,3
\end{equation}
where, $\hat{\textbf{a}}=\begin{pmatrix}\hat{a_1} \\ \hat{a_2}\end{pmatrix}$ and $\hat{\textbf{a}}^\dag=\begin{pmatrix} \hat{a_1}^\dag & \hat{a_2}^\dag\end{pmatrix}$ and $\sigma_i$ being the 2$\times$2 Pauli matrices given by,
\[\sigma_1=\begin{pmatrix} 0 & 1 \\ 1 & 0 \end{pmatrix},\sigma_2=\begin{pmatrix} 0 & -i \\ i & 0 \end{pmatrix},\sigma_3=\begin{pmatrix} 1 & 0 \\ 0 & -1 \end{pmatrix}\]

Then the angular momentum operators become,
\begin{gather}
\hat{J_1}=\frac{1}{2}\big(\hat{a_1}^\dag\hat{a_2}+\hat{a_2}^\dag\hat{a_1}\big) \\
\hat{J_2}=\frac{i}{2}\big(\hat{a_1}\hat{a_2}^\dag-\hat{a_1}^\dag\hat{a_2}\big) \\
\hat{J_3}=\frac{1}{2}\big(\hat{a_1}^\dag\hat{a_1}-\hat{a_2}^\dag\hat{a_2}\big)
\end{gather}
It can be easily seen that the operators $\hat{\vec{J}}$ follows the cyclic commutation algebra $\big[\hat{J_k},\hat{J_l}\big]=i\epsilon_{klm}\hat{J_m}$ for all (k,l,m=1,2,3).
The square of the total angular momentum operator is defined as,
\[\hat{J}^2=\sum_{i=1}^3\hat{J_i}^2=\frac{\hat{N}}{2}\big(\frac{\hat{N}}{2}+\hat{I}\big),  \hat{N}=\hat{a_1}^\dag\hat{a_1}+\hat{a_2}^\dag\hat{a_2}\]
The raising and lowering operators are defined as,
\begin{gather}
    \hat{J_{+}}=\big(\hat{J_1}+i\hat{J_2}\big)=\hat{a_1}^\dag\hat{a_2} \\ \nonumber
    \hat{J_{-}}=\big(\hat{J_1}-i\hat{J_2}\big)=\hat{a_2}^\dag\hat{a_1} \\ \nonumber
    \big[\hat{J_{+}},\hat{J_{-}}\big]=2\hat{J_3}
\end{gather}
And we have, $\big[\hat{J}^2,\hat{J_i}\big]=0$, i=1,2,3. As it commutes with all the components of the angular momentum operator then it is Casimir.
The Eigenvalue equation in the natural unit system i.e. ($\hbar=1$) for $\hat{J}^2$ and $\hat{J}_{3}$ are given by,
\[\hat{J}^2\ket{n_1,n_2}=\frac{(n_1+n_2)}{2}\bigg[\frac{(n_1+n_2)}{2}+1\bigg]\ket{n_1,n_2}=j(j+1)\ket{j,m}\]
\[\hat{J}_{3}\ket{n_1,n_2}=\frac{(n_1-n_2)}{2}\ket{n_1,n_2}=m\ket{j,m}\]
\\ \indent The Schwinger oscillator model establishes the correspondence between the angular momentum states i.e. the mutual eigenstates of $\hat{J}^2$ and $\hat{J_3}$ denoted by $\ket{j,m}$ and the harmonic oscillator states i.e. $\ket{n_1,n_2}$ which is exploited in the above equations with $m=\frac{(n_1-n_2)}{2}$ and $j=\frac{(n_1+n_2)}{2}=\frac{n}{2}$ where, $n=(n_1+n_2)$.

\indent Now the similar idea can be applied in the present context of the SU(2) polarization where the stokes operators \cite{klimov2008quantum} can be constructed by the Jordan Schwinger map.
Let us consider the following Jordan-Schwinger Map defined as follows,
\[\hat{{S}_i}=\big[\hat{\textbf{a}}^\dag\sigma_{i}\hat{\textbf{a}}\big], i=1,2,3\]
and, $\hat{\textbf{a}}=\begin{pmatrix}\hat{a}_{+} \\ \hat{a}_{-}\end{pmatrix}$ and $\hat{\textbf{a}}^\dag=\begin{pmatrix} \hat{a}_{+}^\dag & \hat{a}_{-}^\dag\end{pmatrix}$ and $\sigma_i$ being the 2$\times$2 pauli matrices
\begin{gather}
\hat{S_1}=\begin{pmatrix}\hat{a_{+}}^\dag & \hat{a_{-}}^\dag\end{pmatrix}\begin{pmatrix} 0 & 1 \\ 1 & 0\end{pmatrix}\begin{pmatrix}\hat{a_{+}} \\ \hat{a_{-}}\end{pmatrix}
=\big(\hat{a_{+}}^\dag\hat{a_{-}}+\hat{a_{-}}^\dag\hat{a_{+}}\big) \\
\hat{S_2}=\begin{pmatrix}\hat{a_{+}}^\dag & \hat{a{-}}^\dag\end{pmatrix}\begin{pmatrix} 0 & -i \\ i & 0\end{pmatrix}\begin{pmatrix}\hat{a_{+}} \\ \hat{a_{-}}\end{pmatrix}
=i\big(\hat{a_{-}}^\dag\hat{a_{+}}-\hat{a_{+}}^\dag\hat{a_{-}}\big) \\
\hat{S_3}=\begin{pmatrix}\hat{a_{+}}^\dag & \hat{a_{-}}^\dag\end{pmatrix}\begin{pmatrix} 1 & 0 \\ 0 & -1\end{pmatrix}\begin{pmatrix}\hat{a_{+}} \\ \hat{a_{-}}\end{pmatrix}
=\big(\hat{a_{+}}^\dag\hat{a_{+}}-\hat{a_{-}}^\dag\hat{a_{-}}\big) \\
\hat{S_0}=\begin{pmatrix}\hat{a_{+}}^\dag & \hat{a_{-}}^\dag\end{pmatrix}\begin{pmatrix} 1 & 0 \\ 0 & 1\end{pmatrix}\begin{pmatrix}\hat{a_{+}} \\ \hat{a_{-}}\end{pmatrix}
=\big(\hat{a_{+}}^\dag\hat{a_{+}}+\hat{a_{-}}^\dag\hat{a_{-}}\big)
\end{gather}
The four stokes operators forms a vector, known as stokes vector represented as, $\hat{\vec{S}}=\begin{pmatrix}\hat{S_0} \\ \hat{S_1} \\ \hat{S_2} \\ \hat{S_3}\end{pmatrix} $.
Similar to the angular momentum operators the stokes operators satisfies the similar cyclic commutation given by,
\[\big[\hat{S_k},\hat{S_l}\big]=2i\epsilon_{klm}\hat{S_m}\] and \[\big[\hat{S_0},\hat{S_i}\big]=0\].The commutation algebra of the stokes operators resembles the commutation algebra of the generators of the symmetry group SU(2).
It is worth to note that the Quantum mechanical expectation value of the stokes operator with respect to certain state gives the classical stokes parameters.The expectation values of the stokes operators can be expressed as follows,
\[\braket{\hat{S_i}}=Tr[\Phi\sigma_i]\] for i=1,2,3.
where, $\Phi$ be the $2\times2$ coherence matrix or polarization matrix with elements $\Phi_{kl}=\braket{\hat{a_l}^\dag\hat{a_k}}$, (k,l=1,2,3).
The coherence matrix is,
\begin{equation}
\Phi_{2\times2}=\begin{pmatrix}\braket{\hat{a_1}^\dag\hat{a_1}} & \braket{\hat{a_2}^\dag\hat{a_1}} \\ \braket{\hat{a_1}\hat{a_2}} & \braket{\hat{a_2}^\dag\hat{a_2}}\end{pmatrix}
\end{equation}
The operator $\hat{S_0}$ commutes with all other stokes operators i.e. $\hat{S_i}$ for i=1,2,3 so,evidently $\hat{S_0}$ is the casimir operator. Here also we can use the standard two mode Fock basis which will be the simultaneous eigenstates of the mutually commuting observable $\hat{S_0}$ and $\hat{S_3}$.
\\ \indent The standard two mode fock basis is defined as,
\begin{equation}
    \ket{N,k}=\ket{k}_{+}\otimes\ket{N-k}_{-},\text{k=0,1,2,..,N}
\end{equation}
Here,N denotes the total number of photons and k stands for the number of photons in the first mode denoted by $\{+\}$ and the other (N-k) photons are in the other mode denoted by $\{-\}$ such that the total number of photons is N and they are distributed within the two polarization modes.
\\ \indent Now let us define the raising and lowering operators as,
\[\hat{S_{+}}=\hat{S_1}+i\hat{S_2}=2\hat{a_{+}}^\dag\hat{a_{-}}\ , \hat{S_{-}}=\hat{S_1}-i\hat{S_2}=2\hat{a_{-}}^\dag\hat{a_{+}}\].
Using the properties of the bosonic creation and annihilation operators i.e. $\hat{a}\ket{n}=\sqrt{n}\ket{n-1}$ and $\hat{a}^\dag\ket{n}=\sqrt{n+1}\ket{n+1}$ we can find the action of the creation and annihilation operators on the state $\ket{N,k}$.
\begin{equation}
\begin{split}
    \hat{S_{+}}\ket{N,k}&=2\hat{a_{+}}^\dag\hat{a_{-}}\ket{k}_{+}\otimes\ket{N-k}_{-} \\ 
    &= 2\sqrt{(N-k)} \hat{a_{+}}^\dag\ket{k}_{+}\otimes\ket{N-k-1}_{-} \\
    &= 2\sqrt{(N-k)(k+1)}\ket{k+1}_{+}\otimes\ket{N-k-1}_{-} \\
    &= 2\sqrt{(N-k)(k+1)}\ket{N,k+1}
    \end{split}
\end{equation}
\begin{equation}
        \begin{split}
       \hat{S_{-}}\ket{N,k}&=2\hat{a_{-}}^\dag\hat{a_{+}}\ket{k}_{+}\otimes\ket{N-k}_{-} \\ 
    &= 2\sqrt{k}\hat{a_{-}}^\dag\ket{k-1}_{+}\otimes\ket{N-k}_{-} \\
    &= 2\sqrt{(N-k+1)k}\ket{k-1}_{+}\otimes\ket{N-k+1}_{-} \\
    &= 2\sqrt{(N-k+1)k}\ket{N,k-1} 
    \end{split}
\end{equation}
Similarly we get the eigenvalue equations for $\hat{S_0}$ and $\hat{S_3}$ with $\ket{N,k}$ being their simultaneous eigenstates, using the fact that $\hat{a_{+}}^\dag\hat{a_{+}}\ket{K}_{+}=k\ket{k}_{+}$ and $\hat{a_{-}}^\dag\hat{a_{-}}\ket{N-k}=(N-k)\ket{N-k}_{-}$ we get,
\[\hat{S_0}\ket{N,k}=N\ket{N,k}\]
\[\hat{S_3}\ket{N,k}=(2k-N)\ket{N,k}\]
Required to note that the stokes operators $\hat{\vec{S}}$ with components $\hat{S_{1}},\hat{S_2},\hat{S_3}$ are the generators of the SU(2) group and they are non commuting so,as the total photon number operator i.e. $\hat{S_0}$ commutes with all the generators it is the Casimir operator of that group. So, the polarization has an additional SU(2) invariance which leads to the natural structure of the invariant subspaces and the structure will be preserved under the evolution. The non-commutativity of the stokes operators precludes the simultaneous exact measurement of their physical quantities.
It can also be shown that no field state leaving aside the two mode vacuum state i.e. $\ket{0,0}$ can have definite non-fluctuating value of the stokes operators simultaneously and this is expressed in terms of the uncertainty relation,
\begin{equation}
    \big(\Delta\hat{S_1}\big)^2+\big(\Delta\hat{S_2}\big)^2+\big(\Delta\hat{S_3}\big)^2\geq2\braket{\hat{S_0}}
\end{equation}
Now, the degree of polarization can be defined in terms of the expectation values of the stokes operators as,
\begin{equation}
\label{eq:4.22}
    \mathbb{P}=\frac{\sqrt{\braket{\hat{S_1}}^2+{\braket{\hat{S_2}}^2}+{\braket{\hat{S_3}}^2}}}{\braket{\hat{S_0}}}
\end{equation}
This is the standard definition of the 2D degree of polarization that exclusively deals with the first order moments of the stokes operators but higher order moments of stokes operators can be useful to understand some interesting phenomena \cite{klimov2005b}.For the present work the standard definition is enough to understand the effect of depolarization.
\\ \indent The ideas of the stokes operators is not only useful to study the polarization behaviour of the monochromatic light only but also can be extended for the multi mode electromagnetic fields also. The multi mode generalisation is quite straightforward after the construction of the stokes operators of the single mode fields.
Let, us consider a poly-chromatic quantized transverse electromagnetic beam with $m$ number of spatiotemporal modes so that the positive frequency part of the electric vector can be represented as \cite{klimov2006simple},
\begin{equation}
    \vec{E^{+}}(\vec{r},t)=i\sum_{j=1}^m\sum_{s=\pm}\sqrt{\frac{\hbar\omega_j}{2\epsilon_0}}\hat{a}_{js}\vec{U}_{js}(\vec{r})\exp(-i\omega_jt)
\end{equation}
For the plane waves we have, $\vec{U}_{js}(\vec{r})=\frac{1}{\sqrt{V}}\hat{e}_{js}\exp{(i\vec{k}_j\cdot\vec{r})}$ with the front factor is due to the box normalization within some definite volume V and $\vec{e}_{js}$ being the polarization vector which corresponds to two different circular polarization (left-handed and right-handed) for $s=\pm$ for each single spatiotemporal modes.
Just like the case of the monochromatic beam which has a single spatiotemporal mode and two polarization states so we have two different sets of creation and annihilation operators here also we will have two different sets of creation and annihilation operators for each individual spatiotemporal modes namely, $\{\hat{a}_{js},\hat{a}^\dag_{js}\}$ for j=1,2,...,m and s=$\pm$.
Along with that we have the following commutation relations for the bosonic creation and annihilation operators as follows,
\begin{gather}
    \big[\hat{a}_{js},\hat{a}^\dag_{j\prime s\prime}\big]=\delta_{jj\prime}\delta_{ss\prime}
\end{gather}
The stokes parameters for the multi mode fields can be defines as,
\begin{gather}
    \hat{S}_{0}=\sum_{j=1}^{n}\big(\hat{a}^\dag_{j+}\hat{a}_{j+}+\hat{a}^\dag_{j-}\hat{a}_{j-}\big) \\ 
    \hat{S}_{3}=\sum_{j=1}^{n}\big(\hat{a}^\dag_{j+}\hat{a}_{j+}-\hat{a}^\dag_{j-}\hat{a}_{j-}\big) \\
    \hat{S}_{1}=\sum_{j=1}^m\big(\hat{a}^\dag_{j+}\hat{a}_{j-}+\hat{a}^\dag_{j-}\hat{a}_{j+}\big) 
\end{gather}
\begin{gather}
    \hat{S}_{2}=i\sum_{j=1}^m\big(\hat{a}^\dag_{j-}\hat{a}_{j+}-\hat{a}^\dag_{j+}\hat{a}_{j-}\big) \\ 
    \hat{S}_{+}=\big(\hat{S}_1+i\hat{S}_2\big)=2\sum_{j=1}^m\hat{a}^\dag_{j+}\hat{a}_{j-} \\ 
    \hat{S}_{-}=\big(\hat{S}_1-i\hat{S}_2\big)=2\sum_{j=1}^{m}\hat{a}^\dag_{j-}\hat{a}_{j+}
\end{gather}
Here $\hat{S}_{0}$ being the total number operator i.e. the sum of individual total number operators for each spatiotemporal modes.Similar to the case of the single mode fields we can write that,
$\big[\hat{S}_0,\hat{S}_i\big]=0$ ,  $\big[\hat{S}_i,\hat{S}_j\big]=2i\epsilon_{ijk}\hat{S}_k$ and $\big[\hat{S}_{+},\hat{S}_{-}\big]=4\hat{S}_3$.
Thus for a system containing N number of photons the entire Hilbert space will split into invariant subspaces of dimension $(N+1)$.Thus if we have a single photon state the dimensionality of each invariant subspaces will be 2. Our claim is that the structure of the invariant subspaces must be preserved as a result of evolution. 

\section{SU(2) Depolarization Dynamics for single mode fields}
\setcounter{equation}{0}
\renewcommand{\theequation}{7.1.\arabic{equation}}
The term depolarization has come to mean the effective decrease in the degree of polarization of a light traversing an optical system which is due to the interaction of the electromagnetic beam with the surroundings and the result is an effective anisotropy which leads to the decorrelation of the phase associated with the electric field vector.In the quantum theory the depolarization can be broadly understood in terms of decoherence which corresponds to the loss of coherence in the quantum system due to the interaction with the surroundings i.e. the appearance of uncontrollable quantum correlations due to the dissipation.When the system interacts with the surrounding there will be an exchange of energy due to the interaction which causes dissipation and gives rise to decoherence. 
Here, in the present context we will consider the effect of decoherence in such a way so that the net exchange of energy between the system and the surroundings is very negligible which is also known as the pure dephasing dynamics.
\subsection{Master equation for the pure dephasing process}
First we begin by focusing our attention on the single mode field which is traversing through some optical medium i.e. here the system is the single mode (monochromatic) plane polarized electromagnetic wave coupled to a bath system so that the total Hamiltonian \cite{klimov2006simple} for the global system (system+bath) can be written as the sum of the system Hamiltonian or the field Hamiltonian, the bath Hamiltonian and the interaction Hamiltonian between the field and the bath.
\begin{equation}
    \hat{H}=\hat{H}_{field}+\hat{H}_{bath}+\hat{H}_{int}
\end{equation}
The field Hamiltonian will be identical to the two dimensional isotropic harmonic oscillator Hamiltonian apart from the constant term $\hbar\omega$, where $\omega$ being the frequency of the electromagnetic wave so that we have,
\begin{equation}
    \hat{H}_{field}=\hbar\omega\sum_{s=\pm}\hat{a}^\dag_{s}\hat{a}_{s}
\end{equation}
$\hat{H}_{bath}$ describes the free evolution of the environment, we are not making any assumption regarding the type of bath actually we don't need the precise knowledge of the bath, we assume that the bath with which the field is coupled with is so large that its statistical properties remain unaffected due to the interaction with the system.
In, the interaction picture of quantum mechanics if we consider the global density operator which can be represented as $\hat{\rho}_{Tot}=\hat{\rho}_{field}\otimes\hat{\rho}_{B}$ then the partial trace of the global density operator with respect to the reservoir or, the bath states will give the actual density operator for the system i.e. field. Then $\hat{\rho}_{field}=TR_{B}\big[\hat{\rho}_{Tot}\big]$
Under the weak coupling limit the Born-Markov master equation \cite{gardiner2000quantum} leads to,
\begin{equation}
\label{eq:5.3}
    \Dot{\hat{\rho}}(t)=-\frac{1}{\hbar^2}\int_0^{\infty}d\tau Tr_{B}\big\{\Big[\hat{H}_{int}(t),\big[\hat{H}_{int}(t-\tau),\hat{\rho}(t)\otimes\hat{\rho}_{B}\big]\big]\big\}
\end{equation}
Here,$Tr_{B}$ indicates the partial trace with respect to the bath states and $\hat{\rho}(t)$ being the reduced density operator for the field.
As, we can see that the master equation does not depend on the specific choice of the bath or, it's Hamiltonian rather it only depends on the interaction Hamiltonian i.e. $\hat{H}_{int}$.

\indent One common way to couple the bath and the single mode field is by the interaction Hamiltonian of this type:
\begin{equation}
\label{eq:7.1.4}
    \hat{H}_{int}=\hbar\sum_{\lambda}\sum_{s=\pm}\big(\kappa_{\lambda}\Gamma_{\lambda}\hat{a}^\dag_{s}+\kappa_{\lambda}^{*}\Gamma^\dag_{\lambda}\hat{a}_{s}\big)
\end{equation}
Here, $\kappa_{\lambda}$ being the complex coupling constant describing the interaction strength between the field and the bath, the sum over $\lambda$ corresponds to the sum over all possible accessible bath modes and $\{\hat{\Gamma_{\lambda}},\hat{\Gamma_{\lambda}^\dag
}\}$ corresponds to the annihilation and creation operators corresponding to the bath quanta for each bath modes.A photon can create a bath quanta by loosing energy and vice versa.
Now, for simplicity we assume the zero temperature bath in order to neglect the process of stimulated emission and with those underlying assumptions the master equation~[\ref{eq:5.3}] with the interaction Hamiltonian of the above form becomes,
\begin{equation}
\label{eq:5.5}
   \Dot{ \hat{\rho}}(t)=\sum_{s=\pm}\frac{\gamma_{s}}{2}\mathcal{L}\big[\hat{a}_s\big]\hat{\rho}(t)
\end{equation}
where,$\gamma_{s}$ being the decoherence rates associated with each mode of polarization and We have the lindblad superoperators \cite{lindblad1976generators} $\mathcal{L}\big[\hat{C}_{s}\big]$ defined as,
\begin{equation}
\label{eq:5.6}
    \mathcal{L}\big[\hat{C}_{s}\big]=2\hat{C}_{s}\hat{\rho}\hat{C}^\dag_s-\{\hat{C}_{s}^\dag\hat{C}_s,\hat{\rho}\}
\end{equation}
The decoherence rate can be expressed in terms of the coupling constant $\kappa_{\lambda}$. $\Dot{\hat{\rho}}(t)$ is hermitian and this form of master equation is always positive \cite{kraus1971general} describing the dynamical evolution of an open system ensuring that the state of the system remains always valid at all times.
If, we solve this above master equation for a single photon state then, first we need to find the matrix realisation of the operators.Here, the basis can be choosen as the two mode fock basis i.e. $\ket{N,k}=\ket{k}_{+}\otimes\ket{N-k}_{-}$.
For, single photon states the dimension of each invariant subspaces will be 2 and the density operator will be a $2\times2$ hermitian matrix in the orthonormal basis: $\ket{1,0}=\ket{0}_{+}\otimes\ket{1}_{-}=\ket{-}$ and $\ket{1,1}=\ket{1}_{+}\otimes\ket{0}_{-}=\ket{+}$ respectively.
\\ \indent Let, the density matrix being $\hat{\rho}(t)=\begin{pmatrix} \rho_{11}(t) & \rho_{12}(t) \\ \rho_{21}(t) & \rho_{22}(t)\end{pmatrix}$.
Now, the finite dimensional matrix representation of the creation and annihilation operators can be obtained by knowing the action of the operators on the basis $\{\ket{+},\ket{-}\}$.
we have,
\begin{gather}
    \hat{a}_{+}\ket{+}=\ket{0}_{+}\otimes\ket{0}_{-} \\
    \hat{a}_{-}\ket{-}=\ket{o}_{+}\otimes\ket{0}_{-} \\
    \hat{a}_{+}\ket{-}=0=\hat{a}_{-}\ket{+}.
\end{gather}
so, the matrix representation for the operator $\hat{a}_{+}$ and its hermitian adjoint will be,
\[\hat{a}_{+(2\times2)}=\begin{pmatrix}\bra{+}\hat{a}_{+}\ket{+} & \bra{+}\hat{a}_{+}\ket{-} \\ \bra{-}\hat{a}_{+}\ket{+} & \bra{-}\hat{a}_{+}\ket{-} \end{pmatrix}=\begin{pmatrix} 0 & 0 \\ 0 & 0 \end{pmatrix}=\hat{a}^\dag_{+(2\times2)}\]
similarly,the matrix for the $\hat{a}_{-}$ and its hermitian adjoint will be the same i.e. $\hat{a}_{-(2\times2)}=\begin{pmatrix} 0 & 0 \\ 0 & 0 \end{pmatrix}=\hat{a}^\dag_{-(2\times2)}$.
In, the similar way we can find the matrix representation of the individual number operators i.e. $\big(\hat{a}_{+}^\dag\hat{a}_{+}\big)_{ij}$ and $\big(\hat{a}_{-}^\dag\hat{a}_{-}\big)_{ij}$ which leads to,
\[\big(\hat{a}_{+}^\dag\hat{a}_{+}\big)_{ij}=\begin{pmatrix}\bra{+}\hat{a}_{+}^\dag\hat{a}_{+}\ket{+} & \bra{+}\hat{a}_{+}^\dag\hat{a}_{+}\ket{-} \\ \bra{-}\hat{a}_{+}^\dag\hat{a}_{+}\ket{+} & \bra{-}\hat{a}_{+}^\dag\hat{a}_{+}\ket{-} \end{pmatrix}=\begin{pmatrix} 1 & 0 \\ 0 & 0 \end{pmatrix}\]
and similarly we can write,$\big(\hat{a}_{-}^\dag\hat{a}_{-}\big)_{ij}=\begin{pmatrix} 0 & 0 \\ 0 & 1 \end{pmatrix}$. 
With the definition given by~[\ref{eq:5.6}] simple matrix multiplication leads to,
\[\mathcal{L}\big[\hat{a}_s\big]\hat{\rho}=2\hat{a}_{s}\hat{\rho}\hat{a}^\dag_s-\{\hat{a}_{s}^\dag\hat{a}_s,\hat{\rho}\}=-\begin{pmatrix}2\rho_{11} & \rho_{12} \\ \rho_{21} & 0\end{pmatrix},\hspace{0.1cm}s=+\] and,
\[\mathcal{L}\big[\hat{a}_s\big]\hat{\rho}=2\hat{a}_{s}\hat{\rho}\hat{a}^\dag_s-\{\hat{a}_{s}^\dag\hat{a}_s,\hat{\rho}\}=-\begin{pmatrix}0 & \rho_{12} \\ \rho_{21} & 2\rho_{22}\end{pmatrix},\hspace{0.1cm} s=-\]
Then with $\gamma_{+}$ and $\gamma_{-}$ denoting the decoherence rates for the individual modes of polarization the master equation~[\ref{eq:5.5}] becomes,
\begin{equation}
    \begin{pmatrix}\Dot{\rho_{11}}(t) & \Dot{\rho_{12}}(t) \\ \Dot{\rho_{21}}(t) & \Dot{\rho_{22}}(t)\end{pmatrix}=\frac{\gamma_{+}}{2}\begin{pmatrix}-2\rho_{11}(t) & -\rho_{12}(t) \\ -\rho_{21}(t) & 0\end{pmatrix}+\frac{\gamma_{-}}{2}\begin{pmatrix}0 & -\rho_{12}(t) \\ -\rho_{21}(t) & -2\rho_{22}(t)\end{pmatrix}
\end{equation}
The above matrix equation leads to simple ordinary differential equations in terms of the time derivative of the matrix elements of the density operator given by,
\begin{gather}
    \Dot{\rho_{11}}(t)=-\gamma_{+}\rho_{11}(t) \\
    \Dot{\rho_{22}}(t)=-\gamma_{-}\rho_{22}(t) \\
    \Dot{\rho_{12}}(t)=-\frac{\gamma_{+}}{2}\rho_{12}(t)-\frac{\gamma_{-}}{2}\rho_{12}(t) \\
    \Dot{\rho_{21}}(t)=-\frac{\gamma_{+}}{2}\rho_{21}(t)-\frac{\gamma_{-}}{2}\rho_{21}(t)
\end{gather}
The solution to the set of ordinary differential equations leads to,
\begin{gather}
\rho_{11}(t)=\rho_{11}(0)\exp{\{-\gamma_{+}t\}} \\
\rho_{22}(t)=\rho_{22}(0)\exp{\{-\gamma_{-}t\}} \\
\rho_{12}(t)=\rho_{12}(0)\exp{\{-(\gamma_{+}+\gamma_{-})t/2\}} \\
\rho_{21}(t)=\rho_{21}(0)\exp{\{-(\gamma_{+}+\gamma_{-})t/2\}}
\end{gather}
But,the above master equation can not describe the depolarization process correctly as along with the interaction Hamiltonian of the form given by equation describes two standard independent Lindblad decaying processes for individual polarization states with the decoherence parameters $\gamma_{\pm}$ as if we are studying the decay process for the individual polarisation modes.The master equation can be written as the sum of two lindbladian describing the decay of individual polarisation modes i.e. $s=\pm$. Since the coupling transfers energy between the system and the bath the result will be a damping of the beam intensity.
The density matrix is completely known along with the given initial conditions and is given by,
\begin{equation}
    \hat{\rho}(t)=\begin{pmatrix}\rho_{11}(0)\exp{\{-\gamma_{+}t\}} & \rho_{12}(0)\exp{\{-(\gamma_{+}+\gamma_{-})t/2\}} \\ \rho_{21}(0)\exp{\{-(\gamma_{+}+\gamma_{-})t/2\}} & \rho_{22}(0)\exp{\{-\gamma_{-}t\}}\end{pmatrix}
\end{equation}
\\ \indent In fact from the solutions of the master equation we can see that the matrix elements $\rho_{ij}(t)$ decays with time and irrespective of the given initial conditions i.e. $\rho_{ij}(0)$, the stationary states is always be $\hat{\rho}(t)=\ket{0,0}\bra{0,0}$ at $t\rightarrow\infty$ with $\ket{0,0}=\ket{0}_{+}\otimes\ket{0}_{-}$.
It is the density operator of the vacuum state for both circular polarization i.e. $s=\pm$. So, obviously this cannot describe the depolarization process.We expect the density matrix to be a diagonal matrix as opposed to the present situation where the density matrix corresponds to the Vacuum state for every polarization modes.
\\ \indent As we have anticipated earlier that we are interested to study the pure dephasing dynamics so that the amount of energy exchange between the system and the bath is negligible.In mathematical terms we can say that the Interaction Hamiltonian must commute with the system Hamiltonian i.e. $\big[\hat{H}_{field},\hat{H}_{int}\big]=0$ which implies no exchange of energy and only the phase changes.
\\ \indent One way to model the pure dephasing dynamics is by choosing the interaction Hamiltonian of the following form \cite{gardiner2000quantum} so that it commutes with the field Hamiltonian.
\begin{equation}
\label{eq:7.1.20}
    \hat{H}_{int}=\hbar\sum_{\lambda}\sum_{s=\pm}\big(\kappa_{\lambda}\Gamma_{\lambda}\hat{a}^\dag_{s}\hat{a}_{s}+\kappa_{\lambda}^{*}\Gamma^\dag_{\lambda}\hat{a}^\dag_{s}\hat{a}_{s}\big)
\end{equation}
With this choice of the interaction term the Lindblad Master equation~[\ref{eq:5.3}] becomes,
\begin{equation}
\label{eq:5.21}
     \Dot{ \hat{\rho}}(t)=\sum_{s=\pm}\frac{\gamma_{s}}{2}\mathcal{L}\big[\hat{a}_s^\dag\hat{a}_{s}\big]\hat{\rho}(t)
\end{equation}
The above interaction Hamiltonian can be viewed as a scattering process, in which a bath quanta is absorbed or,emitted but the number of photons in each polarization is preserved.We will solve the master equation for the single and two photon states.
\\ \indent As we have already worked out the matrix representation of the Number operators with single photon states for each polarization states i.e. $\hat{a}_{s}^\dag\hat{a}_{s}$ for, $s=\pm$ along with the fact that $(\hat{a}_s^\dag\hat{a}_s)^2=\hat{a}_s^\dag\hat{a}_s$, with the definition~[\ref{eq:5.6}] simple matrix multiplication leads to,
\[\mathcal{L}\big[\hat{a}_{s}^\dag\hat{a}_s\big]\hat{\rho}=2\hat{a}_{s}^\dag\hat{a}_s\hat{\rho}\hat{a}^\dag_s\hat{a}_s-\{\hat{a}_{s}^\dag\hat{a}_s,\hat{\rho}\}=\begin{pmatrix}0 & -\rho_{12}(t) \\ -\rho_{21}(t) & 0\end{pmatrix} ,\hspace{0.2cm} s=+ \]
\[\mathcal{L}\big[\hat{a}_{s}^\dag\hat{a}_s\big]\hat{\rho}=2\hat{a}_{s}^\dag\hat{a}_s\hat{\rho}\hat{a}^\dag_s\hat{a}_s-\{\hat{a}_{s}^\dag\hat{a}_s,\hat{\rho}\}=\begin{pmatrix}0 & -\rho_{12}(t) \\ -\rho_{21}(t) & 0\end{pmatrix} ,\hspace{0.2cm} s=- \]
Substituting them back into the master equation~[\ref{eq:5.21}] leads to,
\begin{equation}
  \begin{pmatrix}\Dot{\rho_{11}}(t) & \Dot{\rho_{12}}(t) \\ \Dot{\rho_{21}}(t) & \Dot{\rho_{22}}(t)\end{pmatrix}=\frac{\gamma_{+}}{2}\begin{pmatrix}0 & -\rho_{12}(t) \\ -\rho_{21}(t) & 0\end{pmatrix}+\frac{\gamma_{-}}{2}\begin{pmatrix}0 & -\rho_{12}(t) \\ -\rho_{21}(t) & 0\end{pmatrix}  
\end{equation}
Compairing the two matrices element-wise we get the following ordinary differential equations,
\begin{gather}
    \Dot{\rho_{11}}(t)=\Dot{\rho_{22}}(t)=0 \\
    \Dot{\rho_{12}}(t)=-\frac{1}{2}\big(\gamma_{+}+\gamma_{-}\big)\rho_{12}(t) \\
    \Dot{\rho_{21}}(t)=-\frac{1}{2}\big(\gamma_{+}+\gamma_{-}\big)\rho_{21}(t)
\end{gather}
From the above equations it is confirmed that the positivity of the density operator is preserved as Tr$\big[\Dot{\hat{\rho}}(t)\big]=0$ and $\Dot{\hat{\rho}}(t)$ is hermitian.
One can write the $2\times2$ density matrix as a linear combination of one $2\times2$ identity matrix and three Pauli matrices i.e. $\sigma_{i}$, i=1,2,3. So, we may write $\hat{\rho}=\frac{1}{2}\big(\hat{I}+\vec{s}\cdot\vec{\hat{\sigma}}\big)$ which leads to,
\begin{equation}
\label{eq:5.26}
    \hat{\rho}(t)=\frac{1}{2}\begin{pmatrix} 1+s_z(t) & s_x(t)-is_y(t) \\ s_x(t)+is_y(t) & 1-s_z(t)\end{pmatrix} \\
    , \Dot{\hat{\rho}}(t)=\frac{1}{2}\begin{pmatrix}\Dot{s_z}(t) & \Dot{s_x}(t)-i\Dot{s_y}(t) \\ \Dot{s_x}(t)+i\Dot{s_y}(t) & -\Dot{s_z}(t)\end{pmatrix}
\end{equation}
which corresponds to the following equations,
\begin{gather}
    \Dot{s_z}(t)=0 \\
    \Dot{s_x}(t)-i\Dot{s_y}(t)=-\frac{1}{2}\big(\gamma_{+}+\gamma_{-}\big)\big(s_{x}(t)-is_{y}(t)\big) \\
    \Dot{s_x}(t)=-\frac{1}{2}\big(\gamma_{+}+\gamma_{-}\big)s_x(t) \\
    \Dot{s_y}(t)=-\frac{1}{2}\big(\gamma_{+}+\gamma_{-}\big)s_y(t) \\
\end{gather}
The solution pertains to,
\begin{gather}
\label{eq:5.32}
    s_{z}(t)=s_{z}(0) \\
    s_{x}(t)=s_x(0)\exp{\{-(\gamma_{+}+\gamma_{-})t/2\}} \\
    s_{y}(t)=s_y(0)\exp{\{-(\gamma_{+}+\gamma_{-})t/2\}}
\end{gather}
The solution of the Lindblad master equation leads to the fact that the population of the system does not change with time as the diagonal matrix elements does not change over time i.e. $\rho_{11}(t)=\rho_{11}(0)$ and $\rho_{22}(t)=\rho_{22}(0)$. And the off diagonal elements of the density matrix which successfully describe the effect of decoherence in the system as a result of the interaction with the surroundings as the off diagonal matrix elements describe the quantum coherence and the solutions suggests that the correlation is exponentially decaying.
The model preserves the invariant subspaces and can also explain that the stationery state is a diagonal state i.e. the off diagonal matrix elements of the density matrix will vanish at $t\rightarrow\infty$ so that the stationery state becomes,
\[\hat{\rho}(t)=\begin{pmatrix}\rho_{11}(0) & 0 \\ 0 & \rho_{22}(0)\end{pmatrix} ,\hspace{0.2cm} t\rightarrow\infty\]
So, the stationery state which is described by the density operator is a diagonal state.
\\ \indent Similarly, the solution of the Lindblad Master equation has been carried out for the two photon states.In the case of two photon system we have $N=2$ so, the invariant subspaces will be 3 dimensional and as a result the density matrix will be a $3\times3$ hermitian matrix. Again we choose the two mode fock basis i.e. $\ket{N,k}=\ket{k}_{+}\otimes\ket{N-k}_{-}$ with $N=2$ and $k=0,1,2$, being the eigenstates of $\hat{S}_{0}$ and $\hat{S}_3$ for the matrix representation of the necessary operators.
Let, the density matrix be, $\hat{\rho}=\begin{pmatrix}
\rho_{11} & \rho_{12} & \rho_{13} \\ \rho_{21} & \rho_{22} & \rho_{23} \\ \rho_{31} & \rho_{32} & \rho_{33}
\end{pmatrix}$.
\\ \indent The basis states are, $\ket{2,2}=\ket{2}_{+}\otimes\ket{0}_{-}=\ket{1}$,$\ket{2,0}=\ket{0}_{+}\otimes\ket{2}_{-}=\ket{2}$ and $\ket{2,1}=\ket{1}_{+}\otimes\ket{1}_{-}=\ket{3}$.The matrix representations for the number operators $\hat{a}_{+}^\dag\hat{a}_{+}$ and $\hat{a}_{-}^\dag\hat{a}_{-}$ in the basis set $\{\ket{1},\ket{2},\ket{3}$ are as follows,
\[\big(\hat{a}_{+}^\dag\hat{a}_{+}\big)_{ij}=\begin{pmatrix}
\bra{2,2}\hat{a}_{+}^\dag\hat{a}_{+}\ket{2,2} & \bra{2,2}\hat{a}_{+}^\dag\hat{a}_{+}\ket{2,0} & \bra{2,2}\hat{a}_{+}^\dag\hat{a}_{+}\ket{2,1} \\ \bra{2,0}\hat{a}_{+}^\dag\hat{a}_{+}\ket{2,2} & \bra{2,0}\hat{a}_{+}^\dag\hat{a}_{+}\ket{2,0} & \bra{2,0}\hat{a}_{+}^\dag\hat{a}_{+}\ket{2,1} \\ \bra{2,1}\hat{a}_{+}^\dag\hat{a}_{+}\ket{2,2} & \bra{2,1}\hat{a}_{+}^\dag\hat{a}_{+}\ket{2,0} & \bra{2,1}\hat{a}_{+}^\dag\hat{a}_{+}\ket{2,1}
\end{pmatrix}=\begin{pmatrix}2 & 0 & 0 \\ 0 & 0 & 0 \\ 0 & 0 & 1\end{pmatrix}\]
similarly the matrix representation of $\hat{a}_{-}^\dag\hat{a}_{-}$ is given by,
\[\big(\hat{a}_{-}^\dag\hat{a}_{-}\big)_{ij}=\begin{pmatrix}
\bra{2,2}\hat{a}_{-}^\dag\hat{a}_{-}\ket{2,2} & \bra{2,2}\hat{a}_{-}^\dag\hat{a}_{-}\ket{2,0} & \bra{2,2}\hat{a}_{-}^\dag\hat{a}_{-}\ket{2,1} \\ \bra{2,0}\hat{a}_{-}^\dag\hat{a}_{-}\ket{2,2} & \bra{2,0}\hat{a}_{-}^\dag\hat{a}_{-}\ket{2,0} & \bra{2,0}\hat{a}_{-}^\dag\hat{a}_{-}\ket{2,1} \\ \bra{2,1}\hat{a}_{-}^\dag\hat{a}_{-}\ket{2,2} & \bra{2,1}\hat{a}_{-}^\dag\hat{a}_{-}\ket{2,0} & \bra{2,1}\hat{a}_{-}^\dag\hat{a}_{-}\ket{2,1}
\end{pmatrix}=\begin{pmatrix}0 & 0 & 0 \\ 0 & 2 & 0 \\ 0 & 0 & 1\end{pmatrix}\]
The stokes operator $\hat{S_3}$ will be $\hat{S}_3=\big[\hat{a}_{+}^\dag\hat{a}_{+}-\hat{a}_{-}^\dag\hat{a}_{-}\big]=2\begin{pmatrix}
1 & 0 & 0 \\ 0 & -1 & 0 \\ 0 & 0 & 0\end{pmatrix}$
Now simple ,matrix multiplication leads to,
\[\mathcal{L}\big[\hat{a}_{s}^\dag\hat{a}_s\big]\hat{\rho}=2\hat{a}_{s}^\dag\hat{a}_s\hat{\rho}\hat{a}^\dag_s\hat{a}_s-\{\hat{a}_{s}^\dag\hat{a}_s\hat{a}_{s}^\dag\hat{a}_s,\hat{\rho}\}=\begin{pmatrix} 0 & -4\rho_{12}(t) & -\rho_{13}(t) \\ -4\rho_{21}(t) & 0 & -\rho_{23}(t) \\ -\rho_{31}(t) & -\rho_{32}(t) & 0\end{pmatrix} \hspace{0.2cm},s=+\]
and,
\[\mathcal{L}\big[\hat{a}_{s}^\dag\hat{a}_s\big]\hat{\rho}=2\hat{a}_{s}^\dag\hat{a}_s\hat{\rho}\hat{a}^\dag_s\hat{a}_s-\{\hat{a}_{s}^\dag\hat{a}_s\hat{a}_{s}^\dag\hat{a}_s,\hat{\rho}\}=\begin{pmatrix} 0 & -4\rho_{12}(t) & -\rho_{13}(t) \\ -4\rho_{21}(t) & 0 & -\rho_{23}(t) \\ -\rho_{31}(t) & -\rho_{32}(t) & 0\end{pmatrix} \hspace{0.2cm},s=-\]
Along with this the Master equation [\ref{eq:5.21}] becomes,
\begin{equation}
\begin{split}
\Dot{\hat{\rho}}(t)& =\begin{pmatrix}
\Dot{\rho_{11}}(t) & \Dot{\rho_{12}}(t) &\Dot{ \rho_{13}}(t) \\ \Dot{\rho_{21}}(t) & \Dot{\rho_{22}}(t) & \Dot{\rho_{23}}(t) \\ \Dot{\rho_{31}}(t) & \Dot{\rho_{32}}(t) &\Dot{ \rho_{33}}(t)
\end{pmatrix} \\
&= \frac{\gamma_{+}}{2}\begin{pmatrix} 0 & -4\rho_{12}(t) & -\rho_{13}(t) \\ -4\rho_{21}(t) & 0 & -\rho_{23}(t) \\ -\rho_{31}(t) & -\rho_{32}(t) & 0\end{pmatrix}+\frac{\gamma_{-}}{2}\begin{pmatrix} 0 & -4\rho_{12}(t) & -\rho_{13}(t) \\ -4\rho_{21}(t) & 0 & -\rho_{23}(t) \\ -\rho_{31}(t) & -\rho_{32}(t) & 0\end{pmatrix}
\end{split}
\end{equation}
Compairing the matrices element-wise we get the set of first order differential equations as follows,
\begin{eqnarray}
\Dot{\rho_{11}}(t)&=&0, \\
\Dot{\rho_{22}}(t)&=&0, \\
\Dot{\rho_{33}}(t)&=&0, \\
\Dot{\rho_{12}}(t)&=&-2\big(\gamma_{+}+\gamma_{-}\big)\rho_{12}(t),\\
\Dot{\rho_{21}}(t)&=&-2\big(\gamma_{+}+\gamma_{-}\big)\rho_{21}(t), \\
\Dot{\rho_{13}}(t)&=&-\frac{1}{2}\big(\gamma_{+}+\gamma_{-}\big)\rho_{13}(t), \\
\Dot{\rho_{31}}(t)&=&-\frac{1}{2}\big(\gamma_{+}+\gamma_{-}\big)\rho_{31}(t), \\
\Dot{\rho_{23}}(t)&=&-\frac{1}{2}\big(\gamma_{+}+\gamma_{-}\big)\rho_{23}(t), \\
\Dot{\rho_{32}}(t)&=&-\frac{1}{2}\big(\gamma_{+}+\gamma_{-}\big)\rho_{32}(t)
\end{eqnarray}
The Solution to the above equations leads to,
\begin{eqnarray}
\label{eq:5.45-5.50}
\rho_{11}(t)&=&\rho_{11}(0), \\
\rho_{22}(t)&=&\rho_{22}(0), \\
\rho_{33}(t)&=&\rho_{33}(0), \\
\rho_{12}(t)&=&\rho_{12}(0)\exp{\{-2(\gamma_{+}+\gamma_{-})t\}}, \\
\rho_{21}(t)&=&\rho_{21}(0)\exp{\{-2(\gamma_{+}+\gamma_{-})t\}}, \\
\rho_{13}(t)&=&\rho_{13}(0)\exp{\{-(\gamma_{+}+\gamma_{-})t/2\}}, \\
\rho_{31}(t)&=&\rho_{31}(0)\exp{\{-(\gamma_{+}+\gamma_{-})t/2\}}, \\
\rho_{23}(t)&=&\rho_{23}(0)\exp{\{-(\gamma_{+}+\gamma_{-})t/2\}}, \\
\rho_{32}(t)&=&\rho_{32}(0)\exp{\{-(\gamma_{+}+\gamma_{-})t/2\}}
\end{eqnarray}
Let us calculate the degree of polarization $P$ as a function of time for the single photon state. 
\\ \indent By definition of degree of polarization we have, $P(t)=\frac{\sqrt{\braket{\hat{S_1}}^2+{\braket{\hat{S_2}}^2}+{\braket{\hat{S_3}}^2}}}{\braket{\hat{S_0}}}$ with, $\braket{\hat{S}_{i}(t)}$=Tr$\big[\hat{\rho}(t)\hat{S_i}\big]$. \\
\\ \indent Now for the single photon state the Degree of polarization has been calculated as follows,
In case of two mode polarization for the single photon state, the matrix representation of $\hat{S}_{i}$ for, i=1,2,3 will resemble the forms of the Pauli matrices i.e. $\sigma_{i}$ which means,
\[\hat{S}_1=\begin{pmatrix}0 & 1 \\ 1 & 0\end{pmatrix},\hat{S}_2=\begin{pmatrix}0 & -i \\ i & 0\end{pmatrix},\hat{S}_3=\begin{pmatrix}1 & 0 \\ 0 & -1\end{pmatrix}\]
Then for single photon state from~[\ref{eq:5.26}] and [\ref{eq:5.32}] we have,
\[\braket{\hat{S}_1}=Tr[\hat{\rho}(t)\hat{S}_1]=(\rho_{12}(t)+\rho_{21}(t))=s_x(t)=s_x(0)\exp{\{-\gamma^\prime t/2\}}\]
\[\braket{\hat{S}_2}=Tr[\hat{\rho}(t)\hat{S}_2]=i(\rho_{12}(t)-\rho_{21}(t))=s_y(t)=s_y(0)\exp{\{-\gamma^\prime t/2\}}\]
\[\braket{\hat{S}_3}=Tr[\hat{\rho}(t)\hat{S}_3]=(\rho_{11}(t)-\rho_{22}(t))=s_z(0)\]
\[\braket{\hat{S}_0}=Tr[\hat{\rho}(t)\hat{S}_3]=(\rho_{11}(t)+\rho_{22}(t))=Tr[\hat{\rho}(t)]=Tr[\hat{\rho}(0)]=1.\]
Where, we have defined the total decoherence rate: $\gamma^\prime=\sum_{s=\pm}\gamma_s$.
By using~[\ref{eq:4.22}] the degree of polarization will be,
\begin{equation}
\mathbb{P}(t)=\sqrt{\big(s_x(0)\big)^2\exp{\{-\gamma^\prime t\}}+\big(s_y(0)\big)^2\exp{\{-\gamma^\prime t\}}+\big(s_z(0)\big)^2}
\end{equation}
As,we can see that the degree of polarization decreases with time so,it can explain the depolarization process.
\subsection{Calculation of Geometric Phase}
Once the matrix elements of the density operator are found~[\ref{eq:5.45-5.50}] then the Geometric phase can be calculated easily. We have obtained the expression of the Geometric phase for a three level quantum system undergoing an quasicyclic evolution given by,
\begin{equation}
   \begin{split}
\gamma_g(\mathcal{C})&=-\Im\oint_{\mathcal{C}}\frac{\braket{\psi|d|\psi}}{\braket{\psi|\psi}} \\
&= -\oint_{\mathcal{C}}\Big\{\sin^2{\theta}[d(\alpha-\gamma)\cos^2{\phi}+d(\beta-\chi)\sin^2{\theta}]+\cos^2{\theta}d\xi\Big\}
\end{split} 
\end{equation}
The Geometric phase is expressed in terms of the Bloch parameters related to the matrix elements of the density operator by the equations~[\ref{eq:2.1.30a-2.1.30e}] used to parameterise the $8$ dimensional Bloch sphere. Now as the Matrix elements are time dependent the Bloch parameters will also be time dependent \cite{jiang2010geometric} so, by expressing the matrix elements in terms of the Bloch parameters and by using the formulae one can determine the Geometric Phase for the system undergoing Depolarization.But here in the present situation the Bloch parameters are time independent although the matrix elements are time dependent because of the choice of Parametraisation used in the present context.

\section{Modeling Depolarization with a non resonant randomly distributed atomic bath}
\setcounter{equation}{0}
\renewcommand{\theequation}{8.\arabic{equation}}
\indent In the classical picture we describe the phenomena of depolarization resulting in an effective anisotropy of the material medium due to the decorrelation of the phases associated with the Electric filed vector. Now, we will consider a more realistic model where we will assume that a plane polarized electromagnetic wave is propagating through a material medium represented by a randomized distribution of two level atoms along with a dispersive coupling as the amount of energy exchange between the field and atoms is very negligible.
\\ \indent Let us consider the frequency of the electromagnetic wave be $\omega$ and the atomic transition frequency between the two levels of the ath atom be $\omega_{a}$.In the natural unit system ($\hbar=1$) the Hamiltonian of the super system \cite{klimov2008quantum}\cite{klimov2006simple} (system coupled with the atomic reservoir) can be written as the sum of the field Hamiltonian,the bath Hamiltonian and the interaction Hamiltonian describing the coupling between the system and the reservoir.So that,
\begin{equation}
\label{eq:8.1}
    \hat{H}_{Tot}=\hat{H}_{field}+\hat{H}_{atomic}+\hat{H}_{int}
\end{equation}
where,
\begin{eqnarray}
\label{eq:8.2-8.4}
\hat{H}_{field}&=&\omega\sum_{\lambda=\pm}\hat{a}_{\lambda}^\dag\hat{a}_{\lambda}, \\
\hat{H}_{atomic}&=&\frac{1}{2}\sum_{a}\omega_{a}\hat{\sigma}_{a}^{z}, \\
\hat{H}_{int}&=&\sum_{a}\sum_{\lambda=\pm}\big(g_{a\lambda}\hat{\sigma}_{a}^{-}\hat{a}_{\lambda}^\dag+g_{a\lambda}^*\hat{\sigma}_{a}^{+}\hat{a}_{\lambda}\big)
\end{eqnarray}
Where, the field operator resembles the Hamiltonian of the two dimensional isotropic harmonic oscillator and the Hamiltonian of a single two level atom is expressed in terms of standard Pauli operators.The Interaction Hamiltonian $(\hat{H}_{int})$ has been written in Dipole and Rotating Wave approximations \cite{born2013principles}.The sum over $a$ goes for all the atoms in the atomic reservoir. The term $g_{a\lambda}$ represents the coupling of the ath atom with the field mode $\lambda=\pm$ which is in general a complex quantity.As the atoms are randomly distributed it is reasonable to use random phase approximation \cite{ashcroft2010solid} for which $g_{a\lambda}=|g_{a}|e^{i\lambda\phi_{a}/2}$ for $\lambda=\pm$ which means that the coupling constants carries random phases. It is pointed out that the random phase approximation works well in the long wavelength limit.
\\ \indent It is well known that the atoms decay irreversibly.This is usually due to the interaction of the atom with the thermo-electromagnetic environment if we assume a thermal distribution of the randomly distributed atoms. Then in the Interaction picture of Quantum Mechanics the time evolution of the Density operator for the super-system will be given by the Master equation of this type:
\begin{equation}
\label{eq:8.5}
    \Dot{\hat{\rho}}(t)=-i\big[\hat{H}_{Tot},\hat{\rho}_{Tot}\big]+\sum_{a}\frac{\gamma_{a}}{2}\{\big(\Bar{n_{a}}+1\big)\mathcal{L}[\hat{\sigma_{a}}^{-}]\hat{\rho}_{Tot}+\Bar{n_{a}}\mathcal{L}[\hat{\sigma_{a}}^{+}]\hat{\rho}_{Tot}\}
\end{equation}
Here,$\hat{\sigma}_{a}^{\pm}$ being the spin raising and lowering operators defined in terms of the Pauli matrices such that, $\hat{\sigma}_{a}^{\pm}=\hat{\sigma}_{a}^{x}\pm\hat{\sigma}_{a}^{y}$.
Along with the definition of the Lindblad Superoperators given by,
\begin{equation}
    \mathcal{L}\big[\hat{C}\big]=2\hat{C}\hat{\rho}\hat{C}^\dag-\{\hat{C}^\dag\hat{C},\hat{\rho}\}
\end{equation}
Here, $\gamma_a$ being the decay constant of the ath atom due to its coupling with the thermal environment containing $\Bar{n_{a}}$ excitations, here we have different Bosonic reservoirs associated with individual atoms in order to avoid any collective effect.
For further simplicity we will consider the high temperature limit for which $\bar{n_{a}}\gg 1$ \cite{kofman2001relaxation}, so that the effect of the spontaneous emission can be disregarded compared to the process of stimulated emission.Emission into the reservoir and the absorption from the reservoir then becomes identical and finally they balance each other in the stationery state. The rate of absorption or emission entirely depends on the initial population of the atomic states.
Lets define the off resonant detuning parameter $\Delta_{a}=\omega_{a}-\omega$ and in case of far off resonant regime we have $|g_a|\ll|\Delta_{a}|$ and in this limit the elimination of the atomic variables using the adiabatic approximation and the averaging over the random phases \cite{klimov2006simple} leads to the Master equation~[\ref{eq:8.5}] reduces to the following form,
\begin{equation}
\label{eq:8.7}
    \Dot{\hat{\rho}}(t)=-i\omega\big[\hat{S}_{0},\hat{\rho}(t)\big]+2\gamma\mathcal{L}\big[\hat{S_{0}}\big]\hat{\rho}(t)+\gamma\mathcal{L}\big[\hat{S_{+}}\big]\hat{\rho}(t)+\gamma\mathcal{L}\big[\hat{S_{-}}\big]\hat{\rho}(t)
\end{equation}
Where,$\gamma$ is the decoherence rate and $\hat{\rho}(t)$=Tr$_{at}\big[\hat{\rho}_{Tot}\big]$ is the reduced density operator describing the evolution of the field mode and $\hat{S}_{\pm}=\hat{S}_{1}\pm i\hat{S}_{2}$.
\begin{equation}
    \gamma=\sum_{a}\frac{|g_{a}|^4}{\gamma_a\Delta^{2}\Bar{n_a}}
\end{equation}
The term $\mathcal{L}[\hat{S}_{\pm}]$ in the Master equation~[\ref{eq:8.7}] describes the depolarization rates in each invariant subspaces.The first term on the R.H.S of the Master equation describes the unitary or the free evolution of the density operator known as the Louivillian  where as, the second term describes the non-unitary evolution of $\hat{\rho}(t)$, known as the Lindbladian.
We will solve the Master Equation for the single photon state.For the single photon state we have $N=1$ so, the dimensionality of the invariant subspace will be $2$ so, the entire Hilbert space will split into invariant subspaces of dimension 2. So, we have to solve the master equation in the 2 dimensional subspace and the density operator will be a $2\times2$ matrix in the basis $\ket{N,k}$ as been done earlier.
\\ \indent For, the single photon state the Stokes operators and the raising lowering operators are given by,
\[\hat{S}_1=\begin{pmatrix}0 & 1 \\ 1 & 0\end{pmatrix},\hat{S}_2=\begin{pmatrix}0 & -i \\ i & 0\end{pmatrix},\hat{S}_3=\begin{pmatrix}1 & 0 \\ 0 & -1\end{pmatrix}\]
\[\hat{S}_0=\begin{pmatrix}1 & 0 \\ 0 & 1\end{pmatrix},\hat{S}_{+}=\begin{pmatrix}0 & 2 \\ 0 & 0\end{pmatrix},\hat{S}_{-}=\begin{pmatrix}0 & 0 \\ 2 & 0\end{pmatrix}\]
Simple Matrix multiplication with $\hat{\rho}(t)=\begin{pmatrix} \rho_{11}(t) & \rho_{12}(t) \\ \rho_{21}(t) & \rho_{22}(t)\end{pmatrix}$ leads to,
\[\big[\hat{S}_{0},\hat{\rho}\big]=0\]
\[\mathcal{L}\big[\hat{S}_{0}]\hat{\rho}=2\hat{S}_{0}\hat{\rho}\hat{S}_{0}-\{\hat{S}_{0}\hat{S}_{0},\hat{\rho}\}=0\]
\[\mathcal{L}\big[\hat{S}_{+}]\hat{\rho}=2\hat{S}_{+}\hat{\rho}\hat{S}_{-}-\{\hat{S}_{-}\hat{S}_{+},\hat{\rho}\}=\begin{pmatrix}
8\rho_{22}(t) & -4\rho_{12}(t) \\ -4\rho_{21}(t) & -8\rho_{22}(t)\end{pmatrix}\]
\[\mathcal{L}\big[\hat{S}_{-}]\hat{\rho}=2\hat{S}_{-}\hat{\rho}\hat{S}_{+}-\{\hat{S}_{+}\hat{S}_{-},\hat{\rho}\}=\begin{pmatrix}
-8\rho_{11}(t) & -4\rho_{12}(t) \\ -4\rho_{21}(t) & -8\rho_{11}(t)\end{pmatrix}\]
And the Master equation given by~[\ref{eq:8.7}] becomes,
\begin{equation}
    \begin{pmatrix}\Dot{\rho_{11}}(t) & \Dot{\rho_{12}}(t) \\ \Dot{\rho_{21}}(t) & \Dot{\rho_{22}}(t)\end{pmatrix}=\gamma\begin{pmatrix}
    8\rho_{22}(t)-8\rho_{11}(t) & -8\rho_{12}(t) \\ -8\rho_{21}(t) & -8\rho_{22}(t)+8\rho_{11}(t)\end{pmatrix}
\end{equation}
The master equation preserves the positivity of the density operator i.e. Tr$[\Dot{\hat{\rho}}(t)]=0$ and the positive definiteness of $\Dot{\hat{\rho}}(t)$ is guaranteed by the hermiticity of $\Dot{\hat{\rho}}(t)$ i.e. $\Dot{\hat{\rho}}^\dag(t)=\Dot{\hat{\rho}}(t)$ 
\\ \indent Now, with $\hat{\rho}(t)=\frac{1}{2}\begin{pmatrix} 1+s_z(t) & s_x(t)-is_y(t) \\ s_x(t)+is_y(t) & 1-s_z(t)\end{pmatrix}$ the Master equation leads to the set of ordinary differential equations given by,
\begin{eqnarray}
\Dot{s_x}(t)&=&-8\gamma s_x(t) , \\
\Dot{s_y}(t)&=&-8\gamma s_y(t), \\
\Dot{s_z}(t)&=&-16\gamma s_z(t)
\end{eqnarray}
And the solution to the equations with $\gamma^\prime=8\gamma$ gives,
\begin{eqnarray}
s_x(t)&=&s_x(0)e^{-\gamma^\prime t}, \\
s_y(t)&=&s_y(0)e^{-\gamma^\prime t}, \\
s_z(t)&=&s_z(0)e^{-2\gamma^\prime t}
\end{eqnarray}
Then for single photon state we have,
\[\braket{\hat{S}_1}=Tr[\hat{\rho}(t)\hat{S}_1]=(\rho_{12}(t)+\rho_{21}(t))=s_x(t)=s_x(0)\exp{\{-\gamma^\prime t\}}\]
\[\braket{\hat{S}_2}=Tr[\hat{\rho}(t)\hat{S}_2]=i(\rho_{12}(t)-\rho_{21}(t))=s_y(t)=s_y(0)\exp{\{-\gamma^\prime t}\}\]
\[\braket{\hat{S}_3}=Tr[\hat{\rho}(t)\hat{S}_3]=(\rho_{11}(t)-\rho_{22}(t))=s_z(t)= s_z(0)\exp{\{-2\gamma^\prime t}\}\]
\[\braket{\hat{S}_0}=Tr[\hat{\rho}(t)\hat{S}_3]=(\rho_{11}(t)+\rho_{22}(t))=Tr[\hat{\rho}(t)]=Tr[\hat{\rho}(0)]=1.\]
By the definition given by~[\ref{eq:4.22}] the degree of polarization will be,
\begin{equation}
\mathbb{P}(t)=\exp{\{-\gamma^\prime t\}}\sqrt{\big(s_x(0)\big)^2+\big(s_y(0)\big)^2+\big(s_z(0)\big)^2\exp{\{-2\gamma^\prime t\}}}
\end{equation}
So, The degree of polarization for this single mode field for the single photon state will evolve accordingly.This degree tends then to zero with a typical time scale $\gamma^{-1}$,which is exceedingly large. So, this model can successfully describe the phenomena of the Depolarization.

\section{\textbf{Discussions And Conclusions}}
From the calculations of the geometric phase for the three level open system of mixed states for the quasicyclic evolution with quasicyclicity $T=\frac{2\pi}{\Omega}$ it is evident that the becausse of the intercation between the system and the surroundings the system is no longer undergoing a cyclic evolution, where the exponent decay factors are included in the matrix elements of the density operator of the system, the components of the Bloch vector and the nonunit state vectors.It can be easily seen by putting the decaying matrix elements i.e. $\rho_{ij}(t)$ on equation~[\ref{eq:2.1.25}] that the Bloch radius also decays with time and it corresponds to the fact that the physical state of the system is having a transition from the pure state to a mixed state. The mixed degree is independent of the decay rate.
\\ \indent The approach towards the generalisation of the geometric phase for the mixed state requires the extensive usage of the kinematic approach due to Mukunda and Simon altough it was done for the pure states we can use the similar idea here. Interesting to note that the expression of the geometric phase is found to be gauge invariant and independent of the choice of the gauge parameter; for both the global and local gauge transformation and also the reparametraisation invariance.
\\ \indent The off diagonal matrix elements of the density operator $\rho_{mn};\hspace{0.1cm} m\neq n$ in the eigenasis of the hamiltonian which measures the amount of overlap between the mth and nth energy level respectively decays as a function of time and it is obvious from the sign of the phase depends on the population determined by the bloch vector components $w_2(t)$ and $w_3(t)$ which involves the diagonal matrix elements of the density operator which remains constant irrespective of time as we have found above ,only depends on the value of $\delta_1,\delta_2,\delta_3$, $w_1(t)=\rho_{11}(t)-\rho_{22}(t)-\rho_{33}(t)$ and $w_2(t)=\rho_{11}(t)-\rho_{22}(t)$.
\\ \indent So, depending on the sign of the bloch vector components associated with the measure of the population inversion the sign of the phase will change accordingly. The Berry phases of the mixed states given by equation~[\ref{eq:4.20}] doesn't depend on all bloch parameters, it is independent of the radial bloch parameter $r$ and as a result it holds for any mixed state that is being mapped inside the sphere. Berry phase depends on the decay rate, if the decay rate increases the berry phase decreases but the absolute value of the Berry phase increases which is evident from the relation between the Bloch radius and the Berry phase.
\\ \indent In the context of the Lindblad Master equation we have assumed the dephasing noise can be represented by a single type of Lindbald operator or quantum jump operator which is a comparatively easy situation compared to the case where the master equation involves all $8$ lindblad operator, although this will give us a more general situation when $\Gamma_i=\sqrt{\eta_i(t)}\lambda_i$ represents the coupling to the environment.
\\ \indent It is interesting to note that when the population inversions have differing signs i.e. $w_1(t)<0$ and $w_2(t)>0$, the Berry phase also changes as a function of the decaying parameter.The nonvanishing value of the Geometric phase establishes the fact that the quantum system under the consideration retains a memory of its evolution in terms of the Berry phase of the Mixed state.Thus, our definition of the geometric phase for the three level system of mixed state may have a hidden rich physics.
\\ \indent As we have mentioned previously that the parametrisation used in the present context is not the only possible choice we can use any other parametraisation which satisfies the equation of the bloch sphere.We will get the similar kind of results for the geometric phase under those parametraisation and this must happen as the geometric phase does not depend on how we have parametrized the bloch sphere, it will be the same irrespective of the choice of it. \\ \indent In order to derive the expression of the geometric phase we have used the Pancharatnams formulae and we have obtained the general expression of the geometric phase more, specifically the Berry Phase for the mixed state which we have expressed as the closed path integral over the Mead-Berry connection one form defined over the manifold $\mathcal{B}$ which we may say an extension of Berry's framework including the conditions of adiabeticity and cyclicity. And along with that under a specific Gauge transformation the Pancharatnam's formulae for the geometric phase reduces to the well known formulae of  Aharonov-Anandan phase we obtain in case of pure state with the condition of cyclicity.
\\ \indent It is interesting to note the universal feature of the geometric which as expected only depend on the choice of the closed curve $\mathcal{C}$ for the quasicyclic evolution of the system lying in the manifold $\mathcal{B}$. Previously the kinematic approach has been used to calculate the geometric phase using the purification approach, here the geometric phase formulae obtained before for the pure states  has been generalized.
\\ \indent In the present context we can easily identify the expression of the geometric phase by dropping the total phase term which resembles the closed mathematical form of the geometric phase obtained in the Berry's and Aharonov-Anandan's framework under the limiting of pure quantum states.   
\\ \indent  In conclusions, a way is expanded to calculate the geometric phase for the three level open quantum system including the interaction of the system with the environment, which was previously done for the two level quantum system.Previously the calculation for the calculation has been done for the pure state using the kinematic approach, the similar idea have been used  to extend it for the mixed states.By mapping the interior points of the eight-dimensional Poincare sphere onto the field amplitudes.
\\ \indent The mapping of the mixed states in the interior of the sphere can be established  by connecting the nonunit vector rays in $\mathcal{B}\subset\mathcal{H}^3$ with the density matrices. The geometric (Berry) phases, defined according to the vector rays only depends on the geometric structure on this space. Interestingly, any state can be mapped either on the surface of the sphere or,in the interior of the sphere and as discussed earlier each mapped state can be described by (generally, a non-normalized) Bloch vector oriented towards a point at which the state is mapped  from the origin of the Bloch sphere. The formulae obtained is in agreement with the Berry phase, Pancharatnam's phase and Aharonov-Anandan phases.\vspace{0.5cm}
\footnote{The figure showing the variation of the bloch radius as a function of decay parameter at the quasicyclicity $T=2\pi/\Omega$,with the corresponding population inversions of system are $w_1(t)>0$ and $w_2(t)>0$.}
\par\indent In the context of Depolarisation we have discussed the phenomena using simple model, first by considering the interaction Hamiltonian of specific type~[\ref{eq:7.1.4}] [\ref{eq:7.1.20}], discussed their drawbacks and then presented a model which describe the dispersive coupling of the plane polarized quantized electromagnetic wave with the reservoir consisting of the collection of randomly distributed two level atoms described by the Hamiltonian~[\ref{eq:8.1}] [\ref{eq:8.2-8.4}] and the resulting master equation~[\ref{eq:8.7}] has quite an appealing structure as it preserves the Invariant sub spaces as a result of system's dynamic evolution. And finally by solving the master equations i.e. by finding the matrix elements of the density operator we are able to calculate the Geometric phase for such three level systems assuming a quasicyclic evolution incorporating the phenomena of Depolarization modeled with suitable and legitimate approximations for single photon and two photon systems.

\section{Future Works}
We will try to extend the theoretical models describing the  Depolarization dynamics for the Multi mode fields as well as for the non-paraxial beams for which the propagation direction is not well-defined i.e. the case of Quantum Polarization for the Three dimensional fields or three mode fields using the SU(3) polarization picture and also to explain the Depolarization dynamics for such beams.
\section{\textbf{Appendix}}
For the convenience of the reader some aspects of the Quantum states, Idea of homomorphic mapping between the $SU(2)\rightarrow SO(3)$, basics of the $SU(2)$,$SU(3)$ groups,Lindblad operators and the Interaction Picture of Quantum Mechanics has been briefly described. 
\subsection{\textbf{Appendix A: Description of Pure and Mixed Quantum States}}
\setcounter{equation}{0}
\renewcommand{\theequation}{11.1.\arabic{equation}}
The quantum states are classified into two one is called the pure state and the other being the Mixed state. There are, two alternative ways to represent the state of a quantum system one is by associating a state vector with that system and the other is to define a density operator that will represent the state of the system. Very often in the context of quantum optics we do specify the state of a light beam by defining a density operator and when we study for example, the quantum mechanics of an  electron in a constant magnetic field we solve the time dependent schrodinger equation i.e. where the time evolution of the quantum state is observed. So, in the later case it is convenient to represent the state of the system by a state vector while in the former example it is easy to deal with the density operator. Now, there must be some way to distinguish the pure and the mixed states.Let us consider the example of the qubit which is a two level quantum system and the basis states $\ket{0}$ and $\ket{1}$ are the so called 'pure states'.Any state which can be written as a linear combination of two pure states that will also be a pure state.
\\ \indent For example the general state of the qubit say $\ket{\psi}$ can be written as a linear combination of two basis states $\ket{0}$ and $\ket{1}$ and it will be a pure state.So, we have,
\begin{equation}
\ket{\psi}=c_1\ket{0}+c_2\ket{1},\hspace{0.1cm} c_1,c_2\in\mathbb{C}
\end{equation}
which is a pure state. We can even consider other examples like the spin states of an electron $\ket{\uparrow}$ and $\ket{\downarrow}$, the eigenstates of the $\hat{S_z}$ operator conventionally choosen as the basis states which are also pure states so the generic spin state of an electron can be written as a linear combination of the basis states and it will also be a pure state. It is interesting to note that any pure state can be written as the linear combination of two or, more pure states in general, the pure states obey the superposition principle of quantum mechanics.
We have not mention anything so far what is a pure state and what is a mixed state, trying to explain them in terms of simple physical examples. Now, let us consider another example which is a mixed state. Let us consider a system of 3 electrons which are spin $\frac{1}{2}$ particles and they can have only two possible orientation either up or down. Now let us switch on a constant magnetic field towards the z direction and now the electrons can either stay in the up configuration i.e. the alignment of the spin in a direction parallel to the magnetic field or antiparallel with respect to the magnetic field. Due to the spin magnetic moment of the electron it will have some potential energy $V$ given by, $V=-\mu\cdot B$ which can have two different values corresponding to the alignment of the spin with respect to the magnetic field, either $-\mu B$ or, $\mu B$ for a single electron. So, the system can have four possible energy configurations corresponding to the different configurations of the spin and the energies are, $V_1=-3\mu B$,$V_2=3\mu B$,$V_3=-\mu B$,$V_4=\mu B$. From quantum mechanical point of view by considering the macroscopic variable i.e. the energy of the system we can say that it's a four level quantum system.
\\ \indent Now from a microscopic point of view the energy configurations are associated with the microscopic electronic configuration of the 3 electrons. When all the electrons are in the up configuration then the total energy of the system will be $V_1=-3\mu B$, when all the electrons are in the down configuration i.e.antiparellal with respect to the external magnetic field then the energy of the system will be, $V_2=+3\mu B$ and when any two of the electrons are in the up configuration and the other in the down configuration then the total energy of the system will be $V_3=-\mu B$ and when two of the electrons are in the down spin configuration and the other being in the up configuration then the total energy of the system will be $V_4=+\mu B$. So, the pure states are the different energy states of the system lets level them by $\ket{1},\ket{2},\ket{3},\ket{4}$ corresponding to the total energies $V_1,V_2,V_3,V_4$ respectively and this does not require the microscopic information regarding the alignment of the electron spins because the energy state $\ket{3}$ with energy $V_3$ can be achieved with any two electrons in the up spin configurations and the remainder in the down state and similarly for the energy state $\ket{4}$ with energy $V_4$ can be achieved with  any two electrons in the down spin configuration and the remainder in the down but it can be one of the three electrons.
\\ \indent So, the energy states with energy $V_3,V_4$ are examples of mixed states when it is viewed microscopically as the energy states has 3 different microstates having the same energy.
Now, let us introduce the notion of the pure and mixed quantum states mathematically. It is better to define them in terms of the density operators because as in case of the pure states which can be written as a linear combination of the pure states but the same does not hold for the mixed states.A mixed state cannot be written as a linear combination of the pure states.We can say that a quantum state is a pure state if the state can be achieved by an unique configuration i.e. the state doesn't have further microstates. 
\\ \indent In order to quantify the purity of a quantum state we define a quantity $\gamma$ known as the purity of a quantum state as follows,
\begin{equation}
\gamma=Tr[\hat{\rho}^2]
\end{equation}
\begin{itemize}
    \item For the pure states we have $\hat{\rho}^2=\hat{\rho}$ and $\gamma=Tr[\hat{\rho}^2]=Tr[\hat{\rho}]=1$.
    \item For, the mixed states $\hat{\rho}^2\neq\hat{\rho}$ and $\gamma=Tr[\hat{\rho}^2]\neq 1$.
\end{itemize}
The purity of a normalized quantum state satisfies ${\displaystyle {\frac {1}{d}}\leq \gamma \leq 1}$ where $d$ be the dimensionality of the hilbert space upon which the state is defined. The upper and the lower bound corresponds to the completely mixed and the completely pure states. Now, in order to introduce the idea of the Pure and mixed states in quantum mechanics let us revisit the Density operator and the density matrix formalism discussed previously in the section~[\ref{sec:2.1.1}], where we have a system and the large number of virtual or, mental copies of the same system constituting an ensemble. If the size of the ensemble be $N$ and the state vector corresponding to the kth member of the ensemble be denoted by $\ket{\psi^{(k)}(t)}$ then the time evolution of the quantum system is governed by the time dependent schrodinger equation corresponding to the system hamiltonian $\hat{H}$ is  given by,
\begin{equation}
\hat{H}\ket{\psi^{(k)}(t)}=i\hbar\frac{d}{dt}\ket{\psi^{(k)}(t)}
\end{equation}
if the orthonormal eigenstaes of the hamiltonian $\hat{H}$ be denoted by $\{\ket{\phi_n(t)}\}$ for which $\braket{\phi_m(t)|\phi_n(t)}=\delta_{mn}$ then we can write,
\begin{equation}
\ket{\psi^{(k)}(t)}=\sum_{n}a_n^k(t)\ket{\phi_n(t)}
\end{equation}
By, definition, the matrix elements of the density operator can be defined as, 
\begin{eqnarray}
\rho_{mn}(t)=\frac{1}{N}\sum_{k=1}^{N}a_m^k(t)a_n^{k*}(t)\nonumber \\
\hat{\rho}=\frac{1}{N}\sum_k\ket{\psi^{(k)}(t)}\bra{\psi^{(k)}(t)}
\end{eqnarray}
where, the matrix elements of the density operator $\hat{\rho}$ has been defined in the eigenbasis of the hamiltonian.
Now, if the system has only a single microstate then from euation (7.1.5) we can say that the quantum state $\ket{\psi^{(k)}(t)}$ becomes identical for all the members of the ensemble and the matrix elements of the density operator and the operator itself can be defined by,
\begin{eqnarray}
\label{eq:11.1.6}
\rho_{mn}(t)=a_m(t)a_n^{*}(t)\nonumber \\
\hat{\rho}=\ket{\psi(t)}\bra{\psi(t)}
\end{eqnarray}
as,the state becomes independent of the member of the ensemble we can simply drop the label $k$ (for the kth member of the ensemble) and we obtain the expression for the pure state density operator and the matrix element of that operator in the eigenbasis of the hamiltonian.So, if the state of the system can be uniquely specified by a state vector we can say that the system is in the pure state i.e. if the state of the system can be uniquely specified by a single state vector $\ket{\psi(t)}$ then the density operator will be $\hat{\rho}=\ket{\psi(t)}\bra{\psi(t)}$.
so, we can define the matrix element of the density operator $\hat{\rho}$ in the eigenbasis of the Hamiltonian  as, $\rho_{mn}=\braket{\phi_m|\hat{\rho}|\phi_n}$ and exploiting the completeness relation of the normalised eigenbasis of the hamiltonian we van write,
\begin{equation}
\begin{split}
\hat{\rho}&=\hat{I}\hat{\rho}\hat{I} \\
&= \sum_m\ket{\phi_m}\bra{\phi_m}\hat{\rho}\sum_n\ket{\phi_n}\bra{\phi_n} \\
&= \sum_m\sum_n\braket{\phi_m|\hat{\rho}|\phi_n}\ket{\phi_m}\bra{\phi_n} \\
&=\sum_m\sum_n\rho_{mn}\ket{\phi_m}\bra{\phi_n}
\end{split}
\end{equation}
where,$\rho_{mn}=\braket{\phi_m|\hat{\rho}|\phi_n}$ defines  the matrix elements of the density operator.
The time evolution of the density operator is governed by the Liouville equation of Quantum Statistical Mechanics given by,
\begin{equation}
i\hbar\frac{\partial}{\partial{t}}\hat{\rho}(t)=[\hat{\rho},\hat{H}]
\end{equation}
in statistical equilibrium $\frac{\partial{\rho}}{\partial{t}}=\Dot{\rho}=0\Rightarrow{[\hat{\rho},\hat{H}]=0}$ so,The density operator and the Hamiltonian has the simultaneous eigenstates i.e. $\{\ket{\phi_n}\}$ which allows us to write $\rho_{mn}=\braket{\phi_m|\hat{\rho}|\phi_n}=\rho_n\braket{\phi_m|\phi_n}=\rho_n\delta_{mn}$ where, $\rho_n$ be the eigenvalues of the density operator and the density matrix will be diagonal in its own eigenbasis. It is easy to check that the density operator is hermitain i.e.$\hat{\rho}^\dag=\hat{\rho}$ So, finally we have,
\begin{equation}
\hat{\rho}=\sum_n\rho_n\ket{\phi_n}\bra{\phi_n}
\end{equation}
The above equation gives the general expression of the density operator for the Mixed states.It is evident from the expression of the density operator that for the mixed state $\hat{\rho}^2\neq\hat{\rho}$.For, the purpose of normalaization we have $\sum_n\rho_n=1$ with $\rho_n$ be the probability to find the system in the nth eigenstate of the hamiltonian. In the case of the completely mixed states we have, $\rho_n=\frac{1}{d}$ with, d be the dimensionality of the hilbert space over which the state of the system has been defined. So, for completely mixed states the purity gamma will be given by,
\begin{gather}
\hat{\rho}=\sum_m\ket{\phi_m}\bra{\phi_m}=\frac{1}{d}\sum_m\ket{\phi_m}\bra{\phi_m}\nonumber \\
\Rightarrow{\hat{\rho}=\frac{1}{d}I_{d\times d}}\nonumber \\
\text{ which further implies that } Tr[\hat{\rho}^2]=\frac{1}{d^2}Tr[I_{d\times d}]=\frac{1}{d} \label{eq:7.1.6}
\end{gather}
For completely mixed state the system is equally likely to be in any one of its possible eigenstates of the hamiltonian.
It, can be easily checked with the definition of the pure state density operator given by the equation~[\ref{eq:7.1.6}] that if the state $\ket{\psi}$ be normalized then, $Tr[\rho]=1$ and,
\begin{equation}
\hat{\rho}^2=\hat{\rho}\hat{\rho}=\ket{\psi}\braket{\psi|\psi}\ket{\psi}=\ket{\psi}\bra{\psi}=\hat{\rho}
\end{equation}
Although its convenient to work with normalised states and the condition $Tr[\hat{\rho}]=1$ can also hold for the mixed states as well, which implies,
\begin{gather}
\braket{\psi^{(k)}(t)|\psi^{(k)}(t)}=1\nonumber \\
\Rightarrow{\sum_{k=1}^N|a_n^k(t)|^2=1}.
\end{gather}
So, the measure of purity $\gamma=Tr[\hat{\rho}^2]$ clearly gives an idea how to distinguish the pure and the mixed states.If $\gamma=1$ then we will say that it's a pure state and otherwise its a mixed state with different degrees.Alternatively we can say that for a pure state $\hat{\rho}^2=\hat{\rho}$ but this is not true for the mixed state.
\subsection{\textbf{Appendix B: The Basics of the SU(2) and SU(3) Group}}
\setcounter{equation}{0}
\renewcommand{\theequation}{11.2.\arabic{equation}}
\textbf{\underline{The SU(2) Group:}}
The group of all $2\times2$ unitary matrices forms the $U(2)$ or, unitary group.As, we are familiar with the definitions or axioms of the group theory we know that the group of matrices satisfies the axioms of group theory given by,
\begin{itemize}
    \item The Closure Property.
    \item The Associative Property.
    \item The Inverse Property.
    \item The Identity Property.
\end{itemize}
Let, $U$ be the $2\times2$ unitary matrix i.e.$U\in U(2)$.Then,the matrix $U$ will be such that $U^\dag U=I_{2\times2}$. The determinant of the unitary matrix can be either $+1$ or $-1$ i.e. det$(U)=\pm1$.
\\ \indent Now, let is consider a subgroup of the bigger $U(2)$ consist of all $2\times2$ unitary matrices with det$(U)=1$.The subgroup is known as the $SU(2)$ group.So, the group of all $2\times2$ unitary matrices with unit determinant is known as the $SU(2)$ group. In general the elements of the unitary matrices are complex numbers but the order of the group is infinity so, the $SU(2)$ group or even $U(2)$ are continuous groups, sometimes we call it Lie Group. It can be shown that the maximum number of independent parameters required to characterize the $SU(2)$ group is 3 and in fact the number of continuous parameters are the number of generators of that group.$SU(2)$ group has three generators and it's a three parameter compact group.
Let, us consider the matrix $U\in SU(2)$ is of the following form.
\begin{equation}
U=\begin{pmatrix} a & b \\ c & d\end{pmatrix} 
\end{equation}
Where, the matrix elements $a,b,c,d$ are in general complex i.e. $a,b,c,d\in\mathbb{C}$.
By, the condition of unit determinant and unitarity we can write,$U^\dag=U^{-1}$ and det$(U)=(ad-bc)$=1.
\begin{equation}
\begin{pmatrix} a^* & c^* \\ b^* & d^*\end{pmatrix}=\begin{pmatrix} d & -b \\ -c & a\end{pmatrix}.
\end{equation}
compairing the two matrices elementwise we get the following conditions,
\begin{equation}
d=a^*,c^*=-b,c=-b^*,d^*=a.
\end{equation}
Then, the general form of the matrix $U\in SU(2)$ will be,
\begin{equation}
U=\begin{pmatrix} a & b \\ -b^* & a^*\end{pmatrix}
\end{equation}
The condition of unit determinant corresponds to the fact that $|a|^2+|b|^2=1$. Now wriiting $a=a_1+ib_1$ and $b=a_2+ib_2$ where, $a_1,a_2,b_1,b_2\in\mathbb{R}$ and we
get,$a_1^2+a_2^2+b_1^2+b_2^2=1$, which is a constraint equation and clearly signifies the fact that the maximum number of independent real parameters requires to characterize the SU(2) matrix elements being 3.
In general the matrix $U\in SU(2)$ can be written as,
\begin{equation}
U=\exp\left(\Vec{A}\cdot\frac{\hat{\Vec{\sigma}}}{2}\right)=\exp\left(\frac{\sum_{i=1}^3A_i\sigma_i}{2}\right)
\end{equation}
As, we can achieve any finite continuous transformation by sequential application of infinitesimal transformation we can identify $A_i;\hspace{0.1cm}i=1,2,3$ as the parameters of the SU(2) group and $\frac{\sigma_i}{2},\hspace{0.1cm}i=1,2,3$ as the generators of the SU(2) group.As, the number of continuous parameters required to characterize the group is equal to the number of generators of that group the group has 3 generators. The generators of the group are the usual $2\times2$ pauli matrices used to describe the spin $\frac{1}{2}$ particles.The Pauli matrices are,
\begin{equation}
\label{eq:11.2.6}
\sigma_x=\begin{pmatrix} 0 & 1\\ 1 & 0\end{pmatrix},\sigma_y=\begin{pmatrix} 0 & -i\\ i & 0\end{pmatrix},\sigma_z=\begin{pmatrix} 1 & 0\\ 0 & -1\end{pmatrix}
\end{equation}
The generators of the group are traceless hermitian matrices with determinant $-1$.
The lie algebra of the SU(2) is simply be the commutation relations obeyed by the generators of the group are given by,
\begin{eqnarray}
[\frac{\sigma_i}{2},\frac{\sigma_j}{2}]=\frac{\sigma_i}{2}\frac{\sigma_j}{2}-\frac{\sigma_j}{2}\frac{\sigma_i}{2}=i\epsilon_{ijk}\frac{\sigma_k}{2}\nonumber \\
\{\sigma_i,\sigma_j\}=\sigma_i\sigma_j+\sigma_j\sigma_i=2\delta_{ij}\nonumber \\
\sigma_i\sigma_j=i\epsilon_{ijk}\sigma_k+\delta_{ij}\forall i,j=1,2,3\nonumber \\ 
Tr[\sigma_i\sigma_j]=2\delta_{ij}\forall i,j=1,2,3
\end{eqnarray}
The generators follows the cyclic commutation relations and as a result the lie algebra of the $SU(2)$ group is closed and the parameter space is compact it's denoted by $\mathcal{S}^3$. But, the generators of the group are mutually noncommuting. So, we define an operator which commutes with all the gnerators of the $Su(2)$ group and the operator is called the \textbf{Casimir Operator}. The Casimir Operator of the $SU(2)$ group is defined as follows,
\begin{equation}
\sigma^2=\sigma_1^2+\sigma_2^2+\sigma_3^2=I_{2\times2}
\end{equation}
So, the Casimir Operator will commute with all the generators of the SU(2) Group and as a result we have, $[\sigma^2,\sigma_i]=0,\hspace{0.2cm} (i=1,2,3)$.For, the conventional choice in quantum mechanics we choose $\sigma^2,\sigma_3$ as the mutually commuting operators and any two mutually commuting operators have simultaneous eigenstates.The algebra of the spin-$\frac{1}{2}$ particle is identical to the SU(2) Lie algebra.Furthermore, we can define the raising and lowering operator from the pauli matrices as, $\sigma_{+}=\sigma_1+i\sigma_2$ and $\sigma_{-}=\sigma_1-i\sigma_{2}$ and $[\sigma^2,\sigma_{\pm}]=0$.
\vspace{0.5cm}

\textbf{\underline{The SU(3) Group}}
Just like we have defined the SU(2) Group we can define the group of all $3\times3$ unitary matrices with determinant $1$.It is a continuous group and a bigger symmetry group compared to $SU(2)$. The number of real continuous parameters required to represent the group element being equal to the number of generators of that group. In general the group $SU(n)$ has $(n^2-1)$ number of generators.
The SU(3) group has 8 generators.Let $U_{3\times3}\in SU(3)$ be an unitary matrix with determinant $+1$ then in general the matrix $U$ can be represented as follows,
\begin{equation}
U=\exp\left(\Vec{A}\cdot\frac{\Vec{\lambda}}{2}\right)=\exp\left(\sum_{i=1}^8A_i\frac{\lambda_i}{2}\right)
\end{equation}
So, the generators of this group are identified as $T_i=\frac{\lambda_i}{2},\hspace{0.2cm} (i=1,2,..,8)$ are called the Colour matrices which are traceless and hermitian with $\lambda_i,\hspace{0.2cm} (i=1,2,..,8)$ are the usual $3\times3$ Gell-Mann matrices given by,
\[\lambda_1=\begin{pmatrix}0 & 1 & 0\\1 & 0 & 0\\0 & 0 & 0\end{pmatrix} ; \lambda_2=\begin{pmatrix}0 & -i & 0\\i & 0 & 0\\0 & 0 & 0\end{pmatrix} ; \lambda_3=\begin{pmatrix}1 & 0 & 0\\0 & -1 & 0\\0 & 0 & 0\end{pmatrix} ;\lambda_4=\begin{pmatrix}0 & 0 & 1\\0 & 0 & 0\\1 & 0 & 0\end{pmatrix}\]
\[\lambda_5=\begin{pmatrix}0 & 0 & -i\\0 & 0 & 0\\i & 0 & 0\end{pmatrix} ; \lambda_6=\begin{pmatrix}0 & 0 & 0\\0 & 0 & 1\\0 & 1 & 0\end{pmatrix} ; \lambda_7=\begin{pmatrix}0 & 0 & 0\\0 & 0 & -i\\0 & i & 0\end{pmatrix} ; \lambda_8=\frac{1}{\sqrt{3}}\begin{pmatrix}1 & 0 & 0\\0 & 1 & 0\\0 & 0 & -2\end{pmatrix}\]
Some important properties of $\lambda_i$'s ; the generators of  $SU(3)$ are listed below,
\begin{gather}
Tr[\lambda_i]=0\nonumber \\
[\lambda_r,\lambda_s]=2if_{rst}\lambda_t\nonumber \\
f_{123}=1,f_{458}=f_{678}=\frac{\sqrt{3}}{2},f_{147}=f_{246}=f_{257}=f_{345}=f_{516}=f_{637}=\frac{1}{2};\nonumber \\
\{\lambda_r,\lambda_s\}=\frac{4}{3}\delta_{rs}+2d_{rst}\lambda_{t}\nonumber \\
d_{118}=d_{228}=d_{338}=-d_{888}=\frac{1}{\sqrt{3}},d_{448}=d_{558} =d_{668}=d_{778}=-\frac{1}{2\sqrt{3}},\nonumber \\
d_{146}=d_{157}=-d_{247}=d_{256}=d_{344}=d_{355}=-d_{366}=-d_{377}=\frac{1}{2};\nonumber \\
\lambda_r\lambda_s=\frac{2}{3}\delta_{rs}+if_{rst}\lambda_{t}+d_{rst}\lambda_{t},Tr[\lambda_r\lambda_s]=2\delta_{rs}
\end{gather}
Their commutators, anti-commutators and products involves two different three-index symbols or invariant tensors respectively $f_{ijk},d_{rst},\delta_{rs}$.Only the independent components of the completely anti-symmetric f's and the completely symmetric d's have been listed above so that, $d_{ijk}=d_{jik}$ which is completely symmetric and $f_{ijk}=-f_{jik}$ which is completely antisymmetric and the kronecker delta $\delta_{rs}$ is completely symmetric unlike the case of the pauli matrices whose commutation and anti-commutation relations only involve a completely antisymmetric levicivita symbol i.e. $\epsilon_{ijk}$ or, an invariant and completely antisymmetric tensor of rank 3.As we can see that the generators of the SU(3) group i.e. the Gell Mann matrices are mutually noncommuting, so again we will define the \textbf{Casimir Operator} for the SU(3) group. Interestingly, unlike the situation of the SU(2) Group here, we can define two different casimir operators denoted by $C_1$ and $C_2$ such that they commutes with all the generators of the SU(3) group i.e. the Gell-Mann matrices so that,$[C_1,\lambda_i]=[C_2,\lambda_i]=0,\hspace{0.2cm} (i=1,2,..,8)$. The Casimir Operators of the SU(3) group are defined as follows,
\begin{gather}
C_1(\lambda_i)=\sum_{i=1}^8\lambda_i^2 \\
C_2(\lambda_i)=\sum_{i=1}^8\sum_{j=1}^8\sum_{k=1}^8d_{ijk}\lambda_i\lambda_j\lambda_k
\end{gather}
As, we have done in case of SU(2) here we can choose the mutually commuting operators as $\lambda_3,\lambda_8$ and $C_1$ and can form the simultaneous eigenbasis from it.In the present context the SU(3) group has been used extensively.
\subsection{Appendix C: Insights of Lindblad Operators and the Master Equation}
\setcounter{equation}{0}
\renewcommand{\theequation}{11.3.\arabic{equation}}
The Lindblad type of Master equation required to study the time evolution of the density operator corresponding to the quantum system in presence of interaction with the surroundings is given by, \begin{equation}
\label{eq:11.3.1}
\boxed{\frac{\partial{\rho}}{\partial t}=-\frac{i}{\hbar}[\hat{H},\rho]+\gamma\sum_{i}\left(\Gamma_i^\dag\rho\Gamma_i-\frac{1}{2}\{\Gamma_i^\dag\Gamma_i,\rho\}\right)}
\end{equation}
The symbol used in the above equation have their usual significance. The right hand side of the equation consists of Lindblad operators or,quantum jump operators.The Lindblad superoperator models the environmental conditions that make up the open quantum system such as dephasing and relaxation. The operators $\lambda_i$ are also known as the collapse operator and it is important for deciding what the Lindblad superoperator describes. This operator is through which the environment couples to the system. Different collapse operators describe different aspects of the environment.$\gamma$ is an important constant that usually describes dephasing rate, rephasing rate, relaxation rate, etc. It is basically a corresponding rate for the coulping of the environment to the system. It also is important to the master equation. Note that whenever this constant is equal to zero, then we get the quantum Liovillian equation for a closed system without any environmental effects.One more  thing we  would like to add is that one can add as many Lindblad superoperators to the anticommutator to describe for different environmental conditions. One can describe for rephasing, another for dephasing, etc. It all depends on the environment of the quantum system.In summary, the Lindblad superoperator models an environmental coupling to the system. Without it, we get a model for a closed system with no environmental effects. That's why the second term in the right hand side is important. A superoperator is like an operator that acts on other linear operators same for the lindblad operatorts too.
\\ \indent Required to note that,If we consider only the first term on the right hand side of Lindblad Master equation~[\ref{eq:11.3.1}] we obtain the Liouville-von Neumann equation. This term is the Liouvillian and describes the unitary evolution of the density operator. The second term on the right hand side of the equation  is the Lindbladian and it emerges when we take the partial trace - a non-unitary
operation - of the degrees of freedom of reservior.The Lindbladian \cite{alexandre2011simple} describes the non-unitary evolution of the density
operator. By the interaction form adopted here  the physical meaning of the Lindblad operators can be understood: they represent the system S contribution to the System-Bath interaction remembering once more that the Lindblad equation was derived from the Liouville-von Neumann one by tracing the bath degrees of freedom.
If the Lindblad operators $\Gamma_i$ are Hermitian (observables), the Lindblad equation can be used to treat the measurement process. A simple application for a two level quantum system in this sense is the system Hamiltonian $\hat{H}_{S}\propto\hat{\sigma}_z$ where $\hat{\sigma_z}$ be the $z$ component of the $2\times2$ pauli matrices. when we want to measure one specific component of the spin ($\Gamma\propto\sigma_{\alpha},\alpha=x,y,z$ without any summation).If the Lindblad operators are non-hermitain then the master equation can be used to treat dissipation, decay and decoherence.For this type of scenario let us choose the system Hamiltonian same as in the previous case i.e. $\hat{H}_{S}\propto\hat{\sigma}_z$ where $\hat{\sigma_z}$ be the $z$ component of the $2\times2$ Pauli matrices with the Lindblad operators are taken as, $\Gamma\propto\hat{\sigma}_{-},\Big(\hat{\sigma}_{-}=\frac{\hat{\sigma}_x-i\hat{\sigma}_y}{2}\Big)$, where $\gamma$ be the spontaneous emission rate \cite{nielsen2003computaccao}.\vspace{0.5cm}
\subsection{\textbf{Appendix D: Homomorphic Mapping between SU(2) and SO(3) and  extension for SU(3)}}
\setcounter{equation}{0}
\renewcommand{\theequation}{11.4.\arabic{equation}}
We see that although the SU(2) group is different than the SO(3) group, the Lie algebras are homomorphic to each other so there exists a homomorphic mapping between the groups which seems to be disconnected from each other.The idea is that any unitary transformation initiated by some $2\times2$ unitary matrix corresponds to the SO(3) transformation of a vector in the 3 dimensional Euclidean Space which is equivalent to a rotation.
Let us consider the following situation where, a spinor $\ket{\psi}$ which can be represented by  a $2\times1$ coloum matrix (a two componnet vector) is undergoing an SU(2) transformation defined as,
\begin{equation}
\ket{\psi^\prime}=H\ket{\psi},\text{ where } H\in SU(2)
\end{equation}
We, can write $\ket{\psi}=\begin{pmatrix}\eta_1\\ \eta_2\end{pmatrix}$ and under the transformation the inner product or, the length of the vector remains invariant i.e. $\braket{\psi^\prime|\psi^\prime}=\braket{\psi|\psi}$, now lets find how a matrix transforms under the unitary transformation. Let us consider the transformation of the outer product;which itself is a $2\times2$ matrix given as $\ket{\psi}\bra{\psi}$ under the unitary transformation.So the outer product transforms to $\ket{\psi^\prime}\bra{\psi^\prime}$ redefining $\ket{\psi}\bra{\psi}$ by $M$ and $\ket{\psi^\prime}\bra{\psi^\prime}$ by $M^\prime$ we have, 
\begin{equation}
\ket{\psi^\prime}\bra{\psi^\prime}=M^\prime=H\ket{\psi}\bra{\psi}H^\dag=HMH^\dag=HMH^{-1}
\end{equation}
So, we get the transformation property of the the matrix $M$ under the unitary transformation of the spinor $\ket{\psi}$. The matrix M is a traceless hermitain matrix. Here, H is given as, 
\begin{equation}
H=\begin{pmatrix} a & b \\ -b^* & a^*\end{pmatrix}
\end{equation}
As,we know that any traceless $2\times2$ hermitian matrix can be written as a linear combination of the pauli matrices i.e. the generators of the SU(2) group. Then we can write for the matrix M,
\begin{equation}
\label{eq:11.4.4}
M=\Vec{r}\cdot\Vec{\hat{\sigma}}
\end{equation}
with,$\Vec{r}$ being the position vector of a point drawn from the origin of the euclidean coordinate system in 3 dimension.Putting the expressions of the pauli matrices given in equation~[\ref{eq:11.2.6}] in equation~[\ref{eq:11.4.4}] we can write,
\begin{equation}
\label{eq:11.4.5}
M=x\sigma_1+y\sigma_2+z\sigma_3=x\begin{pmatrix} 0 & 1\\ 1 & 0\end{pmatrix}+y\begin{pmatrix} 0 & -i\\ i & 0\end{pmatrix}+z\begin{pmatrix} 1 & 0\\ 0 & -1\end{pmatrix} \\
=\begin{pmatrix} z & x-iy\\ x+iy & -z\end{pmatrix}
\end{equation}
under the transformation the matrix $M$ changes to $M^\prime$ such that, $M^\prime=\Vec{r^\prime}\cdot\hat{\Vec{\sigma}}$ and similarly we can write,
\begin{equation}
M^\prime=x^\prime\sigma_1+y^\prime\sigma_2+z^\prime\sigma_3=x^\prime\begin{pmatrix} 0 & 1\\ 1 & 0\end{pmatrix}+y^\prime\begin{pmatrix} 0 & -i\\ i & 0\end{pmatrix}+z^\prime\begin{pmatrix} 1 & 0\\ 0 & -1\end{pmatrix} \\
=\begin{pmatrix} z^\prime & x^\prime-iy^\prime\\ x^\prime+iy^\prime & -z^\prime\end{pmatrix}
\end{equation}
From the transformation property of the traceless hermitian matrix $M$ we get,$det(M^\prime)=det(HMH^{-1})=det(M)det(H)det(H^{-1})=det(M)$.This gives,.
\begin{equation}
x^{\prime 2}+y^{\prime 2}+z^{\prime 2}=x^2+y^2+z^2
\label{eq:7.4.5}
\end{equation}
So, the length of the position vector remains invariant under the transformation.We, know that the length of a vector remains invariant under the orthogonal transformation which is possible only under the SO(3) transformation.
\\ \indent In fact we can show by using the condition $M^\prime=HMH^\dag$ along with equation~[\ref{eq:7.4.5}] that,$\Vec{r^\prime}=R\Vec{r}\Rightarrow{x_i^\prime=R_{ij}x_{j}}$ where, $R\in SO(3)$.So,every SU(2) transformation of a 2 component vector will correspond to a SO(3) transformation of a 3 component vector in the 3 dimensional Euclidean space which is equivalent to the rotation of a vector in Eucledian space.
More specifically we can write the unitary matrix $H$ as,
\begin{equation}
H=\exp{(i\Vec{\sigma}\cdot{\hat{n}}\frac{\theta}{2})}
\end{equation}
with,$\hat{\sigma}$ be the pauli matrices and $\theta$ be the angle of rotation for the vector $\ket{\psi}$ in the two dimemsional spinor space and $\hat{n}$ be the unit vector along the direction of rotation in the spinor space.For example if the direction of rotation of rotation is along the z direction then $\hat{n}=\hat{z}$ then the unitary matrix $H$ will be, $H=\exp{(i\sigma_z\frac{\theta}{2})}$ and so on.The corresponding SO(3) transformation matrix $R$ can be written as, 
\begin{equation}
R=\exp{(i\Vec{J}\cdot\hat{n}\theta)}
\end{equation}
so,the SU(2) transformation of the vector $\ket{\psi}$ in the spinor space or, the rotation of the vector $\ket{\psi}$ in the spinor space with respect to the z direction corresponds to the rotation of the 3 dimensional vector here, in our case is the position vector of a point(although it can be in general true for any vector) in the three dimensional Euclidean space.Then,
\begin{equation}
H=\exp{(i\sigma_z\frac{\theta}{2})}=\begin{pmatrix} e^{i\frac{\theta}{2}} & 0 \\ 0 & e^{-i\frac{\theta}{2}}\end{pmatrix}\longleftrightarrow R_z(\theta)=\begin{pmatrix}\cos{\theta} & \sin{\theta} & 0\\-\sin{\theta} & \cos{\theta} & 0\\ 0 & 0 & 1\end{pmatrix}
\end{equation}
It means, that under the unitary transformation defined by,$\ket{\psi^{\prime}}=\exp{(i\sigma_z\frac{\theta}{2})}\ket{\psi}$ will correspond to the rotation with respect to the z direction in the Euclidean space i.e. $\begin{pmatrix}x^{\prime} \\ y^{\prime} \\ z^{\prime}\end{pmatrix}=\begin{pmatrix}\cos{\theta} & \sin{\theta} & 0\\-\sin{\theta} & \cos{\theta} & 0\\ 0 & 0 & 1\end{pmatrix}\begin{pmatrix} x \\ y\\ z\end{pmatrix}$
And, similarly we can write the same for the rotation in the spinor space with respect to the X and Y direction along with the rotation about z direction  given by,
\begin{eqnarray}
H=\exp{(i\sigma_z\frac{\theta}{2})}=\begin{pmatrix} e^{i\frac{\theta}{2}} & 0 \\ 0 & e^{-i\frac{\theta}{2}}\end{pmatrix}\longleftrightarrow R_z(\theta)=\exp{(iJ_z\theta)}=\begin{pmatrix}\cos{\theta} & \sin{\theta} & 0\\-\sin{\theta} & \cos{\theta} & 0\\ 0 & 0 & 1\end{pmatrix} \\
H=\exp{(i\sigma_y\frac{\theta}{2})}=\begin{pmatrix} \cos{\frac{\theta}{2}} & \sin{\frac{\theta}{2}}\\ -\sin{\frac{\theta}{2}} & \cos{\frac{\theta}{2}}\end{pmatrix}\longleftrightarrow R_y(\theta)=\exp{(iJ_y\theta)}=\begin{pmatrix}\cos{\theta} & 0 & \sin{\theta}\\0 & 1 & 0\\ \sin{\theta} & 0 & \cos{\theta}\end{pmatrix} \\
H=\exp{(i\sigma_x\frac{\theta}{2})}=\begin{pmatrix} \cos{\frac{\theta}{2}} & i\sin{\frac{\theta}{2}}\\ i\sin{\frac{\theta}{2}} & \cos{\frac{\theta}{2}}\end{pmatrix}\longleftrightarrow R_x(\theta)=\exp{(iJ_x\theta)}=\begin{pmatrix}1 & 0 & 0\\0 & \cos{\theta} & \sin{\theta}\\ 0 & -\sin{\theta} & \cos{\theta}\end{pmatrix}
\end{eqnarray}
In general we can write, $H\in SU(2)=\exp{(i\Vec{\sigma}\cdot\hat{n}\frac{\theta}{2})}\longleftrightarrow R\in SO(3)=\exp{(i\Vec{J}\cdot\hat{n}\theta)}$.
Where,$J$ be the generators of rotation of SO(3) group defined by,
\begin{equation}
J_k(\theta)=\frac{1}{i}\frac{dR_k(\theta)}{d\theta}\Big|_{\theta=0};\text{ where } k=x,y,z
\end{equation}
Simple calculation shows that the generators of the group can be written using the equation (7.4.14) as,
\begin{eqnarray}
J_z=\frac{1}{i}\frac{dR_z(\theta)}{d\theta}\Big|_{\theta=0}=\begin{pmatrix}0 & -i & 0\\i & 0 & 0\\0 & 0 & 0\end{pmatrix}\nonumber \\
J_x=\frac{1}{i}\frac{dR_x(\theta)}{d\theta}\Big|_{\theta=0}=\begin{pmatrix}0 & 0 & 0\\0 & 0 & -i\\0 & i & 0\end{pmatrix},\nonumber \\
J_y=\frac{1}{i}\frac{dR_y(\theta)}{d\theta}\Big|_{\theta=0}=\begin{pmatrix}0 & 0 & i\\0 & 0 & 0\\-i & 0 & 0\end{pmatrix}
\end{eqnarray}
Where, the generators of the So(3) group $J_i,\text{ where }(i=x,y,z)$ satisfies the cyclic commutation relation given by,
\begin{equation}
[J_k,J_m]=i\epsilon_{kmn}J_n.
\end{equation}
The lie algebra of the generators of SO(3) being closed and the commutation relations obeyed by the generators $J_i$ is identical to the Lie algebra of SU(2) Group which establishes the homomorphic mapping between the two groups SO(3) and SU(2).\vspace{0.3cm}
\\ \textbf{An SU(2) transformation on $\ket{\psi}=\begin{pmatrix}\eta_1\\ \eta_2\end{pmatrix} \equiv$ SO(3) transformation on $\begin{pmatrix}x\\y\\z\end{pmatrix}$}.
\\ \indent\textbf{\underline{Extension for SU(3):}} The similar idea can be extended in the context of the SU(3) group which is a bigger group than SU(2). The idea is to establish a homomorphic mapping between the groups SU(3) and SO(8) Groups.
Just like we have considered the unitary transformation of the 2 component vector $\ket{\psi}$ here, we will consider the unitary transformation of a three compomnent nonunit vector ray $\ket{\Psi}=\begin{pmatrix}\eta_1\\\eta_2\\\eta_3\end{pmatrix}$ in the hilbert space $\mathcal{H}^3$ defined as follows,
\begin{equation}
\label{eq:7.4.17}
\ket{\Psi^\prime}=U\ket{\Psi};\text{ where } U\in SU(3)
\end{equation}
Now if we consider the transformation of the projection operator $\ket{\Psi}\bra{\Psi}$ represented by a $3\times3$ matrix say,$M$ under the unitary transformation defined by equation~[\ref{eq:7.4.17}] then we may write similarly,
\begin{equation}
\ket{\Psi^\prime}\bra{\Psi^\prime}=M^\prime=U\ket{\Psi}\bra{\Psi}U^\dag=UMU^\dag=UMU^{-1}
\end{equation}
Keeping analogy with the SU(2) situation we can say that any $3\times3$ traceless Hermitian Matrix can also be written as the linear combination of the generators of SU(3) i.e. the Gell Mann matrices which are traceless hermitian. So, we can write for the matrix M,
\begin{equation}
M=\Vec{A}\cdot\Vec{\lambda}=\sum_{i=1}^8A_i\lambda_i
\end{equation}
Where,$\Vec{A}$ be any arbitrary octet vector in the eight dimensional Euclidean space i.e.$\Vec{A}\in\mathbb{R}^8$.From the transformation property of the traceless hermitian matrix $M$ we get,$det(M^\prime)=det(UMU^{-1})=det(M)det(U)det(U^{-1})=det(M)$ along with $M^\prime=\Vec{A^\prime}\cdot\Vec{\lambda}=\sum_{i=1}^8A_i^\prime\lambda_i$ This gives,
\begin{equation}
\sum_{i=1}^8{A_i^\prime}^2=\sum_{i=1}^8A_i^2
\end{equation}
We get the condition for the invariance of the length of the octet vector $\Vec{A}$ under the unitary transformation, the smimilar result has been obtained in the case of a 3 dimensional vector as a result of SU(2) transformation.So, we can say that,
\\ \indent\textbf{An SU(3) transformation on $\ket{\Psi}=\begin{pmatrix}\eta_1\\ \eta_2\\ \eta_3\end{pmatrix} \equiv$ SO(8) transformation on $\begin{pmatrix}A_1\\A_2\\A_3\\A_4\\A_5\\A_6\\A_7\\A_8\end{pmatrix}$}.\vspace{0.3cm}

On summarising the above results we can conclude that,
\begin{equation}
\ket{\Psi^\prime}=U\ket{\Psi}\longleftrightarrow \Vec{A^\prime}=\textbf{R}\Vec{A}.\text{ Where } U\in SU(3)\text{ and } R\in SO(8).
\end{equation}
where,R be the $8\times8$ transformation matrix which, gives the correspondence of the SU(3) transformation of a three component vector $\ket{\Psi}$ with the rotation of a vector
$\Vec{A}$ in the eight dimensional Euclidean Space which is equivalent to a SO(8) transformation of an octet vector $\Vec{A}$ with the components of the vector \textbf{A} transforms obeying the equation,
\begin{equation}
A_i^\prime=R_{ij}A_j
\end{equation}
With,$R_{ij}$ be the matrix elements of the transformation matrix $R_{8\times8}\in SO(8)$. But the homomorphism for the SU(3) case is more intricate. In general we can write $R=\exp{(iL_i\phi_i)}=\exp{(i\Vec{L}\cdot\Vec{\phi})}$.Where, we can identify $L_i$ as the generators of the SO(8) Group and $\phi_i$ as the parameters of the Group.But there are $\binom{8}{2}=28$ generators for the SO(8) Group and as a result 28 parameters.
All real orthogonal unimodular $8\times8$ matrices taken together form the 28 dimensional group SO(8); the matrices of the octet
representation of SU(3) are a `very small' eight-dimensional subset of SO(8), in fact a subgroup.
\subsection{\textbf{Appendix E: Interaction Picture in Quantum Mechanics:}}
\setcounter{equation}{0}
\renewcommand{\theequation}{11.5.\arabic{equation}}
Let, us briefly introduce the idea of the interaction picture in order to understand the idea of the Lindblad Type of master equation describing the time evolution of the Density operator of a quantum system interacting with the environment which is obtained as a special case of the Markovian Master Equation along with the Local time approximation also known as the Born Markovian Approximation.
We have some ideas over the Schr$\ddot{O}$dinger picture and the Heisenburg Picture and about the equivalence of the pictures. The central idea of this formalism's are to be noted.
\begin{itemize}
    \item In Schrodinger picture the observables i.e. the operators are time independent but the state vector evolves in time.The time evolution of the state vector id governed by the time dependent Schrodinger equation given by,
    \begin{equation}
    i\hbar\frac{d}{dt}\ket{\psi^{(s)}(t)}=\hat{H}^{(s)}(t)\ket{\psi^{(s)}(t)}
    \end{equation}
    where, $\ket{\psi^{(s)}(t)}$ be the state vector in the schrodinger picture and the hamiltonian being in general time dependent.
    \item In the Heisenburg Picture the Operators are time dependent and the State Vectors are freezed i.e. independent of time.The time evolution of the operator say, $\hat{A}(t)$ is governed by the Heisenburg's equation of motion given by,
    \begin{equation}
        \frac{d}{dt}\hat{A}(t)=\frac{1}{i\hbar}[\hat{A}(t),\hat{H}^{(H)}(t)]+\frac{\partial{\hat{A}(t)}}{\partial{t}}
    \end{equation}
    \item The intercation Picture where the hamiltonian can be splitted up as the sum of the free hamiltonian and the interaction hamiltonian,both the state vector and the operator are time dependent.
\end{itemize}
As, in the interaction picture both the state vector and the operators are time dependent we need two separate equations governing the time evolution of the State vector and that for the operator as well. Here, for simplicity we will assume the hamiltonian of the system to be independent of time in the schrodinger picture and under this condition in the Heisenburg Picture the operators are defined by,
\begin{equation}
\hat{A}_{H}(t)=e^{i\hat{H}^{(s)}t}\hat{A_s}e^{-i\hat{H}^{(s)}t}
\end{equation}
here, it is convenient to use the natural unit system
i.e.$\hbar=1$.$\hat{A}_s$ be the operator in the schrodinger picture which is independent of time.
The hamiltonian in the schrodinger picture is written as the sum of the Free hamiltonian and the interaction hamiltonian.So, we write,
\begin{equation}
\hat{H}=\hat{H}_0^{(S)}+\hat{H}_{int}^{(S)}
\end{equation}
Here, $\hat{H}_0^{(S)}$ and $\hat{H}_{int}^{(S)}$ be the Free hamiltonian and the interaction hamiltonian in the Schrodinger picture. Now in the schrodinger picture with a time independent hamiltonian we have,
\begin{equation}
\ket{\psi^{(s)}(t)}=e^{-i\hat{H}^{(s)}t}\ket{\psi^{(s)}(0)}
\end{equation}
In the interaction picture the state vector is denoted by $\ket{\psi(t)}^{IP}$ and it is defined by,
\begin{equation}
\ket{\psi(t)}^{IP}=e^{i\hat{H}_0^{(s)}t}\ket{\psi^{(s)}(t)}=e^{i\hat{H}_0^{(s)}t}e^{-i\hat{H}^{(s)}t}\ket{\psi^{(s)}(0)}
\end{equation}
we can in general write, $\ket{\psi(0)}^{IP}=\ket{\psi^{(s)}(0)}=\ket{\psi^{(H)}(0)}$. So we can write,
\begin{equation}
\ket{\psi(t)}^{IP}=e^{i\hat{H}_0^{(s)}t}\ket{\psi^{(s)}(t)}=e^{i\hat{H}_0^{(s)}t}e^{-i\hat{H}^{(s)}t}\ket{\psi(0)}^{IP}
\end{equation}
using the equation (7.5.4) we can write, $\ket{\psi(t)}^{IP}=e^{-i\hat{H}^{(s)}_{int}t}\ket{\psi(0)}^{IP}$
Now, the operator in the Interaction picture is defined as,
\begin{equation}
\begin{split}
\hat{O}^{(IP)}(t)&=e^{i\hat{H}_0^{(s)}t}\hat{O}^{(s)}e^{-i\hat{H}_0^{(s)}t} \\
&= e^{i\hat{H}_0^{(s)}t}e^{-i\hat{H}^{(s)}t}\hat{O}^{(H)}(t)e^{i\hat{H}^{(s)}t}e^{-i\hat{H}_0^{(s)}t}
\end{split}
\end{equation}
where, we have used $\hat{O}^{(s)}=e^{-i\hat{H}^{(s)}t}\hat{O}^{(H)}(t)e^{i\hat{H}^{(s)}t}$, the relation between the operators in the schrodinger and hisenburg picture. As we can see from equation (7.5.8) the operator in the interaction picture is time dependent.
Here, also we can write that, $\hat{O}^{(IP)}(0)=\hat{O}^{(s)}=\hat{O}^{(H)}(0)$
So, far we have defined,

\begin{equation}
\boxed{\ket{\psi(t)}^{IP}=e^{i\hat{H}_0^{(s)}t}\ket{\psi^{(s)}(t)}=e^{i\hat{H}_0^{(s)}t}e^{-i\hat{H}^{(s)}t}\ket{\psi(0)}^{IP} \\
\text{ and }\hat{O}^{(IP)}(t)=e^{i\hat{H}_0^{(s)}t}\hat{O}^{(s)}e^{-i\hat{H}_0^{(s)}t}}
\end{equation}
The equation governing the time evolution of the state vector in the interaction picture is derived as follows,
\begin{equation}
\begin{split}
i\frac{\partial}{\partial{t}}\ket{\psi(t)}^{IP}&=i\frac{\partial}{\partial{t}}\Big[e^{i\hat{H}_0^{(s)}t}e^{-i\hat{H}^{(s)}t}\ket{\psi(0)}^{IP}\Big] \\
&= -\hat{H}_0^{(s)}e^{i\hat{H}_0^{(s)}t}e^{-i\hat{H}^{(s)}t}\ket{\psi}^{H}+e^{i\hat{H}_0^{(s)}t}\hat{H}^{(s)}e^{-i\hat{H}^{(s)}t}\ket{\psi}^{H} \\
&= -\hat{H}_0^{(s)}\ket{\psi(t)}^{IP}+e^{i\hat{H}_0^{(s)}t}\hat{H}^{(s)}e^{-i\hat{H}_0^{(s)}t}e^{i\hat{H}_0^{(s)}t}e^{-i\hat{H}^{(s)}t}\ket{\psi}^{H} \\
&= -\hat{H}_0^{(IP)}\ket{\psi(t)}^{IP}+e^{i\hat{H}_0^{(s)}t}\hat{H}^{(s)}e^{-i\hat{H}_0^{(s)}t}\ket{\psi(t)}^{IP}  \\
&= -\hat{H}_0^{(IP)}\ket{\psi(t)}^{IP}+\hat{H}^{(IP)}\ket{\psi(t)}^{IP} \\
&= \hat{H}^{(IP)}_{int}\ket{\psi(t)}^{IP}
\end{split}
\end{equation}
So,the equation that describes the time evolution of the state vector is given by,
\begin{equation}
i\frac{\partial}{\partial{t}}\ket{\psi(t)}^{IP}=\hat{H}^{(IP)}_{int}\ket{\psi(t)}^{IP}
\end{equation}
Here, we have used, $\hat{H}_{int}^{(IP)}(t)=e^{i\hat{H}_0^{(s)}t}\hat{H}_{int}^{(s)}e^{-i\hat{H}_0^{(s)}t}$ which gives the Interaction hamiltonian in the interaction picture related to the Interaction hamiltonian in the schrodinger picture by an unitary transformation.
\\ \indent Similarly we can derive the equation governing the time evolution of the operator in the interaction picture which is time dependent.
\begin{equation}
\begin{split}
\frac{\partial}{\partial{t}}\hat{O}^{(IP)}(t)&=\frac{\partial}{\partial{t}}\Big[e^{i\hat{H}_0^{(s)}t}\hat{O}^{(s)}e^{-i\hat{H}_0^{(s)}t}\Big] \\
&= i\hat{H}_0^{(s)}e^{i\hat{H}_0^{(s)}t}\hat{O}^{(s)}e^{-i\hat{H}_0^{(s)}t}-ie^{i\hat{H}_0^{(s)}t}\hat{O}^{(s)}e^{-i\hat{H}_0^{(s)}t}\hat{O}^{(s)} \\
&= i\hat{H}_0^{(s)}\hat{O}^{(IP)}(t)-i\hat{O}^{(IP)}(t)\hat{H}_0^{(s)} \\
&= \frac{1}{i}[\hat{O}^{(IP)}(t),\hat{H}_0^{(IP)}(t)]
\end{split}
\end{equation}
Here, we have used that $\hat{H}_0^{(IP)}(t)=e^{i\hat{H}_0^{(s)}t}\hat{H}_0^{(s)}e^{-i\hat{H}_0^{(s)}t}=\hat{H}_0^{(s)}e^{i\hat{H}_0^{(s)}t}e^{-i\hat{H}_0^{(s)}t}=\hat{H}_0^{(s)}$.
Upon summarising the results obtained for the time evolution of the state vector and the operator in the Interaction picture are as follows,
\begin{gather}
\label{eq:7.5.13b}
i\frac{\partial}{\partial{t}}\ket{\psi(t)}^{IP}  =\hat{H}^{(IP)}_{int}(t)\ket{\psi(t)}^{IP} \\
\frac{\partial}{\partial{t}}\hat{O}^{(IP)}(t) 
=\frac{1}{i}[\hat{O}^{(IP)}(t),\hat{H}_0^{(IP)}(t)]
\end{gather}
Here, the free hamiltonian in the interaction picture i.e. $\hat{H}_0^{(IP)}$ is time dependent.
In our present context the Operator is the global density operator of the super-system i.e.(system+surrounding) and the density operator of the system is obtained upon taking the partial trace of the Global density operator with respect to the Bath states (here, the surrounding can be taken as a Bath). The time evolution of the Global Density operator corresponding to the supersystem will be described by the equation just by replacing $\hat{O}$ by the Global density operator $\hat{\rho}_{SB}(t)$ in the equation~[\ref{eq:7.5.13b}].We get,
\begin{equation}
\frac{\partial}{\partial{t}}\hat{\rho}_{SB}^{(IP)}(t) =\frac{1}{i}[\hat{\rho}_{SB}^{(IP)}(t),\hat{H}_0^{(IP)}(t)]
\end{equation}
With, $\hat{\rho}_{SB}^{(IP)}(t)$ is defined by,
\begin{equation}
\hat{\rho}_{SB}^{(IP)}(t)=e^{i\hat{H}_0^{(s)}t}\hat{\rho}_{SB}^{(s)}e^{-i\hat{H}_0^{(s)}t}
\end{equation}
Here, we have used the above equation as the starting point  to derive the Lindblad Master Equation using the Born-Markovian Approximation.The density operator corresponding to the system is obtained upon taking the partial trace of the Global density operator $\hat{\rho}_{SB}^{(IP)}(t)$ with respect to the Bath states defined as follows,
\begin{equation}
\hat{\rho}_{s}^{IP}(t)=Tr_{B}[\hat{\rho}_{SB}^{(IP)}(t)]
\end{equation}
As, a result of the Born-Markov Approximation the time evolution of the system's Density operator $\hat{\rho}_s(t)$ is governed by the Lindblad Master Equation given below,
\begin{equation}
\boxed{\frac{\partial{\hat{\rho}_{s}(t)}}{\partial t}=-\frac{i}{\hbar}[\hat{H},\hat{\rho}_{s}]+\sum_{i=1}^8\left(\Gamma_i^\dag\hat{\rho}_{s}\Gamma_i-\frac{1}{2}\{\Gamma_i^\dag\Gamma_i,\hat{\rho}_{s}\}\right)}
\end{equation}
The derivation of the Master Equation has been skipped as it requires the detailed knowledge of generators related to Quantum Dynamical Semigroups \cite{lindblad1976generators}.
Alternatively, we can define the state vector and the operator in the interaction picture as follows,
\begin{gather*}
\ket{\psi(t)}^{IP} 
=e^{i\hat{H}_{int}^{(s)}t}\ket{\psi^{(s)}(t)}=e^{i\hat{H}_{int}^{(s)}t}e^{-i\hat{H}^{(s)}t}\ket{\psi(0)}^{IP} \\
\hat{O}^{(IP)}(t)  =e^{i\hat{H}_{int}^{(s)}t}\hat{O}^{(s)}e^{-i\hat{H}_{int}^{(s)}t}
\end{gather*}
Similar calculations leads to the equations governing the time evolution of the state vector and the operator in the interaction picture given by,
\begin{gather*}
i\frac{\partial}{\partial{t}}\ket{\psi(t)}^{IP}  =\hat{H}^{(IP)}_{0}(t)\ket{\psi(t)}^{IP} \\
\frac{\partial}{\partial{t}}\hat{O}^{(IP)}(t) 
=\frac{1}{i}[\hat{O}^{(IP)}(t),\hat{H}_{int}^{(IP)}(t)]
\end{gather*}
Here, the standard definition of the state vector and the Operators in the Interaction picture given by the set of equations~[\ref{eq:7.5.13b}] has been used, but ultimately it's a choice which definition will be convenient to use for the purpose of calculation.\vspace{0.3cm}
\\ \indent \textbf{\underline{Remarks regarding the equivalence of the three pictures:}} Although the three pictures i.e. Schrodinger Picture, Heisenburg Picture and The Interaction Picture the quantum mechanical expectation values at three pictures are identical.
Let us check the equivalence of three pictures by evaluating the expectation value of the operator in the interaction picture as follows,
\begin{equation}
\begin{split}.
\braket{\hat{A}(t)}_{IP}&=\braket{\psi^{IP}(t)|\hat{A}^{IP}(t)|\psi^{IP}(t)} \\
&= \braket{\psi^{(s)}(t)|e^{-i\hat{H}_0^{(s)}t}e^{i\hat{H}_0^{(s)}t}\hat{A}^{(s)}e^{-i\hat{H}_0^{(s)}t}e^{i\hat{H}_0^{(s)}t}|\psi^{(s)}(t)} \\
&= \braket{\psi^{(s)}(t)|\hat{A}^{(s)}|\psi^{(s)}(t)}=\braket{\hat{A}}_{s} \\
&= \braket{\psi^{H}(0)|\hat{A}_{H}(t)|\psi^{(H)}(0)}=\braket{\hat{A}(t)}_{H}
\end{split}
\end{equation}
here, we have used that,$\hat{A}^{(s)}=e^{-i\hat{H}^{(s)}t}\hat{A}^{(H)}(t)e^{i\hat{H}^{(s)}t}$ in the above equation.Then we can conclude from the above calculation that,
\begin{equation}
\boxed{\braket{\hat{A}(t)}_{IP}=\braket{\hat{A}(t)}_{H}=\braket{\hat{A}}_{s}}
\end{equation}
Which establishes the equivalence of three different pictures as the expectation values of the operator $\hat{A}$ in all the three pictures are identical and what we do measure experimentally is the expectation value of some observable.
\bibliographystyle{plain}
\bibliography{citation}
\end{document}